\documentclass[12pt]{article}
\usepackage{amsmath}
\usepackage[colorlinks,allcolors=blue,linktocpage]{hyperref}
\usepackage[nameinlink]{cleveref} % make \cref look like \autoref
\makeatletter
\def\@seccntformat#1{\@ifundefined{#1@cntformat}%
   {\csname the#1\endcsname\quad}%   default
   {\csname #1@cntformat\endcsname}% enable individual control
}
\makeatother
\usepackage{abstract}

\usepackage{bbm}
\usepackage{graphicx,psfrag,epsf}
\usepackage{enumerate}
\usepackage[isbn=false,url=false,eprint=false]{biblatex}
\usepackage{authblk}
\usepackage{enumitem}
\usepackage{mathrsfs}  
\usepackage{algorithm}
\usepackage{algpseudocode}
\usepackage{booktabs}
\usepackage{multirow}
\usepackage{spreadtab}
\usepackage{caption}
\usepackage{lipsum}
\usepackage{hyperref}
\usepackage{cleveref}

\usepackage{tocbibind} %for appendix/SI numbering
\usepackage[toc,page]{appendix} %for appendix/SI numbering
\usepackage{setspace}
\usepackage{afterpage}
\usepackage{rotating}
\usepackage{hyperref}
\usepackage{tabularx}
\usepackage{threeparttable}
\usepackage{tabu}
\usepackage{siunitx}
\usepackage{dsfont}

\usepackage{setspace}
\usepackage{sectsty}
\sectionfont{\fontsize{16}{14}\selectfont}
\subsectionfont{\fontsize{14}{14}\selectfont}

\date{ }

\begin{document}
\def\spacingset#1{\renewcommand{\baselinestretch}%
{#1}\small\normalsize} \spacingset{1}

%%%%%%%%%%%%%%%%%%%%%%%%%%%%%%%%%%%%%%%%%%%%%%%%%%%
\title{A Bayesian Finite Mixture Model Approach \\
for Mixed-type Data Clustering and Variable Selection \\ with Censored Biomarkers}

\author[1]{\small{Yueting Wang}}
\author[1]{\small{Shu Wang}}
\author[1,2]{\small{Jonathan G. Yabes}}
\author[1,2]{\small{Chung-Chou H. Chang}}

\affil[1]{\footnotesize{Department of Biostatistics and Health Data Science, School of Public Health, University of Pittsburgh}}
%\affil[2]{\small{United States Food and Drug Administration}}
\affil[2]{\footnotesize{Department of Medicine, School of Medicine, University of Pittsburgh}}
\renewcommand\Authands{ and }

\maketitle\thispagestyle{empty} 
\label{firstpage}
\begin{abstract} 

\begin{spacing}{1.2}

Clustering mixed-type data remains a major challenge in biomedical research to uncover clinically meaningful subgroups within heterogeneous patient populations. Most existing clustering methods impose restrictive assumptions like local independence, fail to accommodate censored biomarkers, or unable to quantify variable importance. We propose a Bayesian finite mixture model (BFMM) clustering framework that addresses these limitations. BFMM flexibly models both continuous and categorical variables, incorporates three covariance structures to capture cluster-specific dependencies among continuous features, and handles censored observations through likelihood-based imputation. To facilitate feature prioritization, BFMM uses spike-and-slab priors to estimate variable importance on a continuous 0-1 scale. Simulation studies demonstrate that BFMM outperforms existing methods in clustering accuracy, particularly given strong within-cluster correlation or censored variables, and reliably distinguishes informative features from noise under varying conditions. We applied BFMM to two real-world datasets: (1) the SENECA cohort integrating electronic health records from patients with Sepsis; and (2) the EDEN randomized trial of patients with acute lung injury. In both settings, BFMM identified clinically interpretable phenotypes and revealed variable-specific contributions to subgroup differentiation. In the EDEN trial, it also uncovered evidence of treatment heterogeneity. These findings validate BFMM as an effective, interpretable, and practically useful clustering tool for complex biomedical datasets.

\end{spacing}
\end{abstract}

\noindent%
\vfill

\newpage\pagenumbering{arabic}
\spacingset{1.73} % DON'T change the spacing!

%%%%%%%%%%%%%%%%%%%%%%%%%%%%%%%%%%%%%%%%%%%%%%%%%%%
%%% Section 1 - Introduction %%%
%%%%%%%%%%%%%%%%%%%%%%%%%%%%%%%%%%%%%%%%%%%%%%%%%%%
\section{Introduction}
Cluster analysis plays a central role in clinical and biomedical research, enabling discovery of patient subgroups with shared characteristics that can improve risk stratification, guide prognosis, and inform treatment. However, clinical datasets present challenges that conventional clustering methods cannot adequately address. First, they often combine continuous and categorical features, including biomarkers, physiologic measures, medications, and demographics. Second, continuous variables are frequently correlated in ways that differ across latent subgroups. Third, many biomarkers are censored by limits of detection (LODs), a pervasive issue in studies of inflammation, infection, and organ dysfunction. Moreover, the applied researchers often require not only patient clusters but also interpretable assessments of which variables are most important in driving subgroup differentiation, which is not easily obtained by many standard clustering algorithms.

Existing clustering methods fall short in handling this combination of features. Distance-based approaches typically require variables to be of the same type or rely on arbitrary distance metrics that may obscure underlying structure. Model-based approaches may assume independence among variables, ignore censoring, or fail to quantify variable importance in an interpretable manner. Moreover, many clustering methods treat continuous and categorical variables separately, or require transformations that complicate interpretation. Together, these limitations reduce the translational utility of clustering in real-world clinical settings. 

To address these challenges, we propose a Bayesian finite mixture model (BFMM) framework that is tailored for clustering clinical datasets with mixed variable types, flexible dependence structures, and censored continuous measurements. BFMM jointly models continuous and categorical variables within a unified likelihood-based framework. For continuous variables, we assume that observations within each cluster follow a multivariate normal distribution, allowing for dependency structures that vary across clusters. We incorporate three parsimonious covariance structures that enables BFMM to adapt to settings where clinical variables exhibit varying degrees of correlation both within and across clusters.

To handle censoring in continuous biomarkers, BFMM embeds a Gibbs sampling routine that imputes the censored values beyond LODs during each sampling iteration. These censored values can contain critical biological signals that reflect early disease progression or treatment response. By retaining rather than discarding such data, BFMM ensures that phenotypic subgroups are defined using the full spectrum of patient information, extending the translational value.

Variable importance is quantified through spike-and-slab priors incorporated in BFMM: a Gaussian-based prior for continuous variables and a Dirichlet-based prior for categorical variables. These priors produce interpretable weights between 0 and 1 that reflect each variable’s contribution to defining the clusters. To ensure robust variable selection, we develop a novel data-driven procedure to calibrate the variance ratio between the slab and spike components of the Gaussian prior, which is critical for discriminating relevant from irrelevant features. This interpretability transforms clustering from an exploratory statistical exercise into actionable knowledge that directly informs clinical reasoning and decision making.

Our BFMM approach is motivated and validated by two applications: identifying sepsis phenotypes in a large EHR-based cohort and characterizing patient subgroups with acute liver injury from a randomized clinical trial, where an initial enteral feeding intervention showed heterogeneity in subgroup-specific treatment effects. In both cases, existing clustering methods cannot simultaneously accommodate the mixed variable types, within-cluster dependencies, censoring, and variable importance quantification, making BFMM a suitable and innovative solution.

The remainder of this manuscript is organized as follows. Section 2 introduces the model formulation and estimation algorithm. Section 3 presents simulation studies evaluating clustering performance and variable selection. Section 4 applies BFMM to the motivating clinical datasets. Section 5 discusses findings, limitations, and future directions. A detailed review of existing work related to mixed data clustering methods is provided in Appendix~\ref{si.review}.

%%%%%%%%%%%%%%%%%%%%%%%%%%%%%%%%%%%%%%%%%%%%%%%%%%%
%%% Section 2 - Proposed BFMM[VVV] %%%
%%%%%%%%%%%%%%%%%%%%%%%%%%%%%%%%%%%%%%%%%%%%%%%%%%%
\section{Proposed Bayesian Finite Mixture Model (BFMM)}

%%% 2.1 %%%
\subsection{Notation and Model Setup}
We begin by introducing the notation and model formulation for the proposed BFMM, designed for clustering data with mixed variable types and flexible dependence structures. We define three variants: BFMM[EEI], BFMM[EEE], and BFMM[VVV], which respectively impose increasing flexibility in modeling the covariance structure of continuous variables.

Let the dataset consist of $n$ i.i.d. observations, where the $i$th observation is denoted by $\boldsymbol{x}_i = (x_{i1},\dots,x_{iM})^T$, comprising $M$ variables. Among these, the first $q$ are continuous and the remaining $M-q$ are categorical. Denote the continuous portion of the $\boldsymbol{x}_i$ as $\boldsymbol{u}_i = (x_{i1}, x_{i2},\dots,x_{iq})^T$, and define the full continuous data matrix as $\boldsymbol{U} = (\boldsymbol{u}_1, \boldsymbol{u}_2,\dots,\boldsymbol{u}_n)^T \in \mathbbm{R}^{n \times q}$, with each continuous variable standardized to have zero marginal mean. For each categorical variable $m=q+1,\dots,M$, let $x_{im} \in \{1,\dots,L_m\}$ denote its observed category, where $L_m$ is the number of levels.

We assume that each observation arises from one of $G$ latent clusters, and introduce $\boldsymbol{z}_i = (z_{i1}, z_{i2},\dots, z_{iG})^T \in \{0,1\}^{G}$, a one-hot vector indicating cluster membership, where $z_{ig} = 1$ if observation $i$ belongs to cluster $g$, and $\sum_{g=1}^G z_{ig} = 1$. Let $\boldsymbol{Z} = (\boldsymbol{z}_1, \boldsymbol{z}_2,\dots,\boldsymbol{z}_n)^T \in \{0,1\}^{n \times G}$ denote the cluster membership matrix across all observations.

Within each cluster $g$, the continuous variables follow a multivariate normal distribution with cluster-specific mean vector $\boldsymbol{\mu}_g = (\mu_{1g}, \dots, \mu_{qg})^T \in \mathbbm{R}^q$, and covariance matrix $\boldsymbol{\Sigma_g} \in \mathbbm{R}^{q \times q}$. Let $\boldsymbol{\mu} = (\boldsymbol{\mu}_{1}, \dots, \boldsymbol{\mu}_{g})\in \mathbbm{R}^{q \times G}$ denote the matrix of all cluster-specific means, and $\Sigma = \left \{ \boldsymbol{\Sigma}_g: g=1,\dots,G \right \}$ the collection set of all cluster-specific covariance matrices, respectively.

For categorical variables, we define $\boldsymbol{\theta}_m \in \mathbbm{R}^{G \times L_m}$ as the matrix of class probabilities for variable $m$, where each row $\boldsymbol{ \theta}_{mg} =(\theta_{mg1}, \dots,\theta_{mgL_m})^T$ contains the probabilities of levels $\ell =1, \dots, L_m$ in cluster $g$, subject to $\sum_{\ell=1}^{L_m} \theta_{mg\ell} = 1$. The complete set of categorical parameters across variables and clusters is denoted by $\theta = \{ \boldsymbol{\theta}_m: m=q+1, \dots, M\}$, and for a given cluster $g$, we define $\theta_{g} = \{\boldsymbol{\theta}_{(q+1)g} ,\dots, \boldsymbol{\theta}_{Mg} \}$.

Latent cluster membership follows a multinomial distribution $\boldsymbol{z}_i \sim \text{Multinomial}(1; \boldsymbol{\tau})$, where $\boldsymbol{\tau} = (\tau_1, \tau_2, \dots, \tau_G)^T$ are the mixture proportions, satisfying $\tau_g \geq 0$ and $\sum_{g=1}^G \tau_g = 1$. The likelihood of observation $\boldsymbol{x}_i$ conditional on its latent cluster assignment $\boldsymbol{z}_i$ is given by:
\begin{align*}
f\left( \boldsymbol{x}_i \mid \boldsymbol{z}_i, \boldsymbol{\mu}, \Sigma, \theta, \boldsymbol{\tau}  \right) &= \prod_{g=1}^G \left\{\tau_g \phi_g \left( \boldsymbol{u}_i \mid \boldsymbol{\mu}_{g}, \boldsymbol{\Sigma}_g \right) \prod_{m=q+1}^M \prod_{\ell=1}^{L_m} \theta_{mg\ell}^{\mathbbm{1}(x_{im}=\ell)} \right\}^{z_{ig}},
\end{align*} 
where $\phi_g$ is the multivariate normal density for cluster $g$, and $\mathbbm{1}(\cdot)$ is the indicator function.

Under the conditional independence assumption between continuous and categorical variables given cluster membership, the complete-data likelihood becomes:
\begin{align*}
\mathcal{L}_C\left(\Theta, \boldsymbol{Z} \mid \boldsymbol{X} \right) 
= \prod_{i=1}^n \prod_{g=1}^G \left\{ \tau_g \phi_g \left(\boldsymbol{u}_i \mid \boldsymbol{\mu}_g, \boldsymbol{\Sigma}_g \right) \prod_{m=q+1}^M \prod_{\ell=1}^{L_m} \theta_{mg\ell}^{\mathbbm{1}\left(x_{im}=\ell\right)} \right\}^{z_{ig}}, 
\end{align*}
where $\Theta = \{\boldsymbol{\mu}, \Sigma,\theta, \boldsymbol{\tau}\}$ represents the full set of model parameters.

Marginalizing over the latent cluster assignments yields the observed data likelihood:
\begin{align*}
\mathcal{L}_O\left (\boldsymbol{\mu}, \Sigma,\theta, \boldsymbol{\tau} \mid \boldsymbol{x} \right ) 
&= \int \mathcal{L}_C\left (\boldsymbol{\mu}, \Sigma,\theta, \boldsymbol{\tau}, \boldsymbol{z} \mid \boldsymbol{x} \right ) dz \\
&= \prod_{i=1}^n \sum_{g=1}^G \tau_g \phi_g \left( \boldsymbol{u}_i \mid \boldsymbol{\mu}_g, \boldsymbol{\Sigma}_g \right) \prod_{m=q+1}^M \prod_{\ell=1}^{L_m} \theta_{mg\ell}^{\mathbbm{1}\left(x_{im}=\ell\right)}.
\end{align*}
This formulation allows BFMM to model both mixed-type data and latent class structure within a unified probabilistic framework.

%%% 2.2 %%%
\subsection{Covariance Structure Accommodation}
An important feature of the BFMMs framework is its ability to accommodate flexible dependence structures among continuous variables via parsimonious covariance modeling. Drawing from the family of Gaussian mixture models with eigen decomposed covariance matrices \cite{ParsGMMEDDA}, we implement three representative structures: BFMM[EEI], BFMM[EEE], and BFMM[VVV]. These differ in how they model the variability and correlation among continuous variables within and across clusters. Table \ref{tab:Param_CovStruc} shows the parameterization of the three covariance structures accommodated in BFMM. Table \ref{tab:14CovStruct} and Figure \ref{fig:14CovStruct} summarize all 14 available parsimonious covariance structures (see Appendix~\ref{si.pgmm}).
\begin{itemize}
    \item \textbf{EEI structure} ($\boldsymbol{ \Sigma}_g = \lambda \boldsymbol{ A } $): All clusters share a diagonal covariance matrix, implying conditional independence among continuous variables, but allowing variable-specific variances. Formally, $\boldsymbol{\Sigma}_g = \text{diag}(\sigma_1^2, \dots, \sigma_q^2)$ for all $g$. This structure imposes the strongest constraints and is computationally efficient, making it useful when variable correlations are weak or not of primary interest.
    \item \textbf{EEE structure} ($\boldsymbol{ \Sigma}_g  = \lambda \boldsymbol{D A D} ^T$): All clusters share a general (non-diagonal) covariance matrix, $\boldsymbol{\Sigma}_g = \boldsymbol{ \Sigma}_1$ for all $g$, allowing for global correlations among variables while enforcing homogeneity across clusters. This balances model flexibility and parsimony and is particularly suitable when dependence pattern is similar across latent classes.
    \item \textbf{VVV structure} ($\boldsymbol{\Sigma}_g  = \lambda_g \boldsymbol{D}_g \boldsymbol{A}_g \boldsymbol{D}_g ^T$): Each cluster has its own unconstrained covariance matrix $\boldsymbol{\Sigma}_g$, allowing for fully cluster-specific variances and correlations. Though this structure provides the greatest flexibility and modeling power, it also produces the most parameters and requires careful regularization for reliable estimation.
\end{itemize}

Together, these three structures enable BFMM to adapt to varying degrees of correlation complexity in real-world datasets. By selecting among EEI, EEE, and VVV, users can tailor the model to balance interpretability, computational feasibility, and empirical fit, depending on the clinical context and goals of the analysis.

%%% Table - Covariance Structures %%%
\begin{table}[!htbp]
\small\centering
\caption{Parameterizations of the covariance structures accommodated in BFMMs}
\scalebox{0.8}{
\begin{tabular}{l|llllll}	
\toprule	
\textbf{\begin{tabular}[c]{@{}l@{}}Covariance\\ Structure \end{tabular}} &
  \textbf{\begin{tabular}[c]{@{}l@{}}Covariance Matrix\\ within Cluster $g$\end{tabular}} &
  \textbf{Family} &
  \textbf{Volume} &
  \textbf{Shape} &
  \textbf{Orientation} &
  \textbf{\begin{tabular}[c]{@{}l@{}}\# of Parameters for \\ $G$ Clusters by $q$ Variables\end{tabular}} \\  \cmidrule(l){1-7}
  EEI & $\boldsymbol{\Sigma}_g = \lambda \boldsymbol{A}$           & Diagonal & Equal    & Equal    & Axes     & $q$                       \\
EEE & $\boldsymbol{\Sigma}_g = \lambda \boldsymbol{DAD}^T$       & General  & Equal    & Equal    & Equal    & $q(q+1)/2$        \\
VVV & $\boldsymbol{\Sigma}_g = \lambda_g \boldsymbol{D}_g \boldsymbol{A}_g \boldsymbol{D}_g^T$ & General  & Variable & Variable & Variable & $G\times q(q+1)/2$ \\ \cmidrule(l){1-7}
\multicolumn{7}{l}{ \begin{tabular}[c]{@{}l@{}} 
%$\boldsymbol{B}$ is a diagonal matrix satisfying $|\boldsymbol{B}|=1$.\\%
$\lambda$ (or $\lambda_g$): the constant determining the volume of $\boldsymbol{\Sigma}_g$.\\
$\boldsymbol{D}$ (or $\boldsymbol{D}_g$): the orthogonal matrix of the eigenvectors of $\boldsymbol{\Sigma}_g$, determining the orientation.\\
$\boldsymbol{A}$ (or $\boldsymbol{A}_g$): the diagonal matrix of the normalized eigenvalues of $\boldsymbol{\Sigma}_g$, controlling the shape with $|\boldsymbol{A}|=|\boldsymbol{A_g}|=1$.
\label{tab:Param_CovStruc}
\end{tabular}} 
\\ \bottomrule                                              \end{tabular} }
\end{table}

%%% Figure - EEI, EEE, VVV Covariance Structures %%%
\begin{figure}[!htbp]
\centering  \includegraphics[width=.8\linewidth]
  {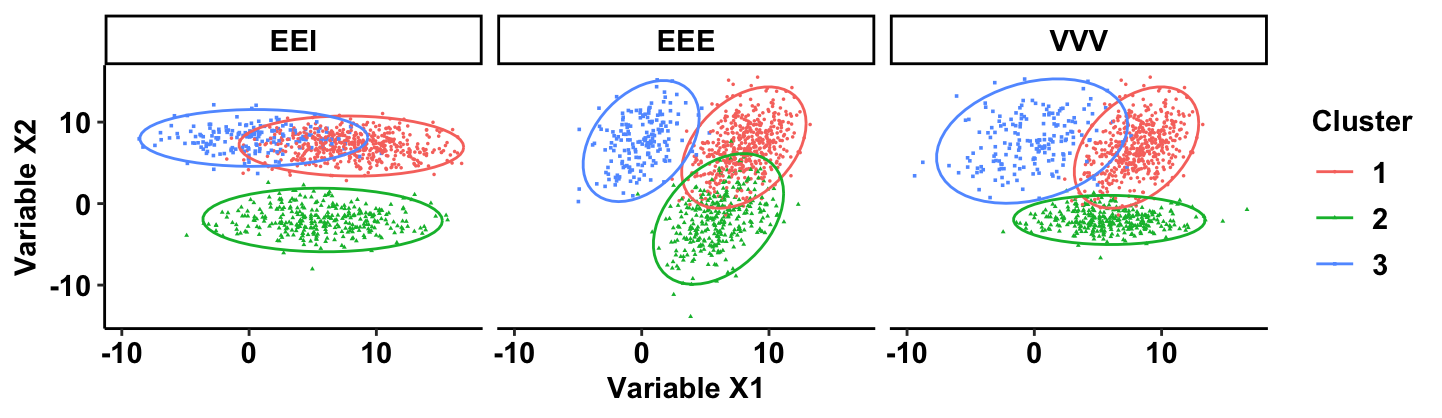}
  \caption{\small Illustrative bivariate 3-cluster mixture density plots for EEI, EEE, and VVV structures.}
  \label{fig:3CovStruct}
\end{figure}

%%% 2.3 %%%
\subsection{Prior Specification}
To enable efficient inference via Gibbs sampling, we assign conjugate prior distributions to the model parameters in BFMM. These priors are selected to support flexible modeling of the variable selection and covariance structure under EEI, EEE, and VVV configurations.

We begin by placing a Dirichlet prior on the cluster mixing proportions. For the latent indicator vector $\boldsymbol{z}_i \sim \text{Multinomial}(1; \tau_1, \dots,\tau_G)$, we assume
$\boldsymbol{\tau} \sim \text{Dir}(\delta_1,\dots,\delta_G)$, where the concentration parameters $\boldsymbol{\delta} =(\delta_1, \dots, \delta_G)^T$ can be chosen to reflect prior beliefs about cluster prevalence or set uniformly for a noninformative prior.

The priors on continuous variables' covariance matrices depend on the selected structure:
\begin{itemize}
    \item \textbf{BFMM[VVV]}: Each cluster's covariance matrix $\boldsymbol{\Sigma}_g$ is assigned an inverse Wishart prior:
    $\boldsymbol{\Sigma}_g  \sim \text{IW}(\nu_g, \boldsymbol{S}_g^{-1} ), \quad g = 1, \dots, G$.
    \item \textbf{BFMM[EEE]}: A single covariance matrix is shared across clusters: 
    $\boldsymbol{\Sigma}_1 \sim \text{IW}(\nu_0, \boldsymbol{S}_0^{-1} )$.
    \item \textbf{BFMM[EEI]}: A single diagonal covariance structure is assumed and the variance of each continuous variable is modeled independently: 
    $\sigma_m^2 \sim \text{Inv}\Gamma(\tilde{a}, \tilde{b}), \quad m = 1, \dots, q$.
\end{itemize}

To support variable selection, we define a binary indicator matrix $\boldsymbol{\Delta} \in \{0,1\}^{M \times G}$, where $\Delta_{mg}=1$ indicates that variable $m$ is important for separating cluster $g$ from others. 

For continuous variable $m = 1,\cdots,q$, we adopt a spike-and-slab prior on its cluster-specific means:
\begin{equation*}
  \mu_{mg} \sim N(0,\sigma_{\Delta_0}^2 \omega^{\Delta_{mg}}) =
      \begin{cases}
      N(0,\sigma_{\Delta_0}^2)  & \text{if $\Delta_{mg}=0$} \\
      N(0,\sigma_{\Delta_0}^2\omega ) & \text{if $\Delta_{mg}=1$},
    \end{cases}       
\end{equation*}
where $\omega >1$ inflates the variance for important variables (those with $\Delta_{mg}=1$). 

The baseline variance parameter is assigned an inverse gamma prior: $\sigma_{\Delta_0}^2 \sim \text{Inv}\Gamma(a_{\Delta_0}, b_{\Delta_0})$.

For categorical variable $m = q+1, \cdots, M$, we adopt a spike-and-slab Dirichlet prior on its cluster-specific level probabilities:
\begin{equation*}
  \boldsymbol{\theta}_{mg} \sim \text{Dir}\left\{\boldsymbol{\alpha}_{m\Delta_0}^{\left( 1- \Delta_{mg}\right)}\boldsymbol{\alpha}_{\Delta_1}^{\Delta_{mg}} \right \} = 
      \begin{cases}
      \text{Dir}(\boldsymbol{\alpha}_{m\Delta_0}) & \text{if $\Delta_{mg}=0$} \\
      \text{Dir}(\boldsymbol{\alpha}_{\Delta_1}) & \text{if $\Delta_{mg}=1$},
    \end{cases}       
\end{equation*}
where $\boldsymbol{\alpha}_{m\Delta_0}$ is proportional to the marginal distribution of variable $m$ with entries $\gg 1$ to shrink cluster-specific variation toward the marginal distribution, while $\boldsymbol{\alpha}_{\Delta_1} = \boldsymbol{1}_{L_m}^T $ ensures a uniform prior to deviate the cluster-specific distribution from the marginal distribution.

The variable importance indicators $\Delta_{mg}$ are themselves modeled with Bernoulli priors, using separate probabilities for continuous and categorical variables:
\begin{equation*}
\Delta_{mg} \sim
\begin{cases}
\text{Ber}(p_{1m}) & \text{if } m = 1, \dots, q \\
\text{Ber}(p_{2m}) & \text{if } m = q+1, \dots, M,
\end{cases}
\end{equation*}
with inclusion probabilities assigned Beta hyperpriors $p_{1m} \sim \text{Beta}(a_{p1}, b_{p1})$, $p_{2m} \sim \text{Beta}(a_{p2}, b_{p2})$, allowing BFMM to account for different degrees of sparsity or relevance across variable types.

\begin{figure}[!htbp] \centering
  \includegraphics[width=.68\linewidth]
  {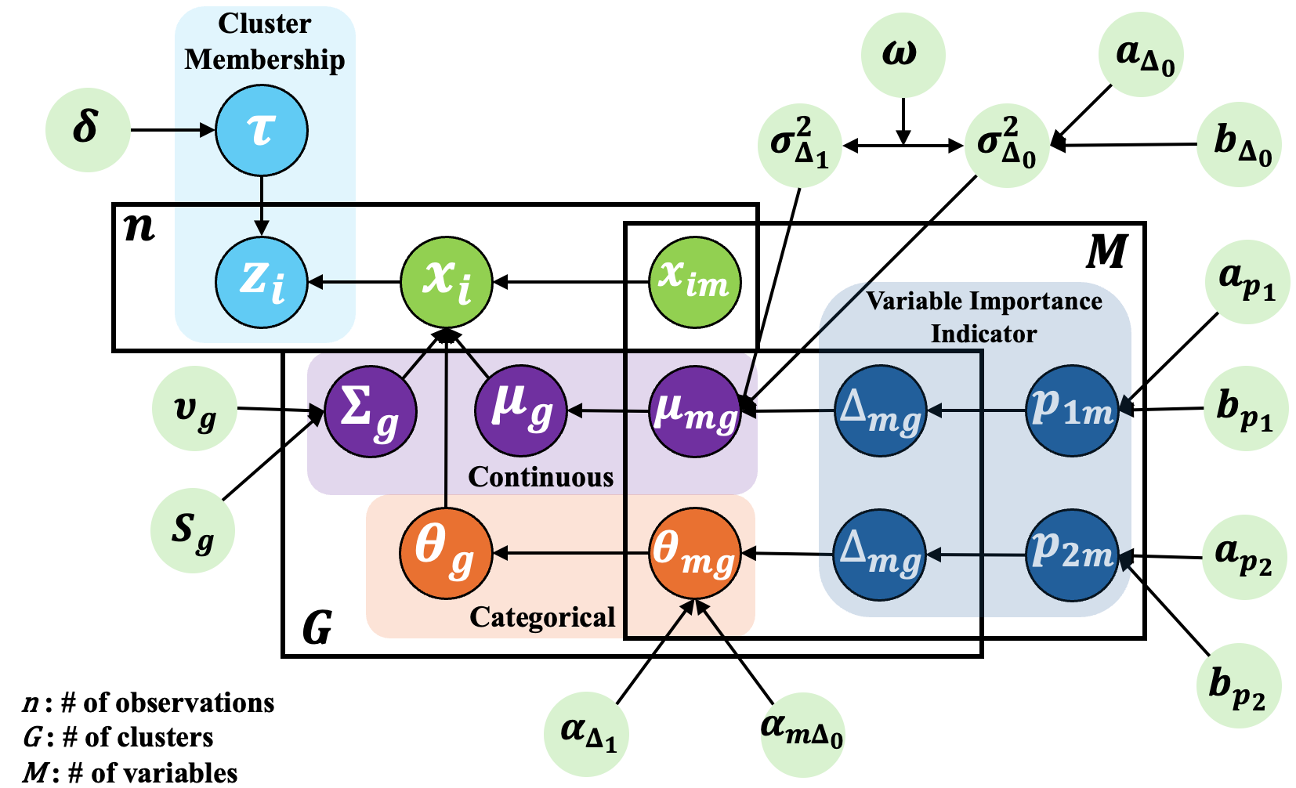}
  \caption{\small Graphical representation of the proposed Bayesian FMM framework given VVV covariance structure - BFMM[VVV]. Subscripts $i,g,m$ denote the $i$th observation ($i=1,\dots,n)$, $g$th cluster ($g=1,\dots, G$), and $m$th variable ($m=1,\dots,q$ if continuous, $m=q+1,\dots,M$ if categorical), respectively.}
  \label{fig:BFMM.VVV}
\end{figure}

A graphical representation of the proposed BFMM framework under the VVV covariance structure is provided in Figure \ref{fig:BFMM.VVV}.

%%% Figure - BFMM[VVV] Framework %%%

%%% 2.4 %%%
\subsection{Posterior Distribution}
The joint posterior distribution of all model parameters is obtained by combining the observed data likelihood with the prior distributions specified above. The posterior distribution under BFMM is proportional to:
\begin{align*}\small
f(\Theta, \boldsymbol{Z} \mid \boldsymbol{X}) \propto& \prod_{i=1}^n \prod_{g=1}^G \left\{ \tau_g \phi_g \left( \boldsymbol{u}_i \mid \boldsymbol{\mu}_g, \boldsymbol{\Sigma}_g \right) \prod_{m=q+1}^M \prod_{\ell=1}^{L_m} \theta_{mg\ell}^{\mathbbm{1}\left(x_{im}=\ell\right)} \right\}^{z_{ig}} \times 
%\prod_{i=1}^n \prod_{g=1}^G \tau_g^{z_{ig}}\times 
\text{Dir}(\boldsymbol{\tau}; \boldsymbol{\delta}) \\
\times& p\left(\boldsymbol{\Sigma}_1,\dots,\boldsymbol{\Sigma}_G; \tilde a, \tilde b, \nu_*, \boldsymbol{S}_* \right)
\times
\prod_{m=1}^q\prod_{g=1}^G N\left(\mu_{mg} ; 0, \omega^{\Delta_{mg}}\sigma_{\Delta_0}^2 \right) \times \text{Inv}\Gamma\left( \sigma_{\Delta_0}^2; a_{\Delta_0}, b_{\Delta_0} \right) \\ 
 \times& \prod_{m=1}^q\prod_{g=1}^G \text{Ber}\left(\Delta_{mg};p_{1m}\right)  \prod_{m=1}^q \text{Beta}\left(p_{1m}; a_{p1}, b_{p1}\right) 
  \prod_{m=q+1}^M\prod_{g=1}^G \text{Dir}\left\{\boldsymbol{\theta}_{mg}; \boldsymbol{\alpha}_{m\Delta_0}^{\left( 1- \Delta_{mg}\right)}\boldsymbol{\alpha}_{\Delta_1}^{\Delta_{mg}}\right\} \\
 \times& \prod_{m=q+1}^M\prod_{g=1}^G \text{Ber}\left(\Delta_{mg};p_{2m}\right)  \prod_{m=1}^q \text{Beta}\left(p_{2m}; a_{p2}, b_{p2}\right).
\end{align*}

The prior distribution for the covariance matrices depends on the selected structure:
\begin{equation*}\small
  p\left(\boldsymbol{\Sigma}_1,\dots, \boldsymbol{\Sigma}_G; \tilde a, \tilde b, \nu_*, \boldsymbol{S}_* \right) =
      \begin{cases}
      \prod_{m=1}^q \text{Inv}\Gamma\left(\sigma_m^2; \tilde a, \tilde b \right) & \text{for BFMM[EEI]}, \\
      \text{IW}\left( \boldsymbol{\Sigma}_1; \nu_0, \boldsymbol{S}_0^{-1}\right)  & \text{for BFMM[EEE]}, \\
      \prod_{g=1}^G \text{IW}\left(\boldsymbol{\Sigma}_g; \nu_g, \boldsymbol{S}_g^{-1}\right)  & \text{for BFMM[VVV]}. \\
    \end{cases}       
\end{equation*}

Let $n_g = \sum_{i=1}^n z_{ig}$ denote the total number of observations assigned to cluster $g$, and let $\bar{x}_m = \frac{1}{n} \sum_{i=1}^n x_{im}$ be the marginal mean of continuous variable $m$. The full conditional posterior distributions for all parameters used in the Gibbs sampling algorithm are given in Appendix~\ref{appendix.postdist}.

%%% 2.5 %%%
\subsection{Hyperparameters}
To initialize the BFMM estimation procedure, values must be specified in advance for several hyperparameters associated with the priors introduced earlier. These choices aim to balance model flexibility with stability, promote interpretability of variable importance measures, and ensure computational tractability across all BFMM variants. We listed below the hyperparameters requiring specifications:
\begin{itemize}
    \item $\boldsymbol{\delta}$ for the Dirichlet prior on mixture proportions $\boldsymbol{\tau}$
    \item $\omega$ as the variance ratio for spike-and-slab prior on cluster-specific mean of continuous variables
    \item $a_{\Delta_0}$, $b_{\Delta_0}$ for the variance parameter $\sigma_{\Delta_0}^2$ in the spike-and-slab prior
    \item $\boldsymbol{\alpha}_{m\Delta_0}$ for the spike Dirichlet prior on cluster-specific distribution of categorical variable $m$
    \item $a_{p1}, b_{p1}$ for Beta priors on inclusion probabilities $p_{1m}$ (continuous variables)
    \item $a_{p2}, b_{p2}$ for Beta priors on $p_{2m}$ (categorical variables)
    \item Covariance structure hyperparameters: 
        \begin{itemize}
            \item $\tilde{a}, \tilde{b}$ for variances $\sigma_m^2$ in BFMM[EEI] \item $\nu_0, \boldsymbol{S}_0$ for the shared covariance matrix $\boldsymbol{\Sigma}_1 $ in BFMM[EEE] 
            \item $\nu_g, \boldsymbol{S}_g$ for the cluster-specific covariance matrices $\boldsymbol{\Sigma_g}$ in BFMM[VVV].
        \end{itemize}    
\end{itemize}

For the mixture proportions, We used a balanced Dirichlet prior $\boldsymbol{\delta} = (1/G, \dots, 1/G)^T$, assuming equal prior weight across clusters. 

For the covariance priors, we set $\tilde{a} = 2$ and $\tilde{b} = 1$ in BFMM[EEI], resulting in a prior mean of 1 for the variances $\sigma_m^2$, assuming the continuous variables have been standardized. In BFMM[EEE] and BFMM[VVV], the prior scale matrix $\boldsymbol{S}_g$ is constructed from the empirical covariance of the observed continous data:
\begin{align*}
\boldsymbol{S}_g = \frac{1}{G^{2/q}} \cdot \frac{1}{n-1} \sum_{i=1}^n (\boldsymbol{u}_i - \bar{\boldsymbol{u}})(\boldsymbol{u}_i - \bar{\boldsymbol{u}})^T,
\end{align*}
where $\boldsymbol{\bar{\boldsymbol{u}}}$ is the $q$-dimensional sample mean vector. We set $\nu_g = q+2$ to allow moderate uncertainty.

For Dirichlet priors on categorical variable probabilities, we define $\boldsymbol{\alpha_{\Delta_1}} = \boldsymbol{1}_{L_m}^T$ for the slab prior, encouraging flexibility in level probabilities. For the spike prior, we set
\begin{align*}
\boldsymbol{\alpha}_{m\Delta_0} = \frac{C}{n} \left\{ \sum_{i=1}^n \mathbbm{1}(x_{im}=1), \dots, \sum_{i=1}^n \mathbbm{1}(x_{im}=L_m) \right\}^T,
\end{align*}
where $C \gg 1$ concentrates the prior around the empirical marginal distribution of categorical variable $m$, encouraging shrinkage toward noninformative inclusion when appropriate.

The variance ratio $\omega$, governing the difference between the slab and spike priors for continuous variables, is selected in a data-adaptive manner using initial estimates of the cluster-specific means $\boldsymbol{\mu^{(0)}}$ from bootstrap K-means clustering. Specifically, we compute:
\begin{equation}
\omega = \left( \frac{\mathbbm{E}\left[ \boldsymbol{\mu}^{(0)} : \boldsymbol{\mu}^{(0)} \geq P_{k}\left\{ \boldsymbol{\mu}^{(0)}\right\} \right] }{\mathbbm{E}\left[ \boldsymbol{\mu}^{(0)} : \boldsymbol{\mu}^{(0)} \leq P_{25}\left\{ \boldsymbol{\mu}^{(0)}\right\} \right]} \right)^2, \quad 60 \leq k \leq 90,
\label{equ:omega}
\end{equation}
where $P_k(\cdot)$ denotes the $k$th percentile and $\mathbbm{E}(\cdot)$ denotes the mean. This rule anchors the higher value of $\mu_{mg}$ under the slab prior and the lower values under the spike prior, helping ensure distinct and interpretable variable inclusion.

Choosing an inappropriate value for $\omega$ can bias variable selection: if $\omega\approx 1$, the spike and slab distributions become indistinguishable, leading to poor separation. Conversely, if $\omega$ is too large, the resulting variance gap may dominate the true signal, falsely pushing all variables toward extreme inclusion or exclusion. Our adaptive strategy mitigates both risks by tailoring $\omega$ to the observed data distribution.

%%% 2.6 %%%
\subsection{Parameter Estimation}
Parameter estimation in BFMM is conducted using Gibbs sampling, leveraging the conjugate priors described above to enable efficient updates of all model parameters. Algorithm~\ref{alg:euclid} outlines the learning procedure, which includes both posterior inference and censored data imputation. To address the label switching problem common in mixture models, we incorporate a relabeling strategy based on minimizing Stephens' Kullback-Leibler (KL) divergence \cite{Stephens2000} across MCMC iterations. 

Censored biomarker imputation is performed for continuous variable subject to lower or upper detection limits. Let $x_{im}$ be a censored observation of subject $i$ with $i \in \left\{i_{mL} \cap i_{mU} \right\}$ under continuous variable $m$, and $\boldsymbol{x}_{i(m)}$ denote the vector of remaining continuous variables. Given cluster assignment $\boldsymbol{z_i}$ and current estimates of $\boldsymbol{\mu}$ and $\boldsymbol{\Sigma}_g$, the conditional distribution of $x_{im}$ is:
\begin{equation}
x_{im} \mid \boldsymbol{x}_{i(m)}, \boldsymbol{z}_i, \boldsymbol{\mu}, \left\{ \boldsymbol{\Sigma}_g \right\}_{g=1}^G \sim \prod_{g=1}^G \left\{ N \left(\mu_{mg}^{\ast}, \sigma^{\ast 2}_{mg} \right) \right\}^{z_{ig}}, \end{equation}

\noindent with mean $\mu_{mg}^{\ast} = \mu_{mg} + \left ( \boldsymbol{\Sigma}_g \right)_{m(m)}\left [ \boldsymbol{\Sigma}_g \right]_{mm}^{-1}\left\{ x_{im}\boldsymbol{1}^T_{(q-1)}  - \boldsymbol{\mu}_{(m)g}\right\}$ and variance $\sigma^{\ast 2}_{mg} = \left ( \boldsymbol{\Sigma}_g \right)_{mm} - \left ( \boldsymbol{\Sigma}_g \right)_{m(m)} \left [ \boldsymbol{\Sigma}_g \right]_{mm}^{-1} \left ( \boldsymbol{\Sigma}_g \right)_{(m)m}$, where subscripts $m(m)$, $(m)m$ and $[\cdot]_{mm}$ denote appropriate row/column deletions and indexing.

To enforce detection limits, we sample from the corresponding truncated normal distribution:
\begin{align}
\begin{split}
  f\left[x_{im} \mid \boldsymbol{x}_{i(m)}, \boldsymbol{z}_{i}, \boldsymbol{\mu}, \left\{ \boldsymbol{\Sigma}_g \right\}_{g=1}^G \right] \\ = &
      \begin{cases}
      \prod_{g=1}^G \left[ \frac{1}{\sigma^*_{mg}} \cdot \frac{ \varphi\left\{ (x_{im}-\mu_{mg}^*)/\sigma^*_{mg}\right\}}{\Phi\left\{ (C_{imL}-\mu_{mg}^*)/ \sigma^*_{mg}\right\}} \right]^{z_{ig}} 
      & \text{if $x_{im} < C_{imL}, i \in i_{mL} $}, \\
      \prod_{g=1}^G \left[\frac{1}{\sigma^*_{mg}} \cdot  \frac{ \varphi\left\{(x_{im}-\mu_{mg}^*)/\sigma^*_{mg}\right\}}{1-\Phi\left\{ (C_{imU}-\mu_{mg}^*)/ \sigma^*_{mg}\right\}} \right]^{z_{ig}} 
      & \text{if $x_{im} > C_{imU}, i \in i_{mU} $},
    \end{cases} 
\label{equ:censor_impute} 
\end{split}      
\end{align}
where $\displaystyle \varphi(\cdot)$ and $\displaystyle\Phi(\cdot)$ are the standard normal density and cumulative distribution functions, respectively; $C_{imL}$ and $C_{imU}$ denote the lower and upper detection limits for censored $x_{im}$, respectively.

For label switching, the invariance of the likelihood to permutations of cluster labels, poses challenges for interpreting MCMC output. To ensure consistent labeling across iterations, we apply the KL relabeling algorithm. This method aligns cluster labels by minimizing the KL divergence between the current clustering assignment probabilities and their posterior average. The relabeling algorithm is detailed in Algorithm \ref{algorithm:KLrelabel} (see Appendix~\ref{si.labelswich}).

\begin{algorithm}
    \caption{\textbf{Proposed BFMM Clustering}}\label{alg:euclid}
    \begin{algorithmic}[1]\small
    \State \textbf{Input:} Observed data $\left \{\boldsymbol{x}_i \right \}_{i=1}^n$ with $q$ standardized continuous variables come first and $(M-q)$ categorical variables in last position, the number of latent clusters $G$.
    \State \textbf{Initialization:} Run bootstrap K-means clustering to set initial cluster membership $\left \{\boldsymbol{z}_i^{(0)} \right \}_{i=1}^n$.
    \State\hskip1.5em Initialize $\boldsymbol{\tau}^{(0)}$, $\boldsymbol{\mu}^{(0)}$, $\Sigma^{(0)}$, $\theta^{(0)}$.
    \State\hskip1.5em Set threshold $60 \leq k \leq 90$ to set $\omega$ given $\boldsymbol{\mu}^{(0)}$ as in \eqref{equ:omega}.   
    \State\hskip1.5em Set $\boldsymbol{\delta}$, $\{ \nu_g, \boldsymbol{S}_g\}_{g=1}^G$, $\boldsymbol{\alpha}_{\Delta_1}$ and $\{\boldsymbol{\alpha}_{m\Delta_0} \}_{m=q+1}^M$ following general choices. 
    \State\hskip1.5em Set other hyperparameters $a_{\Delta_0}$, $b_{\Delta_0}$, $a_{p1}$, $b_{p1}$, $a_{p2}$, $b_{p2}$.
    \State \textbf{Gibbs Sampling}
    \For{$t = 1,\dots,T$}
        \State \textbf{Impute censored continuous observations:}
        \For {$m=1,\dots,q \text{ and } i \in \{i_{mL} \cap i_{mU} \} $}
        	\State Sample $x_{im}^{(t)}$ from $f \left\{ x_{im} \mid \boldsymbol{x}_{i(m)}^{(t-1)},\boldsymbol{z}_i^{(t-1)}, \boldsymbol{\mu}^{(t-1)}, \Sigma^{(t-1)} \right \}$ as in \eqref{equ:censor_impute}.
        \EndFor
        \State \textbf{Sample from the posteriors:}
        \For {$g=1,\dots,G \text{ and } m=1,\dots,q$}
        \State If given EEI structure, sample $\left(\sigma_m^2\right)^{(t)}$, $\mu_{mg}^{(t)}$ as in \eqref{equ:EEI_sigma_g}\eqref{equ:EEI_mu_mg}.
        \State If given EEE structure, sample $\boldsymbol{\Sigma}_1^{(t)}$, $\mu_{mg}^{(t)}$ as in \eqref{equ:EEE_sigma_g}\eqref{equ:EEE_mu_mg}.
        \State If given VVV structure, sample $\boldsymbol{\Sigma}_g^{(t)}$, $\mu_{mg}^{(t)}$ as in \eqref{equ:VVV_sigma_g}\eqref{equ:VVV_mu_mg}.
        \EndFor
        \State Sample $\boldsymbol{\theta}_{mg}^{(t)}$ for $g=1,\dots,G \text{ and } m=1,\dots,q$ as in \eqref{equ:theta_mg}.       
        \State Sample $\boldsymbol{\Delta}^{(t)}$, $\left(\sigma_{\Delta_0}^{2}\right)^{(t)}$, $p_{1m}^{(t)}$, $p_{2m}^{(t)}$ for $g=1,\dots,G \text{ and } m=1,\dots,q$ as in \eqref{equ:delta_mg_p1}\eqref{equ:delta_mg_p2}\eqref{equ:sigma2_delta0}\eqref{equ:p_jm}.
        \State Sample $\tau_g^{(t)}$ for $g=1,\dots,G$ as in \eqref{equ:tau}. 
        \State Sample $\boldsymbol{z}_i^{(t)}$ for $i=1,\dots,n$ as in \eqref{equ:z_i}.
    \EndFor
    \State Store $\boldsymbol{\tau}^{(t)}$, $\boldsymbol{\mu}^{(t)}$, $\Sigma^{(t)}$, $\theta^{(t)}$ for $t=t^*+1,\dots,T$ with first $t^*$ iterations for burn-in.
    \State Calculate $\boldsymbol{P}_{n \times G}^{(t)}$ with elements $p_{i,j}^{(t)} = \frac{\tau_j^{(t)} f_j\left\{ \boldsymbol{x}_i|\boldsymbol{\mu}_j^{(t)}, \boldsymbol{\Sigma}_j^{(t)}, \theta_j^{(t)} \right \} }{\sum_{g=1}^G \tau_g^{(t)} f_g \left\{ \boldsymbol{x}_i|\boldsymbol{\mu}_g^{(t)}, \boldsymbol{\Sigma}_g^{(t)}, \theta_g^{(t)} \right \} } $ for $t=t^*+1,\dots,T$.
    \State \textbf{Kullback-Leibler (KL) Relabeling}
    \State\hskip1.5em Initialize a permutation $\rho_t$ of the columns of $\boldsymbol{P}_{n \times G}^{(t)}$ for $t=t^*+1,\dots,T$.
	\State\hskip1.5em Apply Stephens' KL relabeling \cite{Stephens2000} to find $\left\{ \hat{\rho}_t \right \}_{t = t^*+1}^T$ as in Algorithm \ref{algorithm:KLrelabel}.
	\State\hskip1.5em Update $\boldsymbol{\tau}^{(t)}$, $\boldsymbol{\mu}^{(t)}$, $\Sigma^{(t)}$, $\theta^{(t)}$ and $\boldsymbol{\Delta}^{(t)}$ given $\hat\rho_t$ for $t=t^*+1,\dots,T$.
	\State \textbf{Cluster Assignment by Maximum A Posteriori (MAP)}
	\State\hskip1.5em Calculate $\hat{\boldsymbol{\tau}}, \hat{\boldsymbol{\mu}}, \hat\Sigma, \hat\theta $ by averaging across $t=t^*+1,\dots,T$.
	\State\hskip1.5em Calculate $\hat\Delta_{m} = \frac{1}{G(T-t^*)}\sum_{g=1}^G \sum_{t=t^{*}+1}^T\hat\Delta_{mg}^{(t)}$ for $m=1,\dots,M$.
	\State\hskip1.5em Calculate MAP estimate of $z_{ig}$ for $i=1,\dots,n$ and $g=1,\dots,G$ as: 
	\begin{equation*}
	\hat z_{ig} \mid \boldsymbol{x}_i,\hat{\boldsymbol{\tau}},\hat{\boldsymbol{\mu}},\hat\Sigma,\hat\theta = \mathbbm{1} \left[ g = \underset{k \in \{1,2,\dots,G\}}{\arg\max} \left\{ \frac{\hat\tau_k f_k(\boldsymbol{x}_i \mid \hat{\boldsymbol{\mu}_k},\hat{\boldsymbol{\Sigma}_k},\hat\theta_k)}{\sum_{k=1}^G \hat\tau_k f_k(\boldsymbol{x}_i \mid \hat{\boldsymbol{\mu}_k},\hat{\boldsymbol{\Sigma}_k},\hat\theta_k)}  \right\}  \right ]. 
	\end{equation*}
	\State \textbf{Output:} Cluster membership indicator estimates $\left \{ \hat{\boldsymbol{z}_i}\right \}_{i=1}^n$, FMM distributional parameter estimates $\left \{ \hat\tau_g, \hat{\boldsymbol{\mu}_g}, \hat{\boldsymbol{\Sigma}_g}, \hat\theta_g \right \}_{g=1}^G$, and variable importance estimates $\left\{\hat\Delta_m\right \}_{m=1}^M$.
    \end{algorithmic}
\end{algorithm}

%%% 2.7 %%%
\subsection{Model Selection}
The proposed BFMM framework requires specification of the number of clusters $G$. Since the choice of $G$ can strongly influence the clustering result and its interpretability, selecting an appropriate value is essential. to guide model selection, we consider two widely used criteria in model-based clustering: the Bayesian information criterion (BIC) \cite{BIC} and the Integrated complete-date likelihood criterion (ICL) \cite{ICL}. Both criteria aim to balance model fit with complexity, but differ in how they account for clustering uncertainty.

\subsubsection{Bayesian Information Criterion (BIC)}
The BIC approximates the marginal likelihood of the data under a given model and penalizes for the number of estimated parameters. It is widely used in mixture modeling due to its computational simplicity and asymptotic consistency.

Let $\hat \Theta(G) = \left\{\hat{\boldsymbol{\tau}},\hat{\boldsymbol{\mu}}, \hat \Sigma,\hat \theta \mid G \right\}$ denote the set of estimated parameters for a finite mixture model with $G$ clusters. The BIC is defined as:
\begin{equation*}
\text{BIC}(G) = -2\log \mathcal{L}_O\left\{ \hat\Theta(G) \mid \boldsymbol{x}\right \} + \text{DF} \cdot \log\left(n\right), 
\end{equation*}
where $\mathcal{L}_O$ is the observed data likelihood:  
$\mathcal{L}_O\left\{\Theta(G) \mid \boldsymbol{x} \right \} = \prod_{i=1}^n \left\{ \sum_{g=1}^G \tau_g f_g(\boldsymbol{x}_i \mid \boldsymbol{\mu}_g, \boldsymbol{\Sigma}_g, \theta_g ) \right\}$, and $\text{DF}$ is the total number of free parameters under the specified parsimonious covariance structure:
\begin{equation*}\small
\text{DF} = \begin{cases}
    (G-1) + q + G\cdot \left\{ q + \sum_{m=q+1}^M (L_m - 1) \right\} \text{ under EEI structure; } \\
    (G-1) + \frac{q(q+1)}{2} + G\cdot \left\{ q + \sum_{m=q+1}^M (L_m - 1) \right\} \text{ under EEE structure; } \\
    (G-1) + G\cdot \left\{ q+\frac{q(q+1)}{2} +\sum_{m=q+1}^M (L_m - 1) \right\} \text{ under VVV structure. } \\
    \end{cases}
\end{equation*}
In specific, for the case under the VVV covariance structure, $(G-1)$ accounts for the mixture probabilities of the $G$ clusters; $q + \frac{q(q+1)}{2}$ for the mean and covariance matrix of all the $q$ continuous variables while $(L_m -1)$ for categorical variable $m$ with $L_m$ levels within each cluster. Smaller BIC values indicate better models, favoring parsimonious solutions with strong empirical support.

\subsubsection{Integrated Complete-data Likelihood Criterion (ICL)}
While BIC evaluates marginal likelihood based on the observed data, the ICL extends this idea by incorporating cluster assignment uncertainty. Specifically, it penalizes solutions where posterior probabilities are diffuse across clusters, thus preferring well-separated and more interpretable clusters. ICL is defined as:
\begin{equation*}
\text{ICL}(G) = -2\log \mathcal{L}_C\left\{ \hat\Theta(G) \mid \boldsymbol{x}, \boldsymbol{z}^*\right \} + \text{DF} \cdot \log\left(n\right), 
\end{equation*}
where $\mathcal{L}_C$ is the complete-data likelihood evaluated at the maximum a posteriori (MAP) cluster assignments $\boldsymbol{Z}^*$, defined by:
\begin{equation*}
\mathcal{L}_C\left\{\Theta(G) \mid \boldsymbol{X}, \boldsymbol{Z}^*\right\} = \prod_{i=1}^n \prod_{g=1}^G \left\{ \hat{\tau}_g f_g(\boldsymbol{x}_i \mid \boldsymbol{\hat{\mu}}_g, \boldsymbol{\hat{\Sigma}}_g, \hat{\theta}_g) \right\}^{z_{ig}^*}.
\end{equation*}
\noindent ICL can also be expressed as an adjusted version of BIC that further penalizes posterior entropy:
\begin{equation*}
\text{ICL}(G) = \text{BIC}(G) - \sum_{i=1}^n \sum_{g=1}^G \hat{z}_{ig} \log(\hat{z}_{ig}),
\end{equation*}
where $\hat{z}_{ig}$ is the posterior probability that observation $i$ belongs to cluster $g$:
\begin{equation*}
\hat{z}_{ig} = \frac{\hat{\tau}_g f_g(\boldsymbol{x}_i \mid \hat{\boldsymbol{\mu}}_g, \hat{\boldsymbol{\Sigma}}_g, \hat{\theta}_g)}{\sum_{g=1}^G \hat{\tau}_g f_g(\boldsymbol{x}_i \mid \hat{\boldsymbol{\mu}}_g, \hat{\boldsymbol{\Sigma}}_g, \hat{\theta}_g)}.
\end{equation*}
\noindent Because ICL explicitly accounts for clustering uncertainty, it often favors more distinct and interpretable solutions, and tends to select fewer clusters than BIC in practice \cite{BICvsICL1,BICvsICL2}.

%%%%%%%%%%%%%%%%%%%%%%%%%%%%%%%%%%%%%%%%%%%%%%%%%%%
%%% Section 3 - Simulation Study %%%
%%%%%%%%%%%%%%%%%%%%%%%%%%%%%%%%%%%%%%%%%%%%%%%%%%
\section{Simulations}
To evaluate the performance of the proposed BFMM for clustering mixed-type data with censored variables, we conducted comprehensive simulation studies under varying data-generating mechanisms. We assessed clustering accuracy and variable selection across different covariance structures and censoring levels, and compared BFMM against alternative mixed-data clustering methods.

\subsection{Simulation Design}
We designed nine simulation scenarios to assess the robustness and flexibility of BFMM under realistic conditions. Three baseline scenarios (Data[EEI], Data[EEE], and Data[VVV]) simulate datasets with increasing levels of correlation among continuous variables, corresponding to the EEI, EEE, and VVV covariance structures. The remaining six scenarios incorporate detection-limit censoring in three of the dominant continuous variables ($X_3, X_4, X_5$) at two levels ($20\%, 40\%$) for each structure.

Each dataset includes $n=1,000$ observations and $M=14$ variables: 7 continuous variables ($X_1-X_7$) generated from multivariate normal mixtures and 7 categorical variables ($X_8-X_{14}$) generated from multinomial distributions. We fixed the number of clustrs at $G=3$, with true mixture proportions $\boldsymbol{\tau} = (0.5, 0.3, 0.2)^T$. Across all scenarios, we simulated 100 replicates per setting. The generation protocols are shown in Tables \ref{tab:SimSetCont} and \ref{tab:SimSetCat}.

To mimic complex variable relevance patterns, $X_1-X_4$ (continuous) and $X_8-X_{10}$ (categorical) were set as dominant contributors to clustering. $X_6 - X_7$ (continuous) and $X_{11} - X_{14}$ (categorical) were designed as noise variables, while $X_5$ (continuous) was weakly informative. To introduce correlation between variable types, we modeled $X_8$ and $X_{11}$ using logistic regressions:
$\text{logit}\left(\theta_{X_8}\right )=0.1X_3+0.5X_4$ and $\text{logit}\left(\theta_{X_{11}}\right )=0.1X_6 +0.5X_7$, where $X_8$ and $X_{11}$ were drawn from Bernoulli distributions with respective probabilities $\theta_{X_8}$ and $\theta_{X_{11}}$. 

\begin{table}[!htbp] 
\caption{Simulated three-cluster mixed data continuous variable generation parameters.}
\small\center
\scalebox{0.9}{
\begin{tabular}{lccll} \toprule
\multirow{2}{*}{\textbf{\begin{tabular}[c]{@{}l@{}}Simulation\\ Structure\end{tabular}}} & \multirow{2}{*}{\textbf{\begin{tabular}[c]{@{}c@{}}Cluster \\ $g$\end{tabular}}} & \multirow{2}{*}{\textbf{$\tau_g$}} & \multicolumn{2}{c}{\textbf{Multivariate Normal Distribution for $X_1 - X_7$}} \\ \cmidrule(l){4-5}
 &  &  & \textbf{$\boldsymbol{\mu}_g$} & \textbf{$\boldsymbol{\Sigma}_g$} \\ \cmidrule(l){1-5}
\multirow{3}{*}{\begin{tabular}[c]{@{}l@{}}Data\\ {[}EEI{]}\end{tabular}} & 1 & 0.5 & $\boldsymbol{\mu}_1 = (5,6,0,0,0,0,0)^T$ & \multirow{3}{*}{\begin{tabular}[c]{@{}c@{}}$\boldsymbol{\Sigma}_{ind}=\text{diag}(8,4,4,4,6,6,6) + 0 \times \boldsymbol{J}_7$\\ $\boldsymbol{\Sigma}_1=\boldsymbol{\Sigma}_2=\boldsymbol{\Sigma}_3=\boldsymbol{\Sigma}_{ind}$\end{tabular}} \\
 & 2 & 0.3 & $\boldsymbol{\mu}_2 = (5,0,7,-3,-0.5,-0.2,0)^T$ &  \\
 & 3 & 0.2 & $\boldsymbol{\mu}_3 = (0,6,7,3,0.5,0.2,0)^T$ &  \\
\multirow{3}{*}{\begin{tabular}[c]{@{}l@{}}Data\\ {[}EEE{]}\end{tabular}} & 1 & 0.5 & $\boldsymbol{\mu}_1 = (5,6,0,0,0,0,0)^T$ & \multirow{3}{*}{\begin{tabular}[c]{@{}c@{}}$\boldsymbol{\Sigma}_{cor2}=\text{diag}(2,2,6,2,4,6,8) + 2\times \boldsymbol{J}_7$\\ $\boldsymbol{\Sigma}_1=\boldsymbol{\Sigma}_2=\boldsymbol{\Sigma}_3=\boldsymbol{\Sigma}_{cor2}$\end{tabular}} \\
 & 2 & 0.3 & $\boldsymbol{\mu}_2 = (5,0,7,-3,-0.5,-0.2,0)^T$ &  \\
 & 3 & 0.2 & $\boldsymbol{\mu}_3 = (0,6,7,3,0.5,0.2,0)^T$ &  \\
\multirow{3}{*}{\begin{tabular}[c]{@{}l@{}}Data\\ {[}VVV{]}\end{tabular}} & 1 & 0.5 & $\boldsymbol{\mu}_1 = (5,6,0,0,0,0,0)^T$ & $\boldsymbol{\Sigma}_{cor2}=\text{diag}(2,2,6,2,4,6,8) + 2 \times \boldsymbol{J}_7$ \\
 & 2 & 0.3 & $\boldsymbol{\mu}_2 = (5,0,7,-3,-0.5,-0.2,0)^T$ & $\boldsymbol{\Sigma}_{ind}=\text{diag}(8,4,4,4,6,6,6) + 0 \times \boldsymbol{J}_7$ \\
 & 3 & 0.2 & $\boldsymbol{\mu}_3 = (0,6,7,3,0.5,0.2,0)^T$ & $\boldsymbol{\Sigma}_{cor1}=\text{diag}(3,7,3,3,5,7,7) + 1 \times \boldsymbol{J}_7$ \\ \cmidrule(l){1-5}
\multicolumn{5}{l}{
$\boldsymbol{\Sigma}_{ind} = \begin{bmatrix}
8 & 0 & 0 & 0 & 0 & 0 & 0 \\
0 & 4 & 0 & 0 & 0 & 0 & 0 \\
0 & 0 & 4 & 0 & 0 & 0 & 0 \\
0 & 0 & 0 & 4 & 0 & 0 & 0 \\
0 & 0 & 0 & 0 & 6 & 0 & 0 \\
0 & 0 & 0 & 0 & 0 & 6 & 0 \\
0 & 0 & 0 & 0 & 0 & 0 & 6 
\end{bmatrix}$, 
$\boldsymbol{\Sigma}_{cor1} = \begin{bmatrix}
4 & 1 & 1 & 1 & 1 & 1 & 1 \\
1 & 8 & 1 & 1 & 1 & 1 & 1 \\
1 & 1 & 4 & 1 & 1 & 1 & 1 \\
1 & 1 & 1 & 4 & 1 & 1 & 1 \\
1 & 1 & 1 & 1 & 6 & 1 & 1 \\
1 & 1 & 1 & 1 & 1 & 8 & 1 \\
1 & 1 & 1 & 1 & 1 & 1 & 8 
\end{bmatrix}$,  
$\boldsymbol{\Sigma}_{cor2} = \begin{bmatrix}
4 & 2 & 2 & 2 & 2 & 2 & 2 \\
2 & 8 & 2 & 2 & 2 & 2 & 2 \\
2 & 2 & 8 & 2 & 2 & 2 & 2 \\
2 & 2 & 2 & 4 & 2 & 2 & 2 \\
2 & 2 & 2 & 2 & 6 & 2 & 2 \\
2 & 2 & 2 & 2 & 2 & 8 & 2 \\
2 & 2 & 2 & 2 & 2 & 2 & 10 
\end{bmatrix}$ } \\ \cmidrule(l){1-5}
\multicolumn{5}{l}{
$\tau_g$ denotes the mixture proportion of cluster $g$; $\boldsymbol{J}_q$ denotes the $q\times q$ dimensional all-one matrix. }\\  \bottomrule
\end{tabular} }
\label{tab:SimSetCont}
\end{table}

The three simulated covariance structures were EEI(Data[DDI]): diagonal matrix $\boldsymbol{\Sigma}_{ind}$ with independent continuous variables, EEE (Data[EEE]): moderate correlation matrix $\boldsymbol{\Sigma}_{cor1}$ with correlations between 0.1 and 0.25, and VVV(Data[VVV]): stronger correlation matrix $\boldsymbol{\Sigma}_{cor2}$ with correlations from 0.2 to 0.5. 

\begin{table}[!htbp]
\caption{Simulated three-cluster mixed data categorical variable generation parameters.}
\small\center
\scalebox{0.9}{
\begin{tabular}{l|c|ccc}
\toprule	
%\multirow{2}{*}{\textbf{\begin{tabular}[c]{@{}l@{}}Categorical\\ Variable\end{tabular}}}
\multirow{2}{*}{\textbf{Variable}} & \multirow{2}{*}{\textbf{\#Levels}} & \multicolumn{3}{c}{\textbf{Three Simulated Clusters}} \\
 &  & \textbf{Cluster 1} & \textbf{Cluster 2} & \textbf{Cluster 3} \\ \cmidrule(l){1-5}
$X_8^{*,a}$ & 2 & $\text{Bin}(\theta_{X_8}),\theta_{X_8} \approx 0.5$ & $\text{Bin}(\theta_{X_8}),\theta_{X_8} \approx 0.3$ & $\text{Bin}(\theta_{X_8}),\theta_{X_8} \approx 0.9$ \\
$X_9^*$ & 3 & $M_3(0.6, 0.2, 0.2)$ & $M_3(0.2, 0.6, 0.2)$ & $M_3(0.2, 0.2, 0.6)$ \\
$X_{10}^*$ & 4 & $M_4(0.5, 0.3, 0.1, 0.1)$ & $M_4(0.1, 0.5, 0.3, 0.1)$ & $M_4(0.1, 0.1, 0.5, 0.3)$ \\
$X_{11}^b$ & 2 & $\text{Bin}(\theta_{X_{11}}),\theta_{X_{11}} \approx 0.5$ & $\text{Bin}(\theta_{X_{11}}),\theta_{X_{11}} \approx 0.5$ & $\text{Bin}(\theta_{X_{11}}),\theta_{X_{11}} \approx 0.5$ \\
$X_{12}$ & 3 & $M_3(1/3, 1/3, 1/3)$ & $M_3(1/3, 1/3, 1/3)$ & $M_3(1/3, 1/3, 1/3)$ \\
$X_{13}$ & 4 & $M_4(1/4, 1/4, 1/4, 1/4)$ & $M_4(1/4, 1/4, 1/4, 1/4)$ & $M_4(1/4, 1/4, 1/4, 1/4)$ \\
$X_{14}$ & 4 & \multicolumn{1}{c}{$M_4(0.1, 0.2, 0.3, 0.4)$} & \multicolumn{1}{c}{$M_4(0.1, 0.2, 0.3, 0.4)$} & \multicolumn{1}{c}{$M_4(0.1, 0.2, 0.3, 0.4)$} \\ \cmidrule(l){1-5}
\multicolumn{5}{l}{\begin{tabular}[c]{@{}l@{}}*Dominant categorical variables $X_8, X_9, X_{10}$ truly contribute to cluster assignment.\\$^aX_8$ is generated from true dominant continuous variables $X_3$ and $X_4$ set in Table \ref{tab:SimSetCont}.\\$ X_8 \sim \text{Bin}(\theta_{X_8})$ with $\text{logit} \left(\theta_{X_8}\right )=0.1 \times X_3+0.5\times X_4$.\\ $^bX_{11}$ is generated from true noise continuous variables $X_6$ and $X_7$ set in Table \ref{tab:SimSetCont}.\\$ X_{11} \sim \text{Bin}(\theta_{X_{11}})$ with $\text{logit}\left(\theta_{X_{11}}\right )=0.1 \times X_6+0.5 \times X_7$. \end{tabular}}\\ \bottomrule
\end{tabular} }
\label{tab:SimSetCat}
\end{table}

In the censored scenarios, continuous variables $X_3$, $X_4$, and $X_5$ were censored at either $20\%$ (using the 10th and 90th percentiles) or $40\%$ (using the 20th and 80th percentiles) of their own distributions.

Hyperparameters were selected to reflect minimal prior information while maintaining stability. Specifically, $\boldsymbol{\delta}=(1/3, 1/3, 1/3)^T$ was used for the Dirichlet prior on $\boldsymbol{\tau}$, and the inverse gamma prior for the spike variance parameter $\sigma^2_{\Delta_0}$ used $a_{\Delta_0}=2$ and $b_{\Delta_0}=0.005$. The variance ratio $\omega$ was tuned using the data-adaptive strategy described in Section 2.5 with $k=75$, which was also ensured by convergence diagnostics results. Prior specifications for the Dirichlet spike components were set based on the empirical marginal distributions of the categorical variables.

For convergence assessment, we applied the multivariate potential scale reduction factor (MPSRF) from Brooks and Gelman's diagnostic \cite{ConvDiag}. Convergence was confirmed across all scenarios using 500 Gibbs sampling iterations with a 200-iteration burn-in period, and this setup was used for all simulation replicates.

%%%%%%%%%%%%%%%%%%%%%%%%%%%%%%%%%%%%%%%%%%%%%%%%%%%
%%% 9 simulation scenarios 
%%%%%%%%%%%%%%%%%%%%%%%%%%%%%%%%%%%%%%%%%%%%%%%%%%%
\subsection{Simulation Results}

We evaluated clustering performance of BFMM (under EEI, EEE, and VVV configurations) and compared it to several competing methods for mixed or continuous data, including 
\begin{itemize}
    \item Model-based methods for mixed data: \texttt{clustMD} \cite{clustMD2016}, latent class analysis \texttt{LCA} \cite{LCA}, and \texttt{VarSelLCM} \cite{VarSelLCM};
    \item Distance-based methods: K-prototype \cite{kprototype}, PAM with Gower distance \cite{PAM, Gower}, and HyDaP \cite{HyDaP};
    \item Methods for continuous data (categorical treated as continuous): K-means \cite{KMeans}, and \texttt{mclust} \cite{mclust5}.
\end{itemize}

\begin{figure}[h]
\small\centering
  \includegraphics[width=1\linewidth]{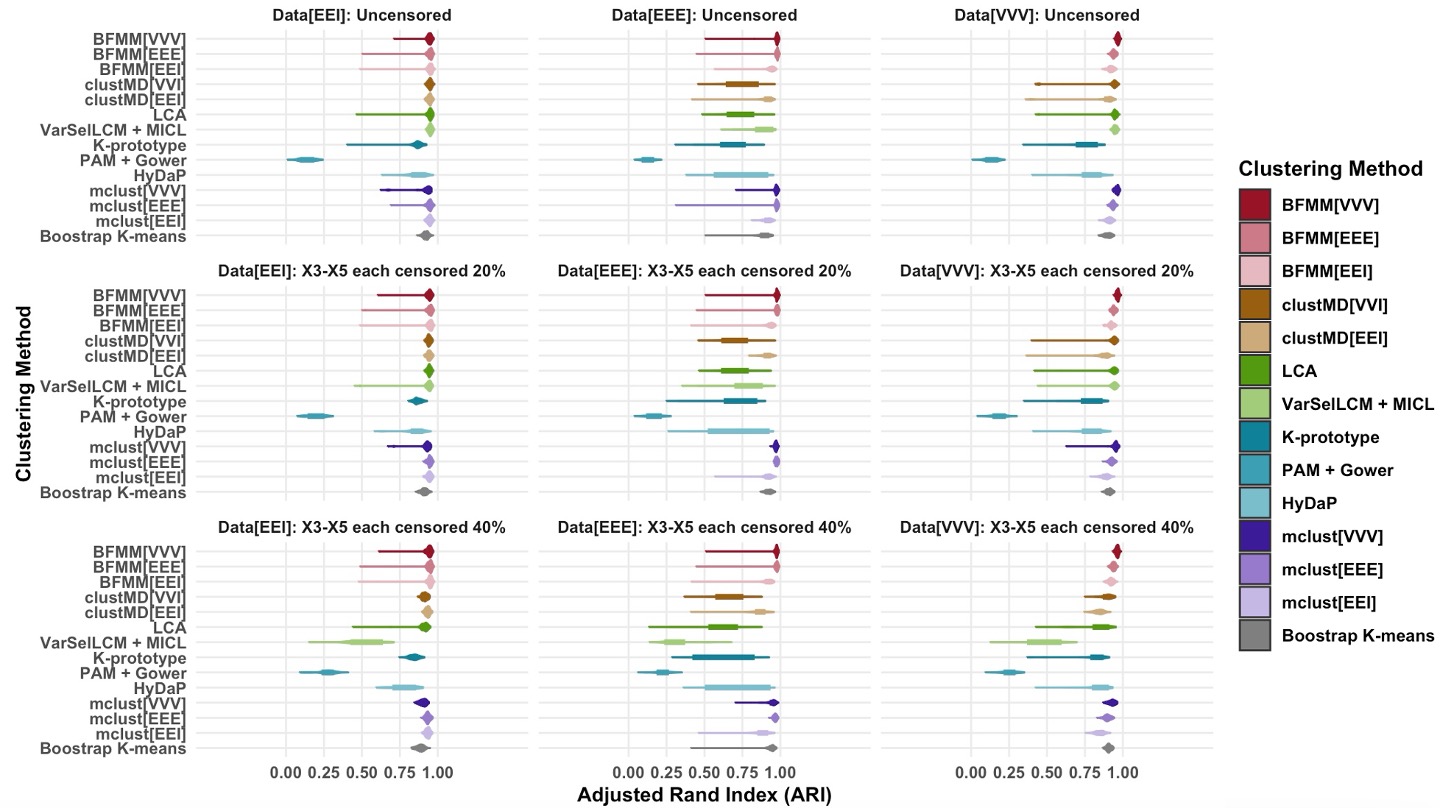}
  \caption{\small Violin plots summarizing adjusted rand index (ARI) by different clustering methods in each simulated scenario. Categorical variables in simulated datasets were treated as continuous when applying the \texttt{mclust} and bootstrap K-means clustering methods. }
  \label{fig:SimAll.ARI}
\end{figure}

All methods were initialized using cluster centers derived from bootstrap K-means clustering results based on continuous variables only. Clustering accuracy was assessed using the Adjusted Rand Index (ARI) \cite{ARI}. Across 100 simulation runs in each scenario, we summarized 
the failure rate in Table~\ref{tab:SimAll.FailCnt}, as well as the median ARI and its corresponding 2.5th-97.5th percentile interval excluding the failed runs in Table~\ref{tab:SimAll.ARI} and Figure~\ref{fig:SimAll.ARI}. The variable importance weights derived by BFMM are shown in Table~\ref{tab:SimAll.VI} (see Appendix~\ref{si.tables}).

%%% Simulation - Clustering Accuracy using ARI %%%
BFMM consistently achieved high clustering accuracy across all settings. For datasets with independent variables (Data[EEI]), BFMM([EEI]) achieved median ARI of 0.950 (0.923 - 0.974) without censoring, and remained robust under 20\% and 40\% censoring, with median ARIs of 0.950 and 0.949, respectively. BFMM[EEE] and BFMM[VVV] also performed well, with median ARIs $\geq$ 0.944 across different censoring levels (see Appendix~\ref{si.tables}, Table~\ref{tab:SimAll.ARI}). 

For datasets with moderate correlations (Data[EEE]), BFMM[EEE] outperformed all methods: median ARI = 0.981 (0.966 - 0.992) without censoring, 0.979 (0.962 - 0.991) with 20\% censoring, and 0.978 (0.957 - 0.991) with 40\% censoring. BFMM[VVV] consistently ranked second, with median ARI $\geq$0.976. In contrast, models assuming local independence (e.g., BFMM[EEI], LCA, K-prototype) showed reduced performance under this structure.  

For strongly correlated datasets (Data[VVV]), BFMM[VVV] provided the most stable and accurate clustering results across all censoring levels, achieving median ARIs of 0.967 (0.951 - 0.982) with no censoring, 0.966 (0.949 - 0.981) with 20\% censoring, and 0.964 (0.941 - 0.978) with 40\% censoring. Its relative advantage increased with the level of censoring. In particular, PAM with Gower distance performed the worst across all scenarios (median ARI $<0.3$, see Appendix~\ref{si.tables}, Table~\ref{tab:SimAll.ARI}).

%%% Simulation - Failure rate %%%
Table \ref{tab:SimAll.FailCnt} reports the failure rates of each method. All BFMM variants successfully converged in all scenarios (0\% failure rate). In contrast, \texttt{mclust}[VVV] exhibited failure rates as high as 65\%, likely due to overparameterization and instability in estimating full covariance matrices via EM. \texttt{clustMD} also showed instability in some cases.

%%% Simulation - Variable Importance %%%
Table \ref{tab:SimAll.VI} presents average variable importance weights estimated by BFMM. Across all scenarios, the model reliably assigned high important weights ($0.72-1.00$) to the true dominant continuous variables ($X_1-X_4$) and low weights ($<0.21$) to noise variables ($X_6-X_7$). The weakly informative variable $X_5$ was assigned intermediate weights ($0.33-0.55$), consistent with its contribution level. Similarly, categorical variables $X_8-X_{10}$ had weights between 0.52 and 0.83, while categorical noise variables $X_{11}-X_{14}$ were correctly down-weighted ($<0.25$). These results confirm BFMM's ability to distinguish informative from noninformative features in both continuous and categorical domains.

%%%%%%%%%%%%%%%%%%%%%%%%%%%%%%%%%%%%%%%%%%%%%%%%%%%
%%% Section 5 - Real Data Application %%%
%%%%%%%%%%%%%%%%%%%%%%%%%%%%%%%%%%%%%%%%%%%%%%%%%%%
\section{Applications}
We illustrate the use of BFMM in two clinical applications with complementary goals. The first leverages the large EHR-based SENECA cohort to identify clinically meaningful sepsis phenotypes. The second applies BFMM to the EDEN randomized trial of feeding strategies in acute lung injury to examine the heterogeneity in treatment effects and explore subgroups that may benefit from trophic feeding. Together, these examples highlight the capability of BFMM to incorporate censored biomarkers, adapt covariance structure choice to dataset size, support phenotype discovery, and prioritize influential variables while down-weighting noise in diverse clinical research contexts.

\subsection{Sepsis Phenotyping using EHR: SENECA Application}

We first applied the BFMM approach to the Sepsis ENdotyping in Emergency Care (SENECA) data, a large EHR-derived cohort capturing detailed clinical information from sepsis patients. This curated dataset contained $N=20,189$ observations, with 26 continuous and 2 categorical variables, representing routinely collected EHR measurements. These include vital signs, laboratory results, demographic information, and comorbidities relevant to early sepsis evaluation. 

Our analytic objectives were twofold: (1) to identify data-driven patient clusters representing clinically meaningful sepsis phenotypes, and (2) to quantify the relative importance of each variable for distinguishing these phenotypes. BFMM's ability to jointly model mixed-type variables is particularly suitable for this high-dimensional EHR context.

We compared BFMM results with those from \texttt{mclust}[VVV], a method that handles continuous variables but not categorical variables or censored values. Before clustering, continuous variables were log-transformed (if skewed) and standardized. The hyperparameters used in this application followed the specification in Section 2.5, including a balanced Dirichlet prior for the mixing proportions $\boldsymbol{\delta} = \frac{1}{G} \cdot \boldsymbol{1}^T_{G}$, and spike Dirichlet priors for categorical variables set in proportion to their observed marginal distributions.

To ensure convergence, we applied Brooks and Gelman's MPSRF to assess 7 independent Markov's chains with varying variance ratio $\omega$. For the most complex model, BFMM[VVV] with 4 clusters, the MPSRF value was 1.181, confirming adequate convergence. We applied this same burn-in strategy (50\% of 10,000 iterations) across all BFMM models.

\begin{figure}[!htbp]
\centering
  \includegraphics[width=.9\linewidth]{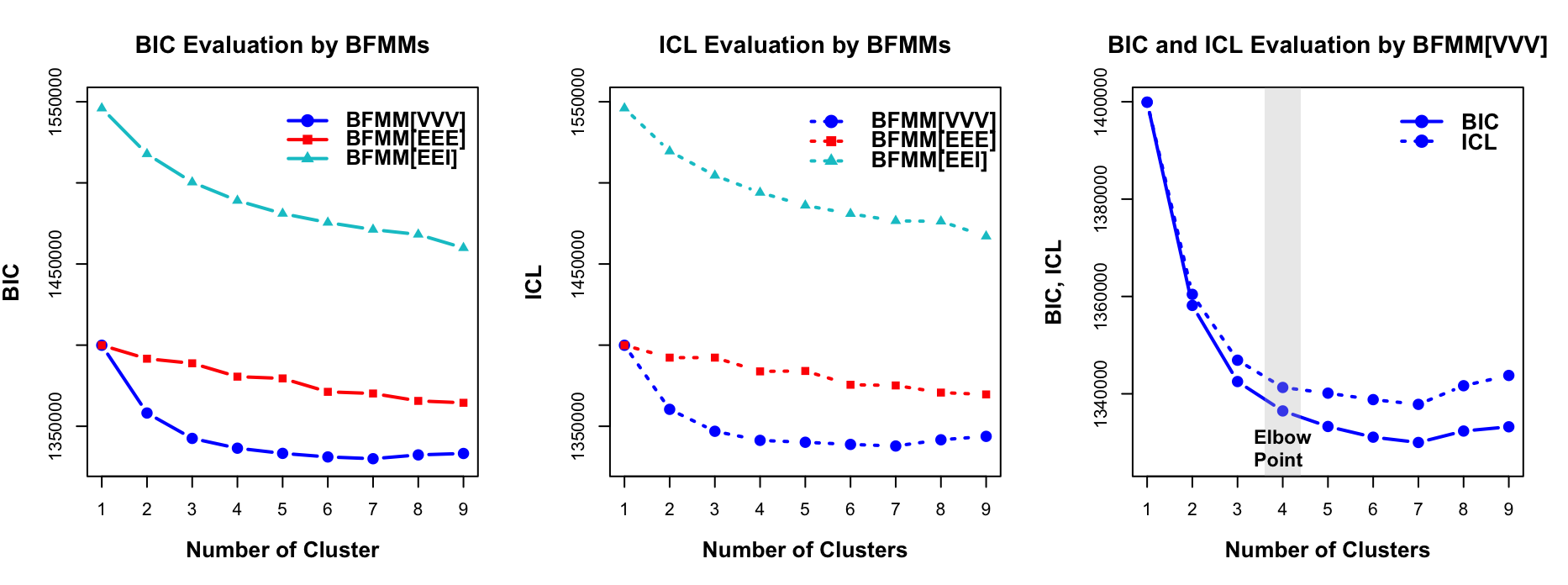}
  \caption{\small BIC and ICL evaluation for SENECA data by BFMM given $G=1, \dots, 9$ clusters. Information criteria evaluation used Markov chains each taking 10000 Gibbs sampling iterations with 5000 for burn-in.}
  \label{fig:BICandICL.seneca}
\end{figure}

Model selection was guided by BIC and ICL. Figure \ref{fig:BICandICL.seneca} shows that BFMM[VVV] consistently achieved the lowest BIC and ICL values among the three covariance structures, with a clear ``elbow" at $G=4$. Based on this and clinical interpretability, we selected BFMM[VVV] with 4 clusters as the optimal model.

BFMM[VVV] identified four clusters of sizes 8,395, 5,628, 3,455, and 2,711. Comparison with \texttt{mclust}[VVV] revealed 81.5\% label concordance and an ARI of 0.675 (see Table~\ref{tab:CompareSeneca}), suggesting consistent but enhanced granularity due to BFMM's inclusion of categorical variables. BFMM also yielded quantitative variable importance weights, identifying six top key contributors to clustering (averaged weights $\geq$0.85) across chains): troponin, AST, lactate, Glasgow Coma Scale (GCS), systolic blood pressure (SBP), and bicarbonate (see Table~\ref{tab:AvgVarImpSeneca}). In contrast, demographic variables (i.e. age, gender) and several laboratory values (i.e. sodium, temperature, creatinine) carried little importance weight $<$0.5, indicating minimal contribution to clustering. The unaveraged importance weights estimated by each Markov chain are provided in Table~\ref{tab:VarImpSeneca} (see Appendix~\ref{si.tables}).

\begin{table}[h]
\small\center
\caption{A comparison between the 4 clusters identified by BFMM[VVV] and by mclust[VVV].}
\scalebox{.95}{\begin{tabular}{ll|cccc}
\toprule	
 &  & \multicolumn{4}{c}{\textbf{BFMM[VVV]}} \\
 &  & Cluster 1 & Cluster 2 & Cluster 3 & Cluster 4 \\ \cmidrule(l){1-6}
\multirow{4}{*}{\textbf{\begin{tabular}[c]{@{}l@{}}mclust{[}VVV{]}\\ (use continuous data)\end{tabular}}} & Cluster 1 & 7133 (85.0\%) & \;\;31 (0.6\%) & 0 (0\%) & \;\;53 (2.0\%) \\
 & Cluster 2 & 137 (1.6\%) & 5274 (93.7\%) & \;416 (12.0\%) & 267 (9.8\%) \\
 & Cluster 3 & 1109 (13.2\%) & 294 (5.2\%) & 1846 (53.4\%) & 181 (6.7\%) \\
 & Cluster 4 & \;\;16 (0.2\%) & \;\;29 (0.5\%) & 1193 (34.5\%) & 2210 (81.5\%) \\ \cmidrule(l){1-6}
\multicolumn{2}{l|}{\textbf{Subtotal}} & 8395 (100\%) & 5628 (100\%) & 3455 (100\%) & 2711 (100\%) \\ \cmidrule(l){1-6}
\multicolumn{6}{l}{Total = 20,189 observations with overall concordance rate=0.815 and ARI=0.675.} \\\bottomrule
\end{tabular}}
\label{tab:CompareSeneca}
%\end{table}
\bigskip
\bigskip
%%%%%%%%%%%%%%%%%%%%%%%%%%
%\begin{table}[]
\caption{SENECA averaged variable importance weights by BFMM[VVV] given 4 clusters.}
\scalebox{0.81}{
\begin{tabular}{lclclc|lc}
\toprule
\multicolumn{6}{c|}{\textbf{Continuous Variables}} & \multicolumn{2}{c}{\textbf{Categorical Variables}} \\
\cmidrule(l){1-8}
\textbf{Variable} & \textbf{Weight$^b$} & \textbf{Variable} & \textbf{Weight$^b$} & \textbf{Variable} & \textbf{Weight$^b$} & \textbf{Variable} & \textbf{Weight$^b$} \\
\cmidrule(l){1-8}
Troponin$^a$ & 0.99 & ALT & 0.76 & White blood cell & 0.56 & \multirow{2}{*}{\begin{tabular}[c]{@{}l@{}}Sex\\ (female/male)\end{tabular}} & \multirow{2}{*}{0.22} \\
AST$^a$ & 0.97 & RR & 0.75 & Chloride & 0.52 &  &  \\
Lactate$^a$ & 0.94 & Heart rate & 0.73 & Hemoglobin & 0.51 & \multirow{2}{*}{\begin{tabular}[c]{@{}l@{}}Race\\ (white/black/others)\end{tabular}} & \multirow{2}{*}{0.07} \\
GCS$^a$ & 0.92 & INR & 0.72 & Glucose & 0.51 &  &  \\
SBP$^a$& 0.89 & CRP & 0.70 & Sodium & 0.43 &  &  \\
Bicarbonate$^a$ & 0.85 & ESR & 0.70 & Temperature & 0.36 &  &  \\
SaO2 & 0.83 & Bilirubin & 0.65 & Creatinine & 0.32 &  &  \\
Albumin & 0.83 & PaO2 & 0.60 & Age & 0.21 &  &  \\
Bands & 0.78 & Platelets & 0.58 &  &  &  &  \\
\cmidrule(l){1-8}
\multicolumn{8}{l}{Abbreviations: AST: aspartate aminotransferase; GCS: Glasgow coma scale; SBP: systolic blood pressure;}\\
\multicolumn{8}{l}{SaO2: oxygen saturation; ALT: alanine aminotransferase; RR: respiration rate; INR: international normalized ratio;}\\
\multicolumn{8}{l}{CRP: C-reactive protein; ESR: erythrocyte sedimentation rate; PaO2: partial pressure of oxygen.} \\  
\multicolumn{8}{l}{$^a$Identified most important variables with the highest averaged importance weights $\geq 0.85$.} \\
\multicolumn{8}{l}{\small $^b$Variable importance weight averaged across 7 different Markov chains that have reached convergence.}\\
 \bottomrule
\end{tabular} }
\label{tab:AvgVarImpSeneca}
\end{table}
%%%%%%%%%%%%%%%%%%%%%%%%%%%%%%%%%%%%%%%%%%%%%%%%%%%%%%%%%%

%%%%%%%%%%%%%%%%%%%%%%%%%%%%%%%%%%%%%%%%%%%%%%%%%%%
%%%%%%%%%%%%%%%%%%%%%%%%%%%%%%%%%%%%%%%%%%%%%%%%%%%
\begin{table}[!htbp]\center
\caption{\fontsize{12pt}{12pt}\selectfont Distributions of all variables used to cluster SENECA data by BFMM[VVV].}
\scalebox{0.8}{
\begin{tabular}{l|ccccc|c}
\toprule
\multirow{2}{*}{\textbf{\begin{tabular}[c]{@{}l@{}}SENECA Data\\ Variable\end{tabular}}} & \multicolumn{5}{c|}{\textbf{Cluster assigned by BFMM{[}VVV{]} with $G = 4$ and $k = 75$}} & \multicolumn{1}{l}{\multirow{2}{*}{\textbf{\begin{tabular}[c]{@{}c@{}}Avg.$^c$\\ Weight\end{tabular}}}} \\
 & \textbf{Cluster 1} & \textbf{Cluster 2} & \textbf{Cluster 3} & \textbf{Cluster 4} & \textbf{Overall} & \multicolumn{1}{l}{} \\
\cmidrule(l){1-7}
\textbf{Cluster size} & \textbf{\begin{tabular}[c]{@{}c@{}}8395\\ (41.6\%)\end{tabular}} & \textbf{\begin{tabular}[c]{@{}c@{}}5628\\ (27.9\%)\end{tabular}} & \textbf{\begin{tabular}[c]{@{}c@{}}3455\\ (17.1\%)\end{tabular}} & \textbf{\begin{tabular}[c]{@{}c@{}}2711\\ (13.4\%)\end{tabular}} & \textbf{\begin{tabular}[c]{@{}c@{}}20189\\ (100\%)\end{tabular}} & \multicolumn{1}{l}{} \\
\cmidrule(l){1-7}
\multicolumn{7}{l}{\textbf{Continuous variable$^a$, mean (SD) or median [IQR] at original scale}} \\
\cmidrule(l){1-7}
Troponin, ng/mL & 0.1 (0.1) & 0.1 (0.1) & 2.0 (5.4) & 0.8 (3.0) & 0.5 (2.6) & 0.99 \\
AST, U/L & 26.0 [22.0] & 26.0 [19.0] & 72.0 [163.5] & 40.0 [57.0] & 30.0 [34.0] & 0.97 \\

Lactate, mmol/L & 1.3 [1.1] & 1.4 [1.1] & 2.0 [2.2] & 2.3 [2.9] & 1.5 [1.4] & 0.94 \\

GCS score & 14.8 (0.4) & 12.1 (2.5) & 11.8 (3.8) & 10.5 (4.31) & 12.9 (3.1) & 0.92 \\

SBP, mm Hg & 115 [32] & 115 [33] & 100 [34] & 95 [37] & 110 [35] & 0.89 \\

Bicarbonate, mEq/L & 25.8 (4.6) & 25.8 (4.4) & 23.2 (4.8) & 23.2 (7.1) & 25.0 (5.1) & 0.85 \\
SaO2, mm Hg & 93.7 (3.9) & 94.3 (3.6) & 94.5 (4.0) & 85.5 (13.3) & 92.9 (6.7) & 0.83 \\
Albumin, g/dL & 3.0 (0.7) & 3.1 (0.7) & 2.7 (0.7) & 2.6 (0.7) & 2.9 (0.7) & 0.83 \\
Band neutrophils, \% & 5.0 [9.5] & 5.0 [9.0] & 9.0 [15.0] & 9.0 [14.2] & 6.0 [12.0] & 0.78 \\
ALT, U/L & 29.0 [22.0] & 28.0 [21.0] & 45.0 [106.5] & 33.0 [35.0] & 31.0 [28.0] & 0.76 \\

RR, breaths/min & 21.3 (4.9) & 20.6 (4.0) & 23.3 (7.0) & 26.6 (9.2) & 22.2 (6.2) & 0.75 \\
Heart rate, beats/min & 94.7 (19.9) & 92.7 (20.2) & 101.3 (23.1) & 109.0 (24.7) & 97.2 (21.9) & 0.73 \\

INR & 1.2 [0.2] & 1.4 [1.1] & 1.6 [1.2] & 1.3 [0.6] & 1.3 [0.5] & 0.72 \\
CRP, mg/L & 5.1 [12.9] & 4.8 [12.4] & 8.9 [21.5] & 8.9 [20.0] & 5.8 [14.6] & 0.70 \\
ESR, mm/h & 48 [58] & 41 [46] & 31 [44] & 48 [60] & 45 [52] & 0.70 \\
Bilirubin, mg/dL & 0.8 [0.8] & 0.7 [0.7] & 1.3 [2.1] & 0.8 [0.7] & 0.8 [0.8] & 0.65 \\
PaO2, mm Hg & 75.0 [42.0] & 84.0 [62.3] & 97.2 [85.0] & 83.4 [88.0] & 81.0 [60.9] & 0.60 \\

Platelets, $\times10^9$/L & 186 [118] & 197 [120] & 137 [119] & 244 [196] & 187 [126] & 0.58 \\

WBC, $\times10^9$/L & 9.5 [6.8] & 9.5 [6.2] & 9.5 [10.2] & 12.6 [10.0] & 9.9 [7.5] & 0.56 \\
Chloride, mEq/L & 102.0 (5.7) & 102.9 (5.6) & 103.7 (6.7) & 104.1 (9.9) & 102.8 (6.6) & 0.52 \\
Hemoglobin, g/dL & 11.6 (2.2) & 11.7 (2.2) & 11.1 (2.4) & 11.3 (2.5) & 11.5 (2.3) & 0.51 \\
Glucose, mg/dL & 4.8 [0.5] &	4.9 [0.5]	& 4.9 [0.5] &	5.1 [0.8]	& 4.9 [0.5] & 0.51 \\

Sodium, mEq/L & 136.7 (4.5) & 137.4 (4.3) & 136.7 (5.6) & 138.68 (8.5) & 137.1 (5.4) & 0.43 \\
Temperature, $^\circ$C & 37.0 (0.9) & 37.0 (0.9) & 37.0 (1.2) & 36.7 (1.4) & 37.0 (1.0) & 0.36 \\
Creatinine, mg/dL & 1.4 [1.1] & 1.3 [1.0] & 1.4 [1.4] & 1.5 [1.5] & 1.4 [1.2] & 0.32 \\
Age, y & 64.6 (17.0) & 65.7 (17.0) & 62.6 (16.8) & 63.4 (17.8) & 64.4 (17.1) & 0.21 \\
\cmidrule(l){1-7}
\multicolumn{7}{l}{\textbf{Categorical variable$^b$, n (\%)}} \\
\cmidrule(l){1-7}
Sex &  &  &  &  &  &  \\
  \;\;\;\;female & 4275 (50.9\%) & 2900 (51.5\%) & 1589 (46.0\%) & 1403 (51.8\%) & 10167 (50.4\%) & \multirow{2}{*}{0.22} \\
  \;\;\;\;male & 4120 (49.1\%) & 2728 (48.5\%) & 1866 (54.0\%) & 1308 (48.2\%) &  10022 (49.6\%) &  \\
Race &  &  &  &  &  &  \\
\;\;\;\;white & 6586 (78.5\%) & 4502 (80.0\%) & 2603 (75.3\%) & 1949 (71.9\%) & 15640 (77.5\%) & \multirow{3}{*}{0.07} \\
\;\;\;\;black & 1037 (12.4\%) & \;654 (11.6\%) & 332 (9.6\%) & \;405 (14.9\%) &  \;2428 (12.0\%) &  \\
\;\;\;\;others & 772 (9.2\%) & 472 (8.4\%) & \;520 (15.1\%) & \;357 (13.2\%) &  \;2121 (10.5\%) &  \\
\cmidrule(l){1-7}
\multicolumn{7}{l}{Abbreviations: SD: standard deviation; IQR: interquartile range, equal to Q3 - Q1; Avg.:averaged;} \\
\multicolumn{7}{l}{AST: aspartate aminotransferase;  GCS: Glasgow coma scale; SBP: systolic blood pressure; } \\
\multicolumn{7}{l}{SaO2: oxygen saturation; ALT: alanine aminotransferase; RR: respiration rate; }\\
\multicolumn{7}{l}{INR: international normalized ratio; CRP: C-reactive protein; ESR: erythrocyte sedimentation rate; }\\
\multicolumn{7}{l}{PaO2: partial pressure of oxygen; WBC: white blood cell.} \\ 
\multicolumn{7}{l}{\begin{tabular}[c]{@{}l@{}}$^{a,b}$Significant difference exists among the means or distributions across the 4 clusters for all analyzed variables; \\
$ $ $ $ $ $ $\textit{P} <.0001$ by one-way ANOVA or Chi-squared test (with log-transformation if needed). \\
\;\;\;\;Median [IQR] is reported for continuous variable with skewed distribution that needs log-transformation. \end{tabular}}\\
\multicolumn{7}{l}{$^c$Variable importance weight averaged across 7 different Markov chains that have reached convergence.}\\ 
\bottomrule
\end{tabular}}
\label{tab:AllVarSeneca}
\end{table}
%%%%%%%%%%%%%%%%%%%%%%%%%%%%%%%%%%%%%%%%
%%%%%%%%%%%%%%%%%%%%%%%%%%%%%%%%%%%%%%%%

Post-hoc analyses further validated these findings. Distributions of all analyzed SENECA variables used for clustering are provided in Table \ref{tab:AllVarSeneca}. Table~\ref{tab:PostHocSeneca} summarizes the distributions of the six external clinical endpoints not used in clustering (see Appendix~\ref{si.tables}).

\begin{figure}[h]
\small\centering
\includegraphics[width=0.9\linewidth]{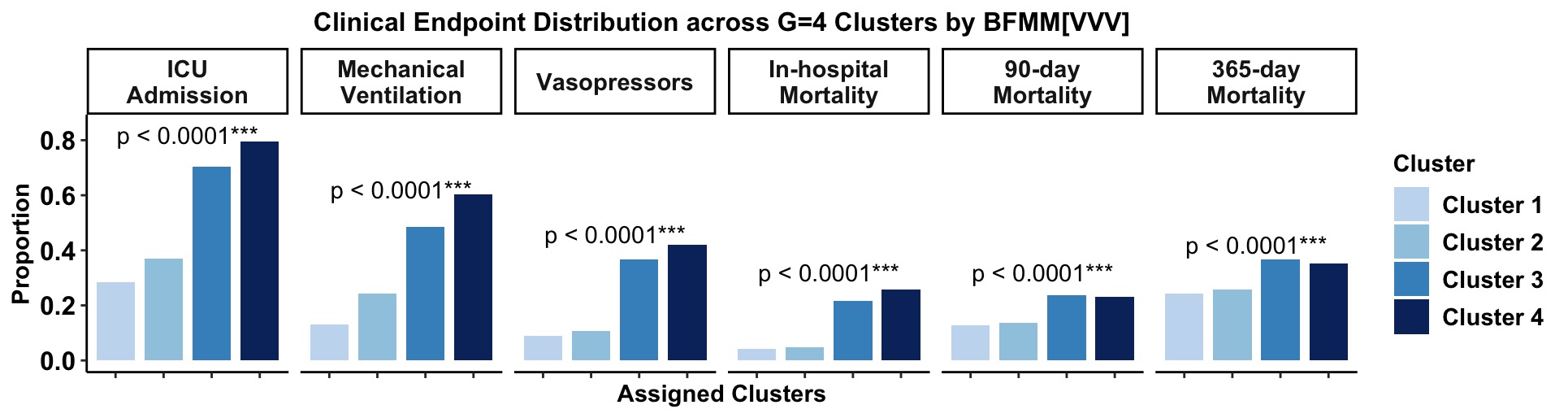}
  \caption{\small Distributions of the six clinical endpoints by BFMM[VVV] clustering results of SENECA dataset. The annotated \textit{p}-values are based on Chi-squared test comparing the proportions among the four clusters.}
  \label{fig:PostHocSeneca}
\end{figure}

Table~\ref{tab:AllVarSeneca} shows significant between-cluster differences in all six influential variables with averaged weights $\geq 0.85$ ($p<$0.0001), supporting their role in defining clinically distinct sepsis phenotypes. Moreover, clsuter-level summaries of six clinical endpoints, including ICU admission, mechanical ventilation, vasoprssor use, and in-hospital, 90-day, and 365-day mortality, revealed significant variation across clusters (all $p<$0.0001, see Table~\ref{tab:PostHocSeneca}). In accordance with the Table ~\ref{tab:PostHocSeneca}, Figure~\ref{fig:PostHocSeneca} visually shows that clinical severity indicators increased progressively from Cluster 1 to Cluster 4, underscoring the prognostic relevance of these phenotypes.

Combining all the post-hoc analyses results, Clusters 1-4 revealed four potential sepsis phenotypes, respectively. Cluster 1 represented a low-risk sepsis phenotype with preserved organ function and favorable outcomes. Cluster 2 characterized a different low-risk sepsis phenotype with impaired neurological status despite largely normal laboratory findings, suggesting neurologic dysfunction. Clusters 3-4 uncovered sepsis phenotypes of higher risk. In specific, Cluster 3 showed hepatic dysfunction, cardiac injury, and tissue hypoperfusion (elevated AST, troponin, lactate; low bicarbonate) and had high ICU admission and mortality rates; while Cluster 4 reflected classic septic shock with hypotension, acidosis, depressed mental status, and the worst prognosis.

This application illustrates the capacity of BFMM to uncover interpretable and prognostically relevant phenotypes from high-dimensional EHR data while identifying influential variables.

%%%%%%%%%%%%%%%%%%%%%%%%%%%%%%%%%%%%%%%%%%%%%%%%%%%%%%%%%%
%%%%%%%%%%%%%%%%%%%%%%%%%%%%%%%%%%%%%%%%%%%%%%%%%%%%%%%%%%
\subsection{Acute Lung Injury Subgroups: EDEN Trial Application}

Next, we applied BFMM to the EDEN tiral dataset from the ARDS Network, a randomized controlled trial of enteral feeding strategies for patients with acute lung injury (ALI). This dataset included $N=889$ patients and 37 variables (29 continuous and 8 categorical), capturing baseline clinical data, comorbidities, and biomarker relevant to ALI pathophysiology.

Seven of the continuous variables were biomarkers subject to limits of detection: angiopoietin, IL-6, IL-8, RAGE, TNF-$\alpha$, PCT, and ST2. Several contained left- or left-censored values. This dataset offered an opportunity to demonstrate BFMM's strength in handling censoring alongside variable selection and mixed data types.

Our primary goals were to (1) identify clinically distinct subgroups within a trial population considered homogeneous at enrollment, and (2) characterize each cluster based on biomarker patterns and clinical outcomes. Hyperparameters followed the default structure used in the SENECA application, with spike Dirichlet priors calibrated to the observed marginal distributions. 

\begin{figure}[!htbp]
\centering
  \includegraphics[width=.9\linewidth]{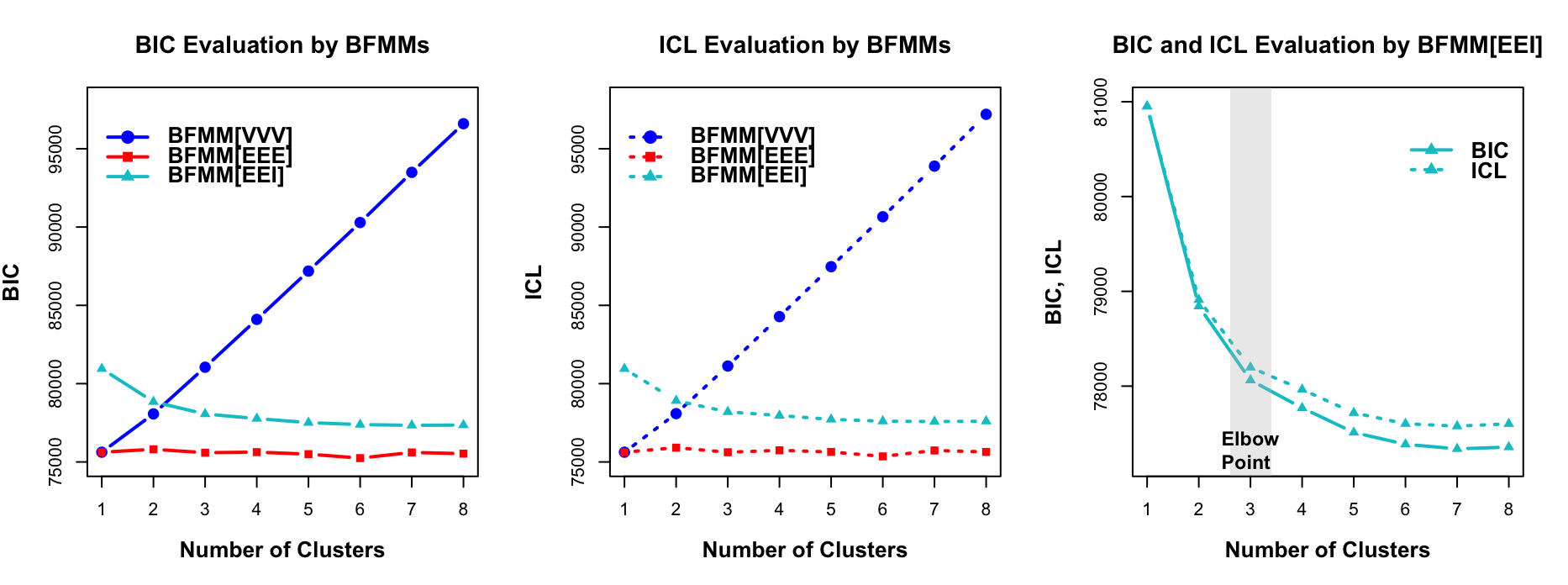}
  \caption{\small BIC and ICL evaluation for EDEN trial data by BFMM given $G=1, \dots, 6$ clusters. Information criteria evaluation used Markov chains each taking 5000 Gibbs sampling iterations with 2500 for burn-in.}
  \label{fig:BICandICL.eden}
\end{figure}

We again verified convergence using MPSRF, achieving a value of 1.107 for BFMM[VVV] with 2 clusters. Model selection using BIC and ICL (Figure~\ref{fig:BICandICL.eden}) suggested no meaningful clustering under BFMM[VVV] or BFMM[EEE], as both criteria increased monotonically with $G$. However, under BFMM[EEI], a clear emerged at $G=3$, indicating improved model fit. We thus selected BFMM[EEI] with 3 clusters as the final model.

BFMM identified clusters of sizes 386, 282, and 221. Table~\ref{tab:AvgVarImpEDEN} reports the averaged variable importance weights across chains. The top nine continuous variables (averaged weights $\geq$0.80) included five biomarkers: RAGE, TNF-$\alpha$, IL-6, PCT, and IL-8; and four clinical parameters: blood urea nitrogen (BUN), creatinine, positive end-expiratory pressure (PEEP), and minute ventilation (MV). Only two categorical variables: diabetes and sepsis-induced lung injury, had averaged weights $\geq$0.65, indicating relevance for clustering. The unaveraged importance weights estimated by each Markov chain are provided in Table~\ref{tab:VarImpEDEN} (see Appendix~\ref{si.tables}).

%%%%%%%%%%%%%%%%%%%%%%%%%%%%%%%%%%%%%%%%
\begin{table}[!htbp]
\small\center
\caption{EDEN Trial averaged variable importance weights by BFMM[EEI] given 3 clusters.}
\scalebox{0.75}{
\begin{tabular}{lclclc|lc}
\toprule	
\multicolumn{6}{c|}{\textbf{Continuous Variables}} & \multicolumn{2}{c}{\textbf{Categorical Variables}} \\
\cmidrule(l){1-8}
\textbf{Variable} & \textbf{Weight$^b$} & \textbf{Variable} & \textbf{Weight$^b$} & \textbf{Variable} & \textbf{Weight$^b$} & \textbf{Variable} & \textbf{Weight$^b$} \\
\cmidrule(l){1-8}
RAGE & 1 & PaO2/FiO2 ratio & 0.77 & Temperature & 0.51 & Diabetes & 0.68 \\
TNF-$\alpha$ & 1 & APACHE III & 0.77 & Albumin & 0.50 & Sepsis-PLI & 0.65 \\
BUN & 1 & ST2$^a$ & 0.77 & Sodium & 0.42 & Pneumonia-PLI & 0.16 \\
Creatinine & 1 & Angiopoietin$^a$ & 0.76 & Total protein & 0.40 & Chronic pulmonary & 0.10 \\
IL-6 & 0.96 & Plateau pressure & 0.74 & Glucose & 0.28 & Sex (female/male) & 0.10 \\
PCT$^a$ & 0.92 & Age & 0.73 & Body mass index & 0.24 & Smoker & 0.08 \\
IL-8$^a$ & 0.90 & Bicarbonate & 0.73 & Tidal volume & 0.20 & Race (white/others) & 0.08 \\
PEEP & 0.85 & Platelets & 0.73 & White blood cell & 0.18 & Aspiration-PLI & 0.06 \\
MV & 0.80 & Heart rate & 0.68 & Mean arterial pressure & 0.18 & \multicolumn{1}{l}{} & \multicolumn{1}{l}{} \\
MAP & 0.78 & Hematocrit level & 0.64 &  &  & \multicolumn{1}{l}{} & \\
\cmidrule(l){1-8}
\multicolumn{8}{l}{\small Abbreviations: RAGE: receptor for advanced glycation end products; TNF: tumor necrosis factor; BUN: blood urea nitrogen; } \\ 
\multicolumn{8}{l}{\small IL: interleukin; PCT: procalcitonin; PEEP: positive end-expiratory pressure; MV: minute ventilation; } \\ 
\multicolumn{8}{l}{\small MAP: mean airway pressure; PaO2: arterial oxygen partial pressure; FiO2: fractional inspired oxygen;}\\ 
\multicolumn{8}{l}{\small APACHE: acute physiology and chronic health evaluation; ST: suppression of tumorigenicity; PLI: primary lung injury.}\\ 
\multicolumn{8}{l}{\small $^a$Biomarkers with censored observations.}\\
\multicolumn{8}{l}{\small $^b$Variable importance weight averaged across 7 different Markov chains that have reached convergence.}\\
\bottomrule
\end{tabular} }
  \label{tab:AvgVarImpEDEN}
%\end{table}

\smallskip

%%%%%%%%%%%%%%%%%%%%%%%%%%%%%%%%%%%%%%%%
%\begin{table}[!htbp]
\small\center
\caption{Distributions of identified important variables by BFMM[EEI] for EDEN trial data.}
\scalebox{0.89}{
\begin{tabular}{lccccc}
\toprule	
\multirow{2}{*}{\textbf{\begin{tabular}[c]{@{}l@{}}EDEN Trial Data\\ Variable\end{tabular}}} & \multicolumn{5}{c}{\textbf{Cluster assigned by BFMM[EEI] with $G=3$ and $k=75$}} \\
 & \textbf{Cluster 1} & \textbf{Cluster 2} & \textbf{Cluster 3} & \textbf{Overall} & \textbf{\textit{P} value$^d$} \\
 \cmidrule(l){1-6}
\textbf{Cluster size} & \textbf{\begin{tabular}[c]{@{}c@{}}386\\ (43\%)\end{tabular}} & \textbf{\begin{tabular}[c]{@{}c@{}}282\\ (32\%)\end{tabular}} & \textbf{\begin{tabular}[c]{@{}c@{}}221\\ (25\%)\end{tabular}} & \textbf{\begin{tabular}[c]{@{}c@{}}889\\ (100\%)\end{tabular}} & \textbf{} \\
\cmidrule(l){1-6}
\multicolumn{6}{l}{\textbf{Identified important continuous variable$^a$, mean (SD) or median [IQR] at original scale}} \\ 
\cmidrule(l){1-6}
RAGE, $\times10^3$ pg/mL & 7.6 [6.3] & 22.5 [25.9] & 19.3 [43.0] & 12.6 [19.1] & \textless{}.0001* \\
TNF-$\alpha$, $\times10^2$ pg/mL & 36.4 [24.9] & 113.0 [85.6] & 41.1 [25.0] & 50.6 [52.6] & \textless{}.0001* \\
Blood urea nitrogen, mg/dL & 16 [13] & 41 [31] & 14 [11] & 21 [23] & \textless{}.0001* \\
Creatinine, mg/dL & 0.9 [0.5] & 2.5 [2.2] & 1.0 [0.7] & 1.1 [1.2] & \textless{}.0001* \\
IL-6, pg/mL & 54.2 [92.8] & 196.6 [604.7] & 184.7 [418.1] & 105.7 [251.1] & \textless{}.0001* \\
PCT, $\times10^2$ pg/mL $^b$ & 6.5 [16.3] & 33.0 [21.6] & 25.3 [26.9] & 19.5 [27.8] & \textless{}.0001* \\
IL-8, pg/mL $^b$ & 13.1 [16.6] & 51.9 [109.8] & 31.4 [46.5] & 23.0 [43.3] & \textless{}.0001* \\
PEEP, cm H$_2$O & 8.1 (2.9) & 8.9 (3.6) & 12.9 (4.8) & 9.5 (4.2) & \textless{}.0001* \\
Minute ventilation, L/min & 9.6 [3.7] & 11.3 [4.0] & 11.8 [4.7] & 10.7 [4.1] & \textless{}.0001* \\

\cmidrule(l){1-6}
\multicolumn{6}{l}{\textbf{Identified important categorical variable$^c$, n (\%)}} \\ 
\cmidrule(l){1-6}
Diabetes & \;\;97 (25.1\%) & 111 (39.4\%) & 34 (15.4\%) & 242 (27.2\%) & \textless{}.0001* \\
Sepsis-induced PLI & 23 (6.0\%) & \;58 (20.6\%) & 28 (12.7\%) & 109 (12.3\%) & \textless{}.0001* \\ \cmidrule(l){1-6}
\multicolumn{6}{l}{Abbreviations: SD: standard deviation; IQR: interquartile range, equal to Q3 - Q1;} \\
\multicolumn{6}{l}{RAGE: receptor for advanced glycation end products; TNF: tumor necrosis factor;} \\ 
\multicolumn{6}{l}{IL: interleukin; PCT: procalcitonin; PEEP: positive end-expiratory pressure.} \\   
\multicolumn{6}{l}{$^a$Identified important continuous variables with weights $\geq 0.80$, summarized at original scale.}\\ 
\multicolumn{6}{l}{\;\;Median [IQR] is reported for continuous variable with skewed distribution that needs log-transformation.}\\
\multicolumn{6}{l}{$^b$Biomarkers with summary statistics calculated using detection limits if given censored observations.}\\
\multicolumn{6}{l}{$^c$Identified important categorical variables with weights $\geq 0.65$.}\\ 
\multicolumn{6}{l}{$^d$One-way ANOVA \textit{p}-value for continuous variable; Chi-squared test \textit{p}-value for categorical variable;}\\
\multicolumn{6}{l}{\;\;(with log-transformation if needed to improve distribution normality).}\\
\multicolumn{6}{l}{*Significant difference exists in the means or distributions among the 3 clusters.}\\ 
 \bottomrule
\label{tab:PostHocEDEN}
\end{tabular} }
\end{table}

%%%%%%%%%%%%%%%%%%%%%%%%%%%%%%%%%%%%%%%%
%%% tables moved to supplemental materials
%%%%%%%%%%%%%%%%%%%%%%%%%%%%%%%%%%%%%%%%

Distributions of all EDEN trial variables used for clustering are provided in Table \ref{tab:AllVarEDEN} (see Appendix~\ref{si.tables}).

\begin{figure}[!htbp]
\small\centering
  \includegraphics[width=0.83\linewidth]{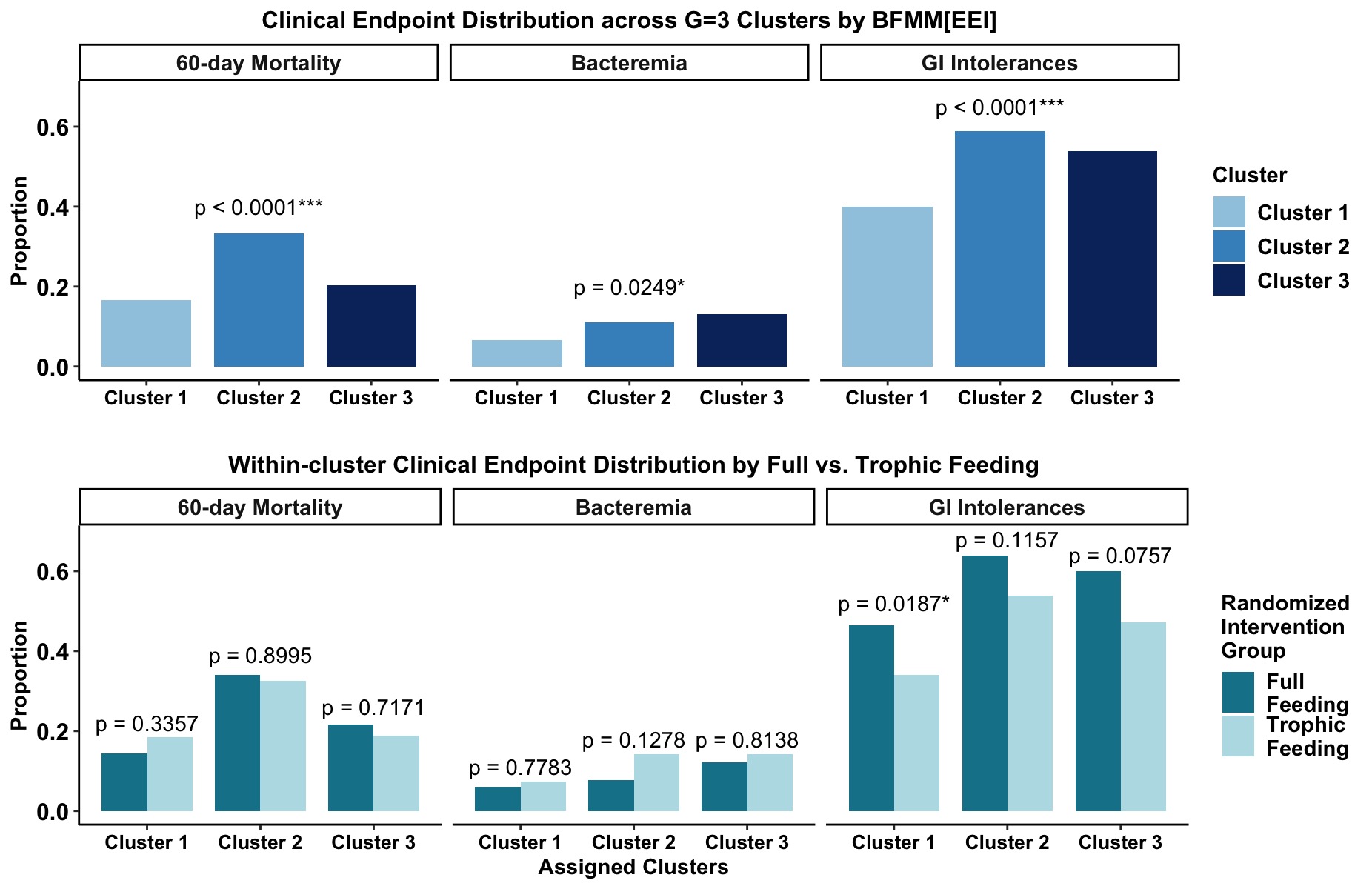}
  \caption{\small Distributions of the three clinical endpoints by BFMM[EEI] clustering results of EDEN trial dataset. The annotated \textit{p}-values are based on Chi-squared test comparing the proportions among the three clusters in the top panel, while between the two feeding groups in each cluster in the bottom panel. }
  \label{fig:PostHocEDEN}
\end{figure}

Post-hoc comparisons confirmed significant differences across clusters in identified variables of the most importance ($p<$0.0001, Table~\ref{tab:PostHocEDEN}). Differences also emerged in key clinical outcomes (Table~\ref{tab:PostHocEDEN2}, Figure~\ref{fig:PostHocEDEN}): 60-day mortality, gastrointestinal intolerance, and bacteremia all varies significantly by cluster ($p<$0.0001), with Cluster 1 showing the lowest risk across outcomes. 

Table~\ref{tab:PostHocEDEN3} examined the within-cluster association between feeding strategies and outcomes (see Appendix~\ref{si.tables}). A significant association between trophic feeding and reduced gastrointestinal intolerance was detected only in Cluster 1 ($p=0.0187$, Figure~\ref{fig:PostHocEDEN}), suggesting treatment benefit in a biologically defined subgroup that would be obscured in overall trial results. This finding illustrates BFMM’s value in identifying the heterogeneous in treatment effects out of the clinical trial populations.

Together, these two applications highlight the versatility of the BFMM clustering approach. In the SENECA EHR dataset, BFMM uncovered prognostically relevant sepsis phenotypes using mixed-type clinical data. In the EDEN trial, BFMM revealed subgroups defined by biomarker patterns and treatment response, identifying potential heterogeneity in treatment effects despite a homogeneous trial population by design. Across both settings, variable importance measures reliably distinguished informative features from noise, and the model flexibly adapted to sample size constraints through covariance structure selection.

To sum up, the major advantages of BFMM applied clinical research include:
\begin{itemize}
    %\item Joint modeling of continuous and categorical variables avoids information loss from marginalizing or dichotomizing features;
    \item Flexible covariance structure accommodation enables adaptation to varying levels of within-cluster correlation;
    \item Censored biomarker imputation improves data utilization given detection limit issues;
    \item Variable importance quantification provides interpretable rankings of predictor relevance.
    
\end{itemize}

These strengths position BFMM as a powerful and flexible tool for clustering complex healthcare datasets, supporting phenotype discovery, clinical stratification, and precision medicine research.

%%%%%%%%%%%%%%%%%%%%%%%%%%%%%%%%%%%%%%%%%%%%%%%%%%%%%%%%%%
%%%%%%%%%%%%%%%%%%%%%%%%%%%%%%%%%%%%%%%%%%%%%%%%%%%%%%%%%%
\section{Discussion}

Clustering is widely used to identify patient subgroups in biomedical research, yet existing methods often struggle to model cluster-specific dependencies, appropriately handle censored biomarker data, or offer interpretable variable selection mechanisms. These challenges restrict their clinical applicability and underscore the need for more flexible and interpretable approaches. 

In this paper, we proposed a Bayesian finite mixture model (BFMM) framework designed to address these gaps. Our approach jointly models continuous and categorical variables, accommodates censored observations through likelihood-based imputation, and quantifies variable importance via spike-and-slab priors. By supporting three distinct covariance structures (EEI, EEE, VVV), BFMM captures varying degrees of within-cluster dependency among continuous variables, offering adaptability across diverse data scenarios. We favored a finite over infinite mixture framework \cite{IFMM} to maintain stability in variable selection, avoiding sensitivity in importance weight estimates due to a non-fixed number components. The model selection process integrates BIC and ICL to jointly determine the optimal number of clusters and appropriate covariance structure.

Simulation studies demonstrated that BFMM consistently outperforms alternative clustering approaches, particularly when the covariance structure appropriately specified. In scenarios involving censored continuous variables, BFMM maintained or improved clustering accuracy, even under high censoring rates, which highlights the effectiveness of its likelihood-based imputation. Notably, when strong within-cluster correlations were present (i.e., under the VVV structure), BFMM[VVV] significantly outperformed methods assuming local independence, such as \texttt{clustMD}, LCA, PAM, and K-prototype. These findings support the robustness of BFMM in high-dimensional, correlated, and partially observed data environments. 

Applications to two distinct clinical datasets further validated the flexibility and translational relevance of BFMM. In the SENECA cohort, BFMM identified four clinically interpretable sepsis phenotypes using both continuous laboratory values and categorical demographic variables. The method's ability to model mixed data types and assign quantitative importance weights facilitated interpretation and provided insight into key drivers of phenotypic heterogeneity. In the EDEN trial with biomarker data subject to detection limits, BFMM successfully identified three latent patient subgroups and highlighted potential heterogeneity in treatment resposnes to feeding interventions. These results illustrate how BFMM can uncover actionable patient subgroups from data sources as diverse as pragmatic EHR systems and randomized controlled trials.

BFMM also offers methodological innovations that enhance interpretabiity. The Bayesian framework supports the use of separate spike-and-slab priors: a normal prior for standardized continuous variables and a Dirichlet prior for categorical variables. In both simulations and application settings, BFMM effectively distinguished dominant predictors from noise variables. We introduced a principled approach to tuning the variance ratio between slab and spike priors (i.e. the slab prior with larger variance than the spike), which was empirically determined to fall between $43-130$ in the SENECA data and $67-135$ in the EDEN data. These ranges were found to produce consistent and robust importance weights across Markov chains. 

Despite its strengths, several limitations remain. First, BFMM assumes multivariate normality within each mixture component for continuous variables and local independence for categorical variables. This assumption may be restrictive in datasets containing outliers or heavy-tailed distributions, Future extensions could incorporate more flexible distributional forms, such as skew-normal or skew-$t$ families \cite{SkewNormal, SkewT}, or adopt latent variable approaches to model dependencies among different variable types. Second, the computational burden of BFMM, especially under the VVV structure, remains substantial. Improvements in MCMC efficiency like blocked Gibbs sampling \cite{blockGibbs}, or more scalable diagnostics besides MPSRF \cite{ConvDiag}, are warranted to support broader implementations.

Third, while BFMM includes three representative covariance structures, it could benefit from integration with a broader class of covariance structures from parsimonious Gaussian mixture models. %More importantly, when the assumed structure deviates from the true one, especially under heavy censoring, the imputation of censored values may become biased. Incorporating doubly robust imputation strategies for censored likelihoods \cite{DoubleRobustEst} could offer protection against model misspecification.% 
Moreover, the use of separate priors for continuous and categorical variables introduces asymmetry, which complicates direct comparisons of importance weights across variable types. While our method includes empirical guidance for tuning continuous shrinkage, parallel strategies for categorical priors remain underdeveloped.

In summary, the proposed BFMM framework offers a unified, flexible, and interpretable solution for clustering mixed-type data, especially in complex biomedical settings. Its ability to accommodate mixed variable types, correlated features with cluster-specific dependencies, censored biomarkers, while offering interpretable variable importance measures, makes it well-suited to modern clinical research. By uncovering clinically relevant phenotypes and informing treatment-responsive subgroups, the proposed BFMM clustering framework has the potential to advance precision medicine and enhance the utility of real-world data for data-driven discovery and decision-making.

\bigskip 

\section*{\large Data Availability}

Data sharing is not applicable to this article as no new data were created or analyzed in this study.

\bigskip

%%%%%%%%%%%%%%%%%%%%%%%%%%%%%%%%%%%%%%%%%%%%%%%%%%%
%%% Bibliography %%%
%%%%%%%%%%%%%%%%%%%%%%%%%%%%%%%%%%%%%%%%%%%%%%%%%%%	
\newpage 

\def\bibfont{\small}
\renewcommand\bibname{\large References}
%\bibliographystyle{agsm}
%\bibliography{Reference.bib}

\printbibliography

%%%%%%%%%%%%%%%%%%%%%%%%%%%%%%%%%%%%%%%%%%%%%%%%%%%
%%% Supplemental Material %%%
%%%%%%%%%%%%%%%%%%%%%%%%%%%%%%%%%%%%%%%%%%%%%%%%%%%
\newpage

%\lipsum[1]
%\renewcommand{\appendixpagename}{\Large Supplementary Information for: \large\textit{A Bayesian Finite Mixture Model Approach for Clustering Correlated Mixed-type Variables and Censored Biomarkers}}
%\appendixpage

\appendix\label{appendix}
\addtocontents{toc}{\bigskip\medskip\noindent%
  \textbf{Appendix Material}\par}
% Activate preparatory code for section-level headers
\makeatletter 
\newcommand{\section@cntformat}{Appendix \thesection\quad}
\makeatother

\renewcommand{\thefigure}{A\arabic{figure}}
\setcounter{figure}{0}
\renewcommand{\thetable}{A\arabic{table}}
\setcounter{table}{0}
\renewcommand{\thealgorithm}{A\arabic{algorithm}}
\setcounter{algorithm}{0}

\renewcommand{\theequation}{A\arabic{equation}}
\setcounter{equation}{0}

%%% Literature Review on Mixed Data Clustering %%%
\section{Literature Review on Related Work}\label{si.review}
Clustering methods for mixed-type data generally fall into three major categories: distance-based, density-based, and model-based approaches.

\subsection{Distance-Based Clustering Methods} 
Distance-based methods typically rely on computing similarity or dissimilarity metrics across variables of different types. One approach computes a uniform distance measure, such as Gower’s distance \cite{Gower}, followed by a clustering algorithm like Partitioning Around Medoids (PAM) \cite{PAM}. Another strategy employs weighted combinations of distance measures specific to variable types, as seen in K-prototypes \cite{kprototype} and K-means-mixed \cite{kmeansmixed}. These methods extend K-means \cite{KMeans} to mixed data but struggle with setting appropriate weightings across different variable types \cite{Hennig2013} and do not account for cluster-specific covariance.

\subsection{Density-Based Clustering Methods} 
Density-based methods identify clusters by their distributions, which work better for clusters with arbitrary shapes and sizes. Density-based spatial clustering of applications with noise (DBSCAN) is the representative for mixed data \cite{DBSCAN}, which assumes constant densities across clusters. Some innovative methods incorporated the density-based methods with other types of clustering algorithms or methods to improve the performance. \textcite{HDBSCAN} proposed the hierarchical DBSCAN (HDBSCAN) based on decision trees, which relaxes the local uniform density assumption of DBSCAN. More recently, \textcite{HyDaP} proposed a two-step hybrid density- and partition-based algorithm (HyDaP) for mixed data clustering, using both density-based and partition-based algorithms to identify the important variables, though inheriting the limitation of local independence assumption.

\subsection{Model-Based Clustering Methods} 
Model-based clustering assumes that the data are generated from a mixture of underlying distributions \cite{mclustFraley2002}. These methods allow for model-based selection of the number of clusters and support probabilistic interpretation. Finite mixture models (FMMs) are a popular choice \cite{FMMbook}, with estimation via maximum likelihood using the EM algorithm \cite{EM0} or Bayesian inference through MCMC \cite{Diebolt1994}. Bayesian models allow for the incorporation of priors, accommodate multimodal likelihoods, and yield stable estimates even in small samples \cite{Robert2007}.

\subsection{Handling Dependencies in Model-Based Clustering} 
Most model-based clustering methods for mixed data assume local independence, particularly in latent class analysis (LCA) \cite{LCA, LCAmix}, which limits their ability to capture correlations among variables. The parsimonious Gaussian mixture models (PGMM) framework \cite{ParsGMM} offers a family of covariance structures for continuous data, including the EEI, EEE, and VVV structures. The \texttt{mclust} R package \cite{mclust5} implements EM algorithms for PGMMs, while \texttt{clustMD} \cite{clustMD2016} and \texttt{regClustMD} \cite{Choi2023} extend this to mixed data but still assume conditional independence of continuous variables.

\subsection{Variable Selection in Model-Based Clustering} 
Few mixed-data clustering methods quantify variable importance. Most variable selection strategies fall into filter or wrapper methods \cite{Dy2004, Law2004}. Wrapper methods integrate variable selection into model fitting but are often limited to single-type data \cite{Fop2017}. Marbac and Sedki \cite{Marbac2017} proposed BIC- and MICL-optimized clustering for mixed data with embedded variable selection, but both assume local independence.

Bayesian spike-and-slab priors are a common wrapper method in continuous data \cite{George1993, Ishwaran2005}, which decomposes variable effects into a spike (irrelevant) and slab (relevant) component with variable-specific indicators. For continuous data, \textcite{bclust} proposed a Bayesian clustering model using a spike-and-slab normal prior. For categorical variables, \textcite{ShuWang} proposed a spike-and-slab Dirichlet prior where the spike is concentrated at the marginal distribution and the slab is uniform, thus extending variable selection to categorical variables in mixed data.

\subsection{Handling Censored Data in Model-Based Clustering} 
Censored data, common in biomedical research, poses an additional challenge. Although methods like the Tobit model \cite{Tobit} and its extensions \cite{Powell1984, Caudill2010} address censored regression, few clustering approaches handle censored observations. In the EM framework, extensions to mixture of factor analyzers (MFA) and mixture of experts (MoE) exist \cite{censMFA, censMoE}. In the Bayesian setting, \textcite{ShuWang} proposed a finite mixture model that incorporates censoring via an imputation step during Gibbs sampling, but it assumes local independence.

\vspace{.3 in}
Despite these advancements, clustering mixed-type data with correlated variables and censored biomarkers remains underdeveloped. This paper builds upon prior work by offering a unified Bayesian framework capable of modeling dependencies, selecting important variables, and handling censored observations.

\bigskip

%%% Parsimonious Gaussian Mixture Model %%%
\newpage 

\section{Covariance Structures}\label{si.pgmm}

Let $\boldsymbol{\Sigma}_g$ denote the covariance matrix specific to the $g$th mixture component in a parsimonious Gaussian mixture model (PGMM), its parsimonious parameterisation is obtained by imposing constraints on eigendecomposition of the form $\boldsymbol{\Sigma}_g = \lambda_g \boldsymbol{D}_g \boldsymbol{A}_g \boldsymbol{D}_g^T$, where  $\lambda_g = |\boldsymbol{\Sigma}_g|^{1/d}$ is a positive scalar controlling the volume; $\boldsymbol{A}_g$ is a diagonal matrix with determinant $|\boldsymbol{A}_g|=1$ and with the normalized eigenvalues of $\boldsymbol{\Sigma}_g$ ordered decreasingly as its diagonal entries, specifying the shape of the density contours; and $\boldsymbol{D}_g$  is $p \times p$ orthogonal matrix of eigenvectors of $\boldsymbol{\Sigma}_g$, governing the corresponding ellipsoid's orientation. This decomposition leads to a wide range of flexible models that can adapt to varied clustering situations. 

Table \ref{tab:14CovStruct} outlines all the 14 distinct covariance structures based on PGMMs, where VVV structure is the most unconstrained and with the highest parameter dimensionality. 

\begin{table}[!htbp]
\centering
\caption{Parameterisations of component-specific covariance matrix in PGMM clustering.}
\scalebox{0.85}{
\begin{tabular}{@{}lcllll@{}}
\toprule
\textbf{Structure} & \textbf{Cluster-specific} $\boldsymbol{\Sigma}_g$ & \textbf{Volume} & \textbf{Shape} & \textbf{Orientation} & \textbf{Covariance Parameters}$^a$ \\ \midrule
EII  & $\lambda \boldsymbol{I}$  & Equal  & Identity & Identity  & $1$                  \\
VII    & $\lambda_g\boldsymbol{I}$ & Variable & Identity & Identity  & $G$                   \\
EEI    & $\lambda \boldsymbol{A}$ & Equal   & Equal    & Identity    & $d$                     \\
VEI    & $\lambda_g \boldsymbol{A}$ & Variable & Equal  & Identity  & $G+(d-1)$               \\
EVI    & $\lambda \boldsymbol{A}_{g}$ & Equal  & Variable & Identity & $1+G(d-1)$              \\
VVI    & $\lambda_g \boldsymbol{A}_g$ & Variable & Variable & Identity  & $Gd$                    \\
EEE    & $\lambda \boldsymbol{DAD}^T$ & Equal & Equal  & Equal & $d(d+1)/2$              \\
EVE    & $\lambda \boldsymbol{D}\boldsymbol{A}_{g}\boldsymbol{D}^T$ & Equal  & Variable & Equal & $1+d(d-1)/2+G(d-1)$     \\
VEE    & $\lambda_g\boldsymbol{DAD}^T$ & Variable & Equal & Equal & $G+d(d-1)/2+(d-1)$      \\
VVE    & $\lambda_g\boldsymbol{D}\boldsymbol{A}_g\boldsymbol{D}^T$ & Variable & Variable & Equal & $1+Gd(d-1)/2+(d-1)$     \\
EEV    & $\lambda \boldsymbol{D}_g\boldsymbol{A}\boldsymbol{D}_g^T$ & Equal  & Equal & Variable & $1+Gd(d-1)/2+(d-1)$     \\
VEV    & $\lambda_g\boldsymbol{D}_g\boldsymbol{A}\boldsymbol{D}_g^T$ & Variable & Equal  & Variable & $G+Gd(d-1)/2+(d-1)$     \\  
EVV    & $\lambda \boldsymbol{D}_g\boldsymbol{A}_g\boldsymbol{D}_g^T$ & Equal & Variable & Variable & $1+Gd(d-1)/2+G(d-1)$    \\
VVV    & $\lambda_g\boldsymbol{D}_g\boldsymbol{A}_g\boldsymbol{D}_g^T$ & Variable & Variable & Variable    & $Gd(d+1)/2$  \\
\midrule
\multicolumn{6}{l}{\small $^a$$\boldsymbol{\Sigma}_g$ denotes the covariance matrix specific to the $g$th cluster, $g=1,\dots,G$.} \\ 
\multicolumn{6}{l}{\small $^b$$G$ denotes total number of clusters, and $d$ the number of continuous variables.} \\ 
\bottomrule          
\label{tab:14CovStruct}
\end{tabular} }
\end{table}

\newpage

\begin{figure}[h] 
\centering\small
  \includegraphics[width=1\linewidth]
  {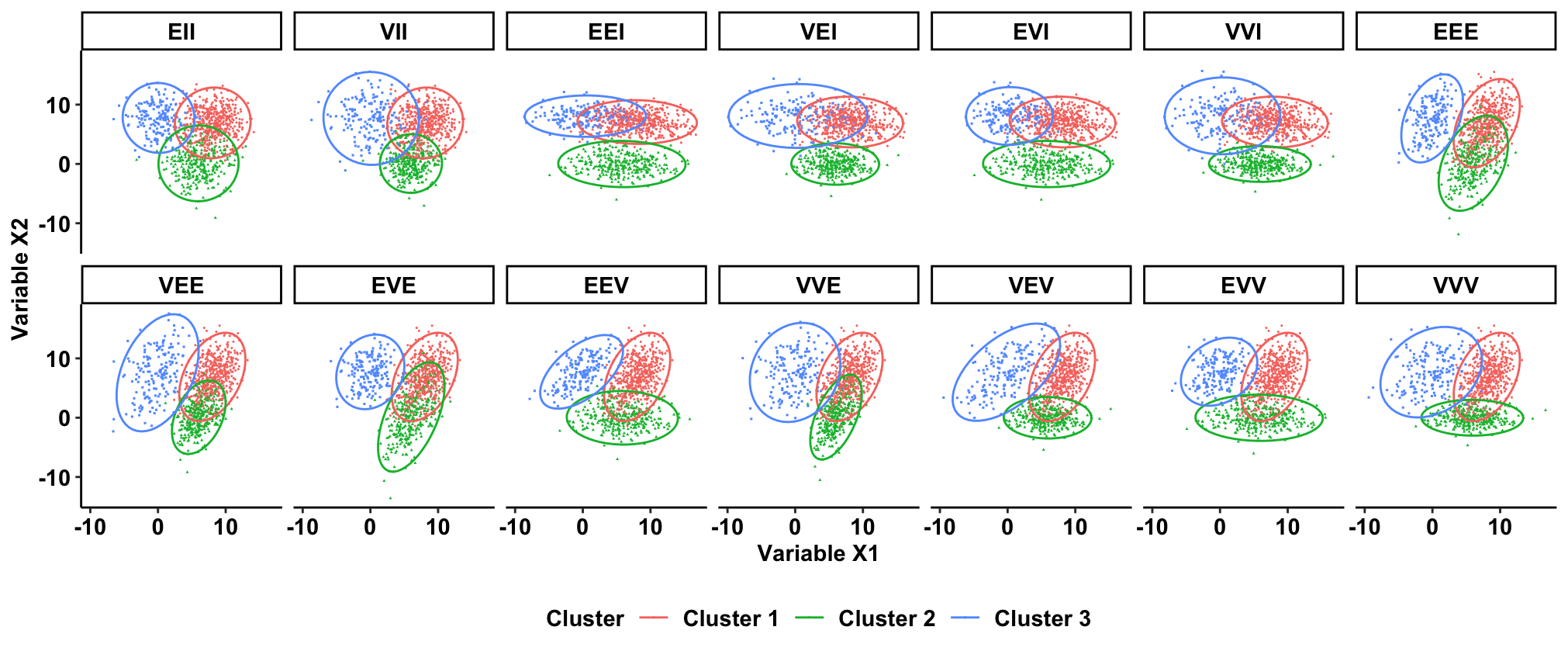}
  \caption{Graphical representations of bivariate mixture density for $G = 3$ clusters for 14 covariance structures in Parsimonious Gaussian mixture models. The first, second and third letters in each three-letter code describe the volume, shape, and orientation of the cluster-specific covariance matrices, where ``I" represents ``Identity", ``E" represents ``Equal", ``V" represents ``Variable". }
  \label{fig:14CovStruct}
\end{figure}

Figure \ref{fig:14CovStruct} accordingly provides graphical representations of a typical two-dimensional mixture densities for $G = 3$ groups to illustrate the differences among the 14 covariance structures. Imposing constraints reduces the number of free covariance parameters from $Gp (p + 1) /2$ in VVV to smaller numbers in other less flexible structures. 

%%% Figure - Covariance Structures %%%

%%%%%%%%%%%%%%%%%%%%%%%%%%%%%%%%%%%%%%%%

\clearpage
\section{Posterior Distributions}\label{appendix.postdist}

The posterior distributions used in Gibbs sampling to update the parameters of the proposed BFMM clustering approach are as follows. Let $ n_g = \sum_{i=1}^n z_{ig}$ denote the number of observations assigned to cluster $g$; and let $\bar{x}_m = \frac{1}{n} \sum_{i=1}^n x_{im}$ denote the marginal mean standardized to 0 of the continuous variable $m$ across all the $G$ clusters. \\

%%% EEI Posterior %%%
\textbf{BFMM[EEI] - Common variance $\sigma_m^2$ of variable $m$ within each cluster, $m=1,\dots,q$}: 
\begin{equation}
\sigma_m^2 \mid \boldsymbol{X},\boldsymbol{Z},\boldsymbol{\mu}_{m*}, \tilde a, \tilde b \sim \text{Inv}\Gamma \left\{ \tilde a + \frac{n}{2}, \tilde b + \frac{1}{2}\sum_{g=1}^G z_{ig}\left(x_{im} - \mu_{mg} \right)^2 \right \}, 
\label{equ:EEI_sigma_g}
\end{equation}
\noindent where $\boldsymbol{\Sigma}_g = \text{diag}(\sigma_1^2,\dots,\sigma_q^2)$ for cluster $g$, $g=1,\dots,G$. \\

\textbf{BFMM[EEI] - Cluster mean ($\mu_{mg}$) for continuous variable $m$ in cluster $g$}: 
\begin{equation}
\mu_{mg} \mid \boldsymbol{X},\boldsymbol{Z},\boldsymbol{\mu}_{(m)g},\sigma_m^2, \omega, \sigma_{\Delta_0}^2,\Delta_{mg} \sim N\left( \mu_{mg}^{\ast}, \frac{\omega^{\Delta_{mg}}\sigma_{\Delta_0}^2 \sigma_m^2 }{\sigma_m^2 + n_g\omega^{\Delta_{mg}}\sigma_{\Delta_0}^2  } \right), \\ 
\label{equ:EEI_mu_mg}   
\end{equation}
\noindent where $
\mu_{mg}^{\ast} = \frac{ \bar x_m \sigma_m^2 + \omega^{\Delta_{mg}}\sigma_{\Delta_0}^2\sum_{i=1}^n z_{ig}x_{im} }{\sigma_m^2 + n_g\omega^{\Delta_{mg}}\sigma_{\Delta_0}^2}$.

%%% EEE Posterior %%%
\textbf{BFMM[EEE] - Common covariance matrix ($\boldsymbol{\Sigma}_1$) within each cluster}: 
\begin{equation}
\Sigma_1 \mid \boldsymbol{X},\boldsymbol{Z},\boldsymbol{\mu}_g, \nu_0, \boldsymbol{S}_0 \sim \text{IW}\left[ \nu_0+n, \left(\boldsymbol{S}_n\right)^{-1} \right ], 
\label{equ:EEE_sigma_g}
\end{equation}
\noindent where 
$\boldsymbol{S}_n = \boldsymbol{S}_0 + \sum_{i=1}^n \left(\boldsymbol{u}_i-\mu \boldsymbol{z}_i \right)\left(\boldsymbol{u}_i-\mu \boldsymbol{z}_i\right)^T$; $\boldsymbol{\Sigma}_g = \boldsymbol{\Sigma}_1, g=2,\dots,G$. \\

\textbf{BFMM[EEE] - Cluster mean ($\mu_{mg}$) for continuous variable $m$ in cluster $g$}: 
\begin{equation}
\mu_{mg} \mid \boldsymbol{X},\boldsymbol{Z},\boldsymbol{\mu}_{(m)g},\boldsymbol{\Sigma}_1, \omega, \sigma_{\Delta_0}^2,\Delta_{mg} \sim N\left\{ \mu_{mg}^{\ast}, \frac{\omega^{\Delta_{mg}}\sigma_{\Delta_0}^2 }{1 + n_g\left(\boldsymbol{\Sigma}_1^{-1}\right)_{mm}\omega^{\Delta_{mg}}\sigma_{\Delta_0}^2  } \right\},\\ 
\label{equ:EEE_mu_mg}   
\end{equation}
\noindent where 
\begin{align*}
& \mu_{mg}^{\ast} = \frac{ \bar x_m + \omega^{\Delta_{mg}}\sigma_{\Delta_0}^2\left\{ \left(\boldsymbol{\Sigma}_1^{-1}\right)_{mm} \sum_{i=1}^n z_{ig}x_{im} + W_g \right\}  }{1 + n_g\left(\boldsymbol{\Sigma}_1^{-1}\right)_{mm}\omega^{\Delta_{mg}}\sigma_{\Delta_0}^2  }; \\
& W_g = \sum_{i=1}^n \left\{ z_{ig}\sum_{p \neq m}^q\left(\boldsymbol{\Sigma}_1^{-1}\right)_{mp}\left(x_{ip} - \mu_{pg}\right ) \right\}.
\end{align*}

%%% VVV Posterior %%%
\textbf{BFMM[VVV] - Covariance matrix ($\boldsymbol{\Sigma}_g$) for cluster $g$}: 
\begin{equation}
\boldsymbol{\Sigma}_g \mid \boldsymbol{X},\boldsymbol{Z},\boldsymbol{\mu}_g, \nu_g, \boldsymbol{S}_g \sim \text{IW}\left\{\nu_g+n_g, (\boldsymbol{S}_g^{\ast})^{-1} \right\}, 
\label{equ:VVV_sigma_g}
\end{equation}
where $\boldsymbol{S}_g^{\ast} = \boldsymbol{S}_g + \sum_{i=1}^n z_{ig}\left(\boldsymbol{u}_i- \boldsymbol{\mu}_g \right)\left(\boldsymbol{u}_i - \boldsymbol{\mu}_g\right)^T$. \\

\textbf{BFMM[VVV] - Cluster mean ($\mu_{mg}$) for continuous variable $m$ in cluster $g$}: 
\begin{equation}
\mu_{mg} \mid \boldsymbol{X},\boldsymbol{Z},\boldsymbol{\mu}_{(m)g},\boldsymbol{\boldsymbol{\Sigma}}_g, \omega, \sigma_{\Delta_0}^2,\Delta_{mg} \sim N\left\{ \mu_{mg}^{\ast}, \frac{\omega^{\Delta_{mg}}\sigma_{\Delta_0}^2 }{1 + n_g\left(\boldsymbol{\Sigma}_g^{-1}\right)_{mm}\omega^{\Delta_{mg}}\sigma_{\Delta_0}^2  } \right\}, 
\label{equ:VVV_mu_mg}   
\end{equation}
\noindent where 
\begin{align*}
& \mu_{mg}^{\ast} = \frac{ \bar x_m + \omega^{\Delta_{mg}}\sigma_{\Delta_0}^2\left\{ \left(\boldsymbol{\Sigma}_g^{-1}\right)_{mm} \sum_{i=1}^n z_{ig}x_{im} + W_g \right\} }{1 + n_g\left(\boldsymbol{\Sigma}_g^{-1}\right)_{mm}\omega^{\Delta_{mg}}\sigma_{\Delta_0}^2}; \\
& W_g = \sum_{i=1}^n z_{ig} \sum_{p \neq m}^q\left(\boldsymbol{\Sigma}_g^{-1}\right)_{mp}\left(x_{ip} - \mu_{pg}\right).
\end{align*}

\textbf{Categorical probability vector} for categorical variable $m$: 
\begin{equation}
\boldsymbol{\theta}_{mg} \mid \boldsymbol{X}, \boldsymbol{Z}, \Delta_{mg},\boldsymbol{\alpha}_{\Delta_1}, \boldsymbol{\alpha}_{m\Delta_0} \sim \text{Dir}\left\{ \boldsymbol{\alpha}_{m\Delta_0}^{1-\Delta_{mg}}\boldsymbol{\alpha}_{m\Delta_1}^{\Delta_{mg}} + \left( \sum_{i=1}^n x_{im1}z_{ig}, \dots, \sum_{i=1}^n x_{imL_{m}}z_{ig} \right)^T \right\}.
\label{equ:theta_mg} 
\end{equation}

\textbf{Importance indicator} for continuous variables ($m=1,\dots,q$):
\begin{equation}
\Delta_{mg} \mid p_{1m},\mu_{mg},\omega,\sigma_{\Delta_0}^2 \sim \text{Ber}\left\{ \frac{p_{1m}N\left(\mu_{mg} - \bar x_m ;0,\omega\sigma_{\Delta_0}^2\right)}{p_{1m}N\left(\mu_{mg} - \bar x_m;0,\omega\sigma_{\Delta_0}^2\right) + (1-p_{1m})N\left(\mu_{mg} - \bar x_m;0,\sigma_{\Delta_0}^2\right)}\right\};
\label{equ:delta_mg_p1}  
\end{equation}

\noindent for categorical variable ($m=q+1,\dots,M$):
\begin{equation}
\Delta_{mg} \mid p_{2m}, \boldsymbol{\theta}_{mg}, \boldsymbol{\alpha}_{\Delta_1}, \boldsymbol{\alpha}_{m\Delta_0} \sim \text{Ber}\left\{ \frac{p_{2m}\text{Dir}\left(\boldsymbol{\theta}_{mg};\boldsymbol{\alpha}_{\Delta_1}\right)}{p_{2m}\text{Dir}\left(\boldsymbol{\theta}_{mg};\boldsymbol{\alpha}_{\Delta_1}\right) + (1-p_{2m})\text{Dir}\left(\boldsymbol{\theta}_{mg};\boldsymbol{\alpha}_{m\Delta_0}\right)} \right\}.
\label{equ:delta_mg_p2}  
\end{equation}

\textbf{Hyperparameter $\sigma_{\Delta_0}^2$, the spike normal prior variance for $\mu_{mg}$}:
\begin{equation}
\sigma_{\Delta_0}^2 \mid \boldsymbol{\Delta},\boldsymbol{\mu}, \omega, a_{\Delta_0}, b_{\Delta_0} \sim \text{Inv}\Gamma(a_{\Delta_0}^{\ast}, b_{\Delta_0}^{\ast}),
\label{equ:sigma2_delta0}   
\end{equation} 
\noindent where $
 a_{\Delta_0}^{\ast} = a_{\Delta_0} + \frac{1}{2} \sum_{m=q}^n \sum_{g=1}^G (1-\Delta_{mg} );
\text{ and } b_{\Delta_0}^{\ast} = b_{\Delta_0} + \sum_{m=1}^q \sum_{g=1}^G \left( 1-\Delta_{mg}\right) \left(\mu_{mg} - \bar x_m \right)^2$.

\textbf{Hyperparameters $p_{1m}$ and $p_{2m}$ in Bernoulli prior for $\Delta_{mg}$}:
\begin{equation}
p_{jm} \mid a_{pj},b_{pj},\Delta_{mg} \sim \text{Beta}\left\{ a_{pj}+\sum_{g=1}^G \Delta_{mg}, b_{pj}+\sum_{g=1}^G (1-\Delta_{mg}) \right\}.
\label{equ:p_jm} 
\end{equation}
\noindent where $j=1$ for continuous variables $(m=1,\dots,q)$, $j=2$ for categorical variables $(m>q)$.

\textbf{Cluster probabilities} for $G$ clusters:
\begin{equation}
\boldsymbol{\tau} \mid \boldsymbol{Z},\delta_1,\dots,\delta_G \sim \text{Dir}\left(\delta_1+n_1, \dots,  \delta_G+n_G\right).
\label{equ:tau} 
\end{equation}

\textbf{Cluster membership} for observation $x_i$:
\begin{equation}
\boldsymbol{z}_i \mid \boldsymbol{x}_i,\boldsymbol{\tau},\boldsymbol{\mu},\Sigma,\theta \sim \text{Multinomial}\left\{ G; \frac{\tau_1 f_1(\boldsymbol{x}_i \mid \boldsymbol{\mu}_{1},\boldsymbol{\Sigma}_1,\theta_{1}) }{\sum_{g=1}^G \tau_g f_g(\boldsymbol{x}_i \mid \boldsymbol{\mu}_{g},\boldsymbol{\Sigma}_g,\theta_{g}) },\dots, \frac{\tau_G f_G(\boldsymbol{x}_i \mid \boldsymbol{\mu}_{G},\boldsymbol{\Sigma}_G,\theta_{G}) }{\sum_{g=1}^G \tau_g f_g(\boldsymbol{x}_i \mid \boldsymbol{\mu}_{g},\boldsymbol{\Sigma}_g,\theta_{g}) } \right \},
\label{equ:z_i} 
\end{equation}
\noindent where $f_g(\boldsymbol{x}_i|\boldsymbol{\mu}_{g},\boldsymbol{\Sigma}_g,\theta_{g}) = \phi_g \left(\boldsymbol{u}_i | \boldsymbol{\mu}_g, \boldsymbol{\Sigma}_g \right)\prod_{m=q+1}^M \prod_{\ell=1}^{L_m}\theta_{mg\ell}^{\mathbbm{1} \left( x_{im} = \ell \right)}.$

%%%%%%%%%%%%%%%%%%%%%%%%%%%%%%%%%%%%%%%%

%%% Kullback-Leibler (KL) Relabeling Algorithm %%%
\section{Kullback-Leibler Relabeling}\label{si.labelswich}
First, store clustering probabilities during the MCMC algorithm. Let $T$ denote the total number of retained MCMC iterations. Let $\mathcal{P}_G$ denote the set consists of all the $G!$ distinct permutations of cluster labeling $\left\{1,\dots, G \right\}$ of all the $G$ clusters. Let $\boldsymbol{P}^{(t)}_{n\times G}$ be the matrix clustering probabilities of $n$ observations of $G$ clusters at iteration $t \in \left\{1,2,\dots,T\right\} $ of the MCMC, that is
\begin{equation*}
\boldsymbol{P}^{(t)}_{n\times G} = \left\{\left\{ p_{i,j}^{(t)} \right\}_{j=1}^G \right\}_{i=1}^n \text{ with } p_{i,j}^{(t)} = \frac{\tau_j^{(t)} f_j\left\{\boldsymbol{x}_i \mid \boldsymbol{\mu}_j^{(t)}, \boldsymbol{\Sigma}_j^{(t)}, \theta_j^{(t)} \right\} }{\sum_{g=1}^G \tau_g^{(t)} f_g \left\{\boldsymbol{x}_i \mid \boldsymbol{\mu}_g^{(t)}, \boldsymbol{\Sigma}_g^{(t)}, \theta_g^{(t)} \right\} },
\end{equation*}
\noindent where $i = 1,\dots, n$ and $j = 1,\dots, G$. 

Second, let $\boldsymbol{Q}_{n\times G}$ be the matrix of true clustering probabilities. Under Stephens' setting \cite{Stephens2000}, the goal is to find a permutation $\rho_t \in \mathcal{P}_G$ of the columns of $\boldsymbol{P}^{(t)}_{n\times G}$: $\rho_t\left\{\boldsymbol{P}^{(t)}\right\}: = \left\{\left\{ p_{i,\rho_t(j)}^{(t)} \right\}_{j=1}^G \right\}_{i=1}^n $, such that the Kullback-Leibler divergence between $\rho_t\left\{ \boldsymbol{P}^{(t)}\right\}$ and $\boldsymbol{Q}$ is minimized across all permutations. The unknown matrix $\boldsymbol{Q}$ is approximated via a recursive algorithm described in Algorithm \ref{algorithm:KLrelabel}.\\

\begin{algorithm}
    \caption{Kullback-Leibler (KL) Relabeling Strategy}\label{euclid}
    \begin{algorithmic}[1]
    \State Initialize $\rho_1,\dots,\rho_T$, set $\rho_t = \left\{ 1,2,\dots,G \right\}$ for all $t$.
   	\For {$i=1,\dots,n \text{ and } j=1,\dots,G$}
    	\State Calculate $\hat q_{i,j} =\frac{1}{T}\sum_{t=1}^Tp^{(t)}_{i,\rho_t(j)}$ to estimate $\boldsymbol{Q}$.
    \EndFor    
    \For {$t=1,\dots,T$}
    	\State Find a permutation $\rho_t^* =  \underset{\rho_t \in \left\{ \mathcal{P}_G \right\}} {\arg\min} \sum_{i=1}^n \sum_{j=1}^G p^{(t)}_{i,j}\log\left\{ \frac{p^{(t)}_{i,j}}{\hat q_{i,j} } \right\}$, update $\rho_t \leftarrow \rho_t^{*}$. 
    \EndFor
    \State If an improvement is made to $\sum_{t=1}^T \sum_{i=1}^n \sum_{j=1}^G p^{(t)}_{i,j}\log\left\{ \frac{p^{(t)}_{i,j}}{\hat q_{i,j} } \right \}$, return to Step 2. \\ Finish otherwise.
    \end{algorithmic}
  \label{algorithm:KLrelabel}
\end{algorithm}

%%%%%%%%%%%%%%%%%%%%%%%%%%%%%%%%%%%%%%%%
%%% Simulations - Supplemental Tables %%%
\clearpage
\section{Supplemental Summary Tables}\label{si.tables}

%%%%%% Simulation: Number of Failed Trials by Methods %%%%%%
\begin{table}[h]
\small\center
\caption{Failure rate of simulation trials by clustering methods.}
\scalebox{0.9}{
\begin{tabular}{ll|ccccc}
\toprule
 &  & \multicolumn{5}{c}{\textbf{Failure rate (\%), out of 100 simulation trials}} \\ 
 \cmidrule(l){1-7}
\multicolumn{2}{l|}{\textbf{Simulation Scenario}} & \textbf{\begin{tabular}[c]{@{}c@{}}clustMD\\ {[}EEI{]}$^a$\end{tabular}} & \textbf{\begin{tabular}[c]{@{}c@{}}clustMD\\ {[}VVI{]}$^a$\end{tabular}} & \textbf{\begin{tabular}[c]{@{}c@{}}mclust\\ {[}EEI{]}$^{a,b}$\end{tabular}} & \textbf{\begin{tabular}[c]{@{}c@{}}mclust\\ {[}EEE{]}$^{a,b}$\end{tabular}} & \textbf{\begin{tabular}[c]{@{}c@{}}mclust\\ {[}VVV{]}$^{a,b}$\end{tabular}} \\ 
 \cmidrule(l){1-7}
\multicolumn{1}{c}{\multirow{3}{*}{\textbf{Data{[}EEI{]}}}} & Uncensored & 4\% & 7\% & 1\% & 1\% & 65\% \\
\multicolumn{1}{c}{} & $X_3-X_5$ each censored 20\% & 4\% & 7\% & 0 & 0 & 47\% \\
\multicolumn{1}{c}{} & $X_3-X_5$ each censored 40\% & 4\% & 8\% & 0 & 0 & 58\% \\ 
 \cmidrule(l){1-7}
\multirow{3}{*}{\textbf{Data{[}EEE{]}}} & Uncensored & 2\% & 6\% & 3\% & 3\% & 13\% \\
 & $X_3-X_5$ each censored 20\% & 2\% & 6\% & 0 & 0 & 11\% \\
 & $X_3-X_5$ each censored 40\% & 3\% & 18\% & 0 & 0 & 12\% \\ 
 \cmidrule(l){1-7}
\multirow{3}{*}{\textbf{Data{[}VVV{]}}} & Uncensored & 0 & 0 & 0 & 2\% & 42\% \\
 & $X_3-X_5$ each censored 20\% & 0 & 0 & 0 & 0 & 33\% \\
 & $X_3-X_5$ each censored 40\% & 0 & 19\% & 0 & 0 & 60\% \\ \cmidrule(l){1-7}
 \multicolumn{7}{l}{\begin{tabular}[c]{@{}l@{}}$^a$Except for \texttt{clustMD} and \texttt{mclust}, other clustering methods all had no failed trial in each scenario. \\ $^b$Categorical variables were treated as continuous when applying the \texttt{mclust} clustering methods.\end{tabular}} \\ \bottomrule
\end{tabular} }
\label{tab:SimAll.FailCnt}
\end{table}

%%%%%%% Simulation: ALL ARI %%%%%%%
\begin{table}[!htbp]
\small\center
\caption{Clustering performance comparison in each simulated scenario.}
\scalebox{0.8}{
\begin{tabular}{ll|cc|cc|cc}
\toprule	
\textbf{} & \textbf{} & \multicolumn{6}{c}{\textbf{ARI (2.5th percentile, 97.5th percentile), across 100 trials}} \\ \cmidrule(l){3-8}
\textbf{\begin{tabular}[c]{@{}l@{}}Simulated\\ Structure\end{tabular}} & \textbf{\begin{tabular}[c]{@{}l@{}}Clustering\\ Approach\end{tabular}} & \multicolumn{2}{c|}{\textbf{Uncensored}} & \multicolumn{2}{c|}{\textbf{\begin{tabular}[c]{@{}c@{}}$X_3-X_5$ each\\ censored 20\%\end{tabular}}} & \multicolumn{2}{c}{\textbf{\begin{tabular}[c]{@{}c@{}}$X_3-X_5$ each\\ censored 40\%\end{tabular}}} \\
  \cmidrule(l){1-8}
\multirow{15}{*}{\textbf{\begin{tabular}[l|]{@{}l@{}}Data\\ {[}EEI{]}\end{tabular}}} 
 & BFMM{[}EEI{]} & \textbf{0.950} & \textbf{(0.923, 0.974)} & \textbf{0.950} & \textbf{(0.917, 0.972)} & \textbf{0.949} & \textbf{(0.912, 0.969)} \\
 & BFMM{[}EEE{]} & 0.951 & (0.922, 0.969) & 0.950 & (0.712, 0.970) & 0.949 & (0.912, 0.969) \\
 & BFMM{[}VVV{]} & 0.947 & (0.922, 0.970) & 0.945 & (0.916, 0.968) & 0.944 & (0.910, 0.966) \\
 & Bootstrap K-means$^a$(init.) & 0.895 & (0.841, 0.931) & 0.893 & (0.658, 0.934) & 0.893 & (0.651, 0.931) \\
 & clustMD{[}EEI{]} & 0.947 & (0.917, 0.967) & 0.943 & (0.914, 0.967) & 0.934 & (0.902, 0.962) \\
 & clustMD{[}VVI{]} & 0.949 & (0.921, 0.971) & 0.941 & (0.916, 0.964) & 0.915 & (0.879, 0.943) \\
 & LCA & 0.951 & (0.924, 0.971) & 0.944 & (0.917, 0.964) & 0.918 & (0.690, 0.948) \\
 & VarSelLCM + MICL & 0.950 & (0.928, 0.971) & 0.944 & (0.485, 0.967) & 0.447 & (0.257, 0.705) \\
 & K-prototype & 0.869 & (0.818, 0.919) & 0.863 & (0.817, 0.916) & 0.845 & (0.780, 0.901) \\
 & PAM + Gower & 0.140 & (0.061, 0.225) & 0.213 & (0.102, 0.298) & 0.275 & (0.108, 0.368) \\
 & HyDaP & 0.886 & (0.746, 0.947) & 0.863 & (0.626, 0.931) & 0.824 & (0.614, 0.898) \\
 & Bootstrap K-means$^b$ & 0.922 & (0.885, 0.956) & 0.910 & (0.872, 0.950) & 0.887 & (0.833, 0.922) \\
 & mclust{[}EEI{]}$^b$ & 0.947 & (0.920, 0.969) & 0.944 & (0.911, 0.967) & 0.934 & (0.902, 0.960) \\ 
 & mclust{[}EEE{]}$^b$ & 0.947 & (0.915, 0.973) & 0.946 & (0.916, 0.966) & 0.933 & (0.907, 0.964) \\
 & mclust{[}VVV{]}$^b$ & 0.934 & (0.661, 0.957) & 0.931 & (0.710, 0.952) & 0.903 & (0.864, 0.931) \\
 \cmidrule(l){1-8}
\multirow{15}{*}{\textbf{\begin{tabular}[l|]{@{}l@{}}Data\\ {[}EEE{]}\end{tabular}}} 
 & BFMM{[}EEI{]} & 0.943 & (0.884, 0.969) & 0.939 & (0.872, 0.967) & 0.923 & (0.472, 0.955) \\
 & BFMM{[}EEE{]} & \textbf{0.981} & \textbf{(0.966, 0.992)} & \textbf{0.979} & \textbf{(0.962, 0.991)} & \textbf{0.978} & \textbf{(0.957, 0.991)} \\
 & BFMM{[}VVV{]} & 0.979 & (0.959, 0.992) & 0.978 & (0.960, 0.991) & 0.976 & (0.959, 0.989) \\
 & Bootstrap K-means$^a$(init.) & 0.435 & (0.366, 0.942) & 0.467 & (0.370, 0.937) & 0.435 & (0.369, 0.917) \\
 & clustMD{[}EEI{]} & 0.921 & (0.815, 0.956) & 0.919 & (0.823, 0.956) & 0.866 & (0.770, 0.953) \\
 & clustMD{[}VVI{]} & 0.740 & (0.536, 0.955) & 0.705 & (0.475, 0.930) & 0.646 & (0.430, 0.834) \\
 & LCA & 0.732 & (0.519, 0.947) & 0.691 & (0.482, 0.894) & 0.626 & (0.242, 0.820) \\
 & VarSelLCM + MICL & 0.916 & (0.641, 0.966) & 0.832 & (0.398, 0.955) & 0.267 & (0.179, 0.623) \\
 & K-prototype & 0.689 & (0.357, 0.871) & 0.756 & (0.334, 0.894) & 0.677 & (0.365, 0.897) \\
 & PAM + Gower & 0.129 & (0.053, 0.201) & 0.172 & (0.063, 0.267) & 0.226 & (0.098, 0.324) \\
 & HyDaP & 0.896 & (0.390, 0.947) & 0.905 & (0.383, 0.944) & 0.634 & (0.382, 0.961) \\
 & Bootstrap K-means$^b$ & 0.896 & (0.661, 0.947) & 0.928 & (0.880, 0.959) & 0.946 & (0.909, 0.971) \\
 & mclust{[}EEI{]}$^b$ & 0.919 & (0.835, 0.954) & 0.925 & (0.816, 0.956) & 0.886 & (0.795, 0.959) \\
 & mclust{[}EEE{]}$^b$ & 0.976 & (0.962, 0.991) & 0.975 & (0.960, 0.989) & 0.966 & (0.946, 0.985) \\
 & mclust{[}VVV{]}$^b$ & 0.973 & (0.942, 0.993) & 0.971 & (0.950, 0.984) & 0.952 & (0.860, 0.977) \\
 \cmidrule(l){1-8}
\multirow{15}{*}{\textbf{\begin{tabular}[l|]{@{}l@{}}Data\\ {[}VVV{]}\end{tabular}}} 
 & BFMM{[}EEI{]} & 0.919 & (0.880, 0.951) & 0.922 & (0.878, 0.950) & 0.922 & (0.886, 0.950) \\
 & BFMM{[}EEE{]} & 0.939 & (0.915, 0.963) & 0.939 & (0.915, 0.964) & 0.939 & (0.911, 0.963) \\
 & BFMM{[}VVV{]} & \textbf{0.967} & \textbf{(0.951, 0.982)} & \textbf{0.966} & \textbf{(0.949, 0.981)} & \textbf{0.964} & \textbf{(0.941, 0.978)} \\
 & Bootstrap K-means$^a$(init.) & 0.739 & (0.458, 0.925) & 0.693 & (0.461, 0.928) & 0.645 & (0.462, 0.927) \\
 & clustMD{[}EEI{]} & 0.907 & (0.380, 0.945) & 0.886 & (0.582, 0.919) & 0.846 & (0.778, 0.894) \\
 & clustMD{[}VVI{]} & 0.943 & (0.429, 0.968) & 0.942 & (0.668, 0.966) & 0.901 & (0.788, 0.944) \\
 & LCA & 0.947 & (0.446, 0.966) & 0.941 & (0.449, 0.966) & 0.881 & (0.579, 0.947) \\
 & VarSelLCM + MICL & 0.949 & (0.924, 0.971) & 0.943 & (0.454, 0.967) & 0.528 & (0.133, 0.668) \\
 & K-prototype & 0.749 & (0.442, 0.874) & 0.822 & (0.400, 0.897) & 0.839 & (0.394, 0.906) \\
 & PAM + Gower & 0.132 & (0.033, 0.204) & 0.190 & (0.079, 0.298) & 0.253 & (0.131, 0.341) \\
 & HyDaP & 0.832 & (0.407, 0.912) & 0.833 & (0.410, 0.897) & 0.841 & (0.544, 0.918) \\
 & Bootstrap K-means$^b$ & 0.901 & (0.850, 0.938) & 0.911 & (0.875, 0.940) & 0.907 & (0.876, 0.933) \\
 & mclust{[}EEI{]}$^b$ & 0.912 & (0.870, 0.944) & 0.889 & (0.825, 0.925) & 0.849 & (0.778, 0.904) \\
 & mclust{[}EEE{]}$^b$ & 0.935 & (0.905, 0.958) & 0.928 & (0.894, 0.955) & 0.899 & (0.854, 0.935) \\
 & mclust{[}VVV{]}$^b$ & 0.961 & (0.936, 0.982) & 0.953 & (0.916, 0.973) & 0.931 & (0.887, 0.963) \\
 \cmidrule(l){1-8}
\multicolumn{8}{l}{Abbreviation: init.: initial.}\\
\multicolumn{8}{l}{\begin{tabular}[c]{@{}l@{}}$^a$Initial clustering setting of BFMM approach using only the continuous variables. \\
$^b$Using all variables with categorical variables treated as continuous. \end{tabular}}\\
 \bottomrule
\end{tabular} }
\label{tab:SimAll.ARI}
\end{table}

%%%%%% Simulation: Avg. Variable Importance Weights %%%%%%
\begin{table}[!htbp]
\small\center
\caption{Variable importance weight estimated by BFMMs in each simulated scenarios.}
\scalebox{0.78}{
\begin{tabular}{lc|ccc|ccc|ccc}
\toprule	
\textbf{} & \textbf{} & \multicolumn{9}{c}{\textbf{Mean Variable Importance Weight, across 100 trials}} \\
\cmidrule(l){3-11}
\textbf{} & \textbf{} & \multicolumn{3}{c|}{\textbf{Uncensored}} & \multicolumn{3}{c|}{\textbf{\begin{tabular}[c]{@{}c@{}}$X_3-X_5$ each\\ censored 20\%\end{tabular}}} & \multicolumn{3}{c}{\textbf{\begin{tabular}[c]{@{}c@{}}$X_3-X_5$ each\\ ensored 40\%\end{tabular}}} \\
\cmidrule(l){3-11}
\multicolumn{2}{l|}{\textbf{Simulated}} & \multirow{2}{*}{\textbf{\begin{tabular}[c|]{@{}c@{}}BFMM\\ {[}EEI{]}\end{tabular}}} & \multirow{2}{*}{\textbf{\begin{tabular}[c]{@{}c@{}}BFMM\\ {[}EEE{]}\end{tabular}}} & \multirow{2}{*}{\textbf{\begin{tabular}[c]{@{}c@{}}BFMM\\ {[}VVV{]}\end{tabular}}} & \multirow{2}{*}{\textbf{\begin{tabular}[c|]{@{}c@{}}BFMM\\ {[}EEI{]}\end{tabular}}} & \multirow{2}{*}{\textbf{\begin{tabular}[c]{@{}c@{}}BFMM\\ {[}EEE{]}\end{tabular}}} & \multirow{2}{*}{\textbf{\begin{tabular}[c]{@{}c@{}}BFMM\\ {[}VVV{]}\end{tabular}}} & \multirow{2}{*}{\textbf{\begin{tabular}[c]{@{}c@{}}BFMM\\ {[}EEI{]}\end{tabular}}} & \multirow{2}{*}{\textbf{\begin{tabular}[c]{@{}c@{}}BFMM\\ {[}EEE{]}\end{tabular}}} & \multirow{2}{*}{\textbf{\begin{tabular}[c]{@{}c@{}}BFMM\\ {[}VVV{]}\end{tabular}}} \\
\textbf{Structure} & \textbf{Variable} &  &  &  &  &  &  &  &  &  \\
\cmidrule(l){1-11}
\multirow{14}{*}{\textbf{\begin{tabular}[l|]{@{}l@{}}Data\\ {[}EEI{]}\end{tabular}}} 
 & $X^*_1$ & 0.992 & 0.991 & 0.993 & 0.985 & 0.979 & 0.984 & 0.980 & 0.981 & 0.982 \\
 & $X^*_2$ & 1.000 & 1.000 & 1.000 & 1.000 & 1.000 & 0.997 & 1.000 & 1.000 & 0.997 \\
 & $X^*_3$ & 1.000 & 1.000 & 1.000 & 1.000 & 1.000 & 0.996 & 1.000 & 1.000 & 0.997 \\
 & $X^*_4$ & 0.785 & 0.781 & 0.783 & 0.811 & 0.810 & 0.807 & 0.849 & 0.844 & 0.840 \\
 & $X_5$ & 0.437 & 0.440 & 0.440 & 0.468 & 0.465 & 0.464 & 0.498 & 0.495 & 0.495 \\
 & $X_6$ & 0.143 & 0.146 & 0.138 & 0.143 & 0.140 & 0.142 & 0.131 & 0.131 & 0.132 \\
 & $X_7$ & 0.092 & 0.091 & 0.091 & 0.098 & 0.102 & 0.088 & 0.091 & 0.092 & 0.084 \\
 & $X^*_8$ & 0.536 & 0.534 & 0.539 & 0.535 & 0.533 & 0.537 & 0.536 & 0.533 & 0.538 \\
 & $X^*_9$ & 0.723 & 0.724 & 0.723 & 0.722 & 0.723 & 0.718 & 0.720 & 0.722 & 0.722 \\
 & $X^*_{10}$ & 0.780 & 0.781 & 0.781 & 0.776 & 0.774 & 0.775 & 0.777 & 0.775 & 0.774 \\
 & $X_{11}$ & 0.236 & 0.237 & 0.232 & 0.244 & 0.247 & 0.232 & 0.241 & 0.242 & 0.233 \\
 & $X_{12}$ & 0.089 & 0.091 & 0.090 & 0.090 & 0.092 & 0.091 & 0.091 & 0.089 & 0.092 \\
 & $X_{13}$ & 0.033 & 0.033 & 0.034 & 0.033 & 0.034 & 0.035 & 0.034 & 0.033 & 0.037 \\
 & $X_{14}$ & 0.027 & 0.026 & 0.026 & 0.028 & 0.027 & 0.030 & 0.025 & 0.026 & 0.030 \\
\cmidrule(l){1-11}
\multirow{14}{*}{\textbf{\begin{tabular}[l|]{@{}l@{}}Data\\ {[}EEE{]}\end{tabular}}} 
 & $X^*_1$ & 0.893 & 0.941 & 0.944 & 0.867 & 0.928 & 0.933 & 0.800 & 0.879 & 0.885 \\
 & $X^*_2$ & 0.994 & 0.995 & 0.995 & 0.991 & 0.992 & 0.993 & 0.975 & 0.981 & 0.981 \\
 & $X^*_3$ & 0.999 & 0.997 & 0.997 & 0.997 & 0.999 & 0.999 & 0.990 & 0.999 & 0.999 \\
 & $X^*_4$ & 0.749 & 0.745 & 0.743 & 0.762 & 0.754 & 0.753 & 0.785 & 0.768 & 0.768 \\
 & $X_5$ & 0.433 & 0.335 & 0.339 & 0.488 & 0.372 & 0.372 & 0.544 & 0.404 & 0.403 \\
 & $X_6$ & 0.204 & 0.161 & 0.163 & 0.197 & 0.155 & 0.153 & 0.192 & 0.146 & 0.142 \\
 & $X_7$ & 0.158 & 0.158 & 0.155 & 0.157 & 0.149 & 0.146 & 0.162 & 0.145 & 0.142 \\
 & $X^*_8$ & 0.541 & 0.523 & 0.524 & 0.542 & 0.521 & 0.524 & 0.548 & 0.520 & 0.523 \\
 & $X^*_9$ & 0.721 & 0.716 & 0.717 & 0.713 & 0.714 & 0.720 & 0.708 & 0.714 & 0.716 \\
 & $X^*_{10}$ & 0.802 & 0.793 & 0.795 & 0.794 & 0.791 & 0.795 & 0.785 & 0.794 & 0.798 \\
 & $X_{11}$ & 0.236 & 0.243 & 0.239 & 0.237 & 0.241 & 0.237 & 0.243 & 0.241 & 0.238 \\
 & $X_{12}$ & 0.091 & 0.088 & 0.091 & 0.091 & 0.089 & 0.091 & 0.089 & 0.090 & 0.089 \\
 & $X_{13}$ & 0.034 & 0.034 & 0.034 & 0.034 & 0.034 & 0.033 & 0.034 & 0.032 & 0.033 \\
 & $X_{14}$ & 0.027 & 0.025 & 0.026 & 0.027 & 0.026 & 0.027 & 0.026 & 0.027 & 0.026 \\
\cmidrule(l){1-11}
\multirow{14}{*}{\textbf{\begin{tabular}[l|]{@{}l@{}}Data\\ {[}VVV{]}\end{tabular}}} 
 & $X^*_1$ & 0.943 & 0.963 & 0.934 & 0.914 & 0.943 & 0.909 & 0.867 & 0.903 & 0.869 \\
 & $X^*_2$ & 0.997 & 0.996 & 0.998 & 0.993 & 0.992 & 0.995 & 0.981 & 0.976 & 0.986 \\
 & $X^*_3$ & 1.000 & 1.000 & 1.000 & 1.000 & 1.000 & 1.000 & 1.000 & 1.000 & 1.000 \\
 & $X^*_4$ & 0.730 & 0.742 & 0.755 & 0.728 & 0.740 & 0.751 & 0.726 & 0.734 & 0.749 \\
 & $X_5$ & 0.458 & 0.366 & 0.405 & 0.488 & 0.397 & 0.431 & 0.513 & 0.424 & 0.463 \\
 & $X_6$ & 0.191 & 0.145 & 0.155 & 0.183 & 0.141 & 0.151 & 0.166 & 0.135 & 0.147 \\
 & $X_7$ & 0.131 & 0.123 & 0.120 & 0.130 & 0.125 & 0.117 & 0.126 & 0.120 & 0.117 \\
 & $X^*_8$ & 0.542 & 0.540 & 0.529 & 0.544 & 0.541 & 0.532 & 0.539 & 0.537 & 0.529 \\
 & $X^*_9$ & 0.709 & 0.708 & 0.712 & 0.710 & 0.707 & 0.711 & 0.711 & 0.705 & 0.712 \\
 & $X^*_{10}$ & 0.830 & 0.814 & 0.810 & 0.828 & 0.816 & 0.812 & 0.828 & 0.815 & 0.812 \\
 & $X_{11}$ & 0.234 & 0.232 & 0.231 & 0.232 & 0.230 & 0.235 & 0.232 & 0.232 & 0.234 \\
 & $X_{12}$ & 0.089 & 0.090 & 0.090 & 0.088 & 0.088 & 0.089 & 0.090 & 0.090 & 0.089 \\
 & $X_{13}$ & 0.033 & 0.034 & 0.032 & 0.033 & 0.033 & 0.032 & 0.033 & 0.033 & 0.033 \\
 & $X_{14}$ & 0.027 & 0.026 & 0.026 & 0.027 & 0.027 & 0.029 & 0.028 & 0.027 & 0.027 \\
 \cmidrule(l){1-11}
\multicolumn{11}{l}{\begin{tabular}[c]{@{}l@{}}*$X_1-X_4$ and $X_8-X_{10}$ are true dominant variables contributing to assign the clusters.\\ Variables $X_1-X_7$ are continuous while $X_8-X_{14}$ are categorical.\\Variable $X_8$ was generated from dominant continuous variables $X_3, X_4$, while $X_{11}$ from noise continuous variables $X_6, X_7$. \end{tabular}}\\ 
 \bottomrule
\end{tabular} }
\label{tab:SimAll.VI}
\end{table}

%%%%%%%%%%%%%%%%%%%%%%%%%%%%%%%%%%%%%%%%

%%%%%%%%%%%%%%%%%%%%%%%%%%%%%%%%%%%%%%%%
%%% SENECA Data Application - Supplemental Tables %%%
\begin{table}[]
\small\center
\caption{SENECA data variable importance estimation by BFMM[VVV] given 4 clusters.}
\scalebox{.93}{
\begin{tabular}{l|ccccccc|c}
\toprule	
\multirow{3}{*}{\textbf{\begin{tabular}[c]{@{}l@{}}SENECA Data\\ Variable\end{tabular}}} & \multicolumn{7}{c|}{\textbf{\begin{tabular}[c]{@{}c@{}}Variable importance weight by different values of threshold $k$\\ used to set slab vs. spike prior variance ratio $\omega$ \end{tabular}}} & \multirow{3}{*}{\textbf{\begin{tabular}[c]{@{}c@{}}Avg.\\Weight\end{tabular}}} \\ \cmidrule(l){2-8}
 & $k=60$ & $k=65$ & $k=70$ & $k=75$ & $k=80$ & $k=85$ & $k=90$ &  \\ \cmidrule(l){2-8} 
 & $\omega=43$ & $\omega=48$ & $\omega=57$ & $\omega=67$ & $\omega=80$ & $\omega=100$ & $\omega=130$ &  \\ \cmidrule(l){1-9}
\multicolumn{9}{l}{\textbf{Continuous variable}}\\ \cmidrule(l){1-9}
Troponin$^a$ & 0.99 & 0.99 & 0.99 & 1.00 & 1.00 & 1.00 & 1.00 & 0.99 \\
AST$^a$ & 0.94 & 0.94 & 0.96 & 0.97 & 0.98 & 0.99 & 0.99 & 0.97\\
Lactate$^a$ & 0.88 & 0.90 & 0.92 & 0.95 & 0.96 & 0.99 & 1.00 & 0.94\\
GCS$^a$ & 0.85 & 0.93 & 0.88 & 0.94 & 0.93 & 0.95 & 0.97 & 0.92 \\
SBP$^a$ & 0.81 & 0.83 & 0.85 & 0.90 & 0.92 & 0.96 & 0.98 & 0.89 \\
Bicarbonate$^a$ & 0.77 & 0.75 & 0.80 & 0.83 & 0.87 & 0.93 & 0.97 & 0.85 \\
SaO2 & 0.76 & 0.74 & 0.81 & 0.83 & 0.86 & 0.92 & 0.97 & 0.83 \\
Albumin & 0.74 & 0.73 & 0.78 & 0.82 & 0.85 & 0.91 & 0.95 & 0.83 \\
Bands & 0.72 & 0.66 & 0.74 & 0.76 & 0.80 & 0.87 & 0.93 & 0.78 \\
ALT & 0.68 & 0.66 & 0.72 & 0.74 & 0.78 & 0.84 & 0.87 & 0.76 \\
RR & 0.63 & 0.68 & 0.67 & 0.74 & 0.78 & 0.84 & 0.91 & 0.75 \\
Heart rate & 0.63 & 0.63 & 0.67 & 0.74 & 0.77 & 0.81 & 0.87 & 0.73 \\
INR & 0.66 & 0.72 & 0.68 & 0.74 & 0.74 & 0.76 & 0.77 & 0.72 \\
CRP & 0.65 & 0.58 & 0.69 & 0.69 & 0.73 & 0.77 & 0.80 & 0.70 \\
ESR & 0.70 & 0.55 & 0.71 & 0.66 & 0.71 & 0.76 & 0.80 & 0.70 \\
Bilirubin & 0.61 & 0.60 & 0.63 & 0.63 & 0.66 & 0.70 & 0.74 & 0.65 \\
PaO2 & 0.53 & 0.53 & 0.58 & 0.59 & 0.62 & 0.65 & 0.67 & 0.60 \\
Platelets & 0.58 & 0.58 & 0.58 & 0.57 & 0.57 & 0.57 & 0.57 & 0.58 \\
WBC & 0.57 & 0.55 & 0.57 & 0.56 & 0.56 & 0.56 & 0.55 & 0.56 \\
Chloride & 0.42 & 0.35 & 0.48 & 0.48 & 0.55 & 0.64 & 0.72 & 0.52 \\
Hemoglobin & 0.48 & 0.36 & 0.51 & 0.46 & 0.52 & 0.58 & 0.66 & 0.51 \\
Glucose & 0.50 & 0.43 & 0.50 & 0.47 & 0.50 & 0.54 & 0.60 & 0.51 \\
Sodium & 0.40 & 0.36 & 0.41 & 0.40 & 0.43 & 0.47 & 0.51 & 0.43 \\
Temperature & 0.36 & 0.32 & 0.37 & 0.36 & 0.37 & 0.38 & 0.39 & 0.36 \\
Creatinine & 0.41 & 0.21 & 0.39 & 0.27 & 0.31 & 0.31 & 0.32 & 0.32 \\
Age & 0.17 & 0.17 & 0.17 & 0.19 & 0.21 & 0.24 & 0.29 & 0.21 \\ \cmidrule(l){1-9}
\multicolumn{9}{l}{\textbf{Categorical variable}}\\ \cmidrule(l){1-9}
Sex$^{b}$ & 0.22 & 0.22 & 0.22 & 0.22 & 0.23 & 0.22 & 0.22 & 0.22 \\
Race$^{b}$ & 0.07 & 0.08 & 0.07 & 0.08 & 0.07 & 0.07 & 0.08 & 0.07 \\  \cmidrule(l){1-9}
\multicolumn{9}{l}{Abbreviations: Avg.: averaged; AST: aspartate aminotransferase; GCS: Glasgow coma scale;}\\
\multicolumn{9}{l}{SBP: systolic blood pressure; SaO2: oxygen saturation; ALT: alanine aminotransferase;}\\
\multicolumn{9}{l}{RR: respiration rate; INR: international normalized ratio; CRP: C-reactive protein;}\\
\multicolumn{9}{l}{ESR: erythrocyte sedimentation rate; PaO2: partial pressure of oxygen; WBC: white blood cell.} \\  
\multicolumn{9}{l}{$^a$Identified most important variables with the highest averaged importance weights $\geq 0.85$.} \\
\multicolumn{9}{l}{$^b$Gender is categorized by 2 levels (female/male), race by 3 levels (white/black/others).}\\
 \bottomrule
\end{tabular}
}
\label{tab:VarImpSeneca}
\end{table}

%%%%%%%%%%%%%%%%%%%%%%%%%%%%%%%%%%%%%%%%
%%%%%%%%%%%%%%%%%%%%%%%%%%%%%%%%%%%%%%%%

%%%%%%%%%%%%%%%%%%%%%%%%%%%%%%%%%%%%%%%%%%%%%%%%%%%%%%%%%%

\begin{table}[!htbp]
\small\center
\caption{Distributions of clinical endpoints by BFMM[VVV] clustering of SENECA data.}
\scalebox{.95}{
\begin{tabular}{lcccccc}
\toprule	
\multirow{2}{*}{\textbf{\begin{tabular}[c]{@{}l@{}}Variable and\\ Clinical Endpoint\end{tabular}}} & \multicolumn{6}{c}{\textbf{Cluster assigned by BFMM[VVV] with $G=4$ and $k=75$}} \\
 & \textbf{Cluster 1} & \textbf{Cluster 2} & \textbf{Cluster 3} & \textbf{Cluster 4} & \textbf{Overall} & \textbf{\textit{P} value$^d$} \\ \cmidrule(l){1-7}
\textbf{Cluster size} & \textbf{\begin{tabular}[c]{@{}c@{}}8395\\ (41.6\%)\end{tabular}} & \textbf{\begin{tabular}[c]{@{}c@{}}5628\\ (27.9\%)\end{tabular}} & \textbf{\begin{tabular}[c]{@{}c@{}}3455\\ (17.1\%)\end{tabular}} & \textbf{\begin{tabular}[c]{@{}c@{}}2711\\ (13.4\%)\end{tabular}} & \textbf{20189} & \textbf{} \\ \cmidrule(l){1-7}
\multicolumn{7}{l}{\textbf{Clinical endpoint$^a$, n (\%)} } \\ \cmidrule(l){1-7}
ICU admission & \begin{tabular}[c]{@{}c@{}}2391\\ (28.5\%)\end{tabular} & \begin{tabular}[c]{@{}c@{}}2080\\ (37.0\%)\end{tabular} & \begin{tabular}[c]{@{}c@{}}2434\\ (70.4\%)\end{tabular} & \begin{tabular}[c]{@{}c@{}}2158\\ (79.6\%)\end{tabular} & \begin{tabular}[c]{@{}c@{}}9063\\ (44.9\%)\end{tabular} & \textless{}.0001* \\
Mechanical ventilation & \begin{tabular}[c]{@{}c@{}}1100\\ (13.1\%)\end{tabular} & \begin{tabular}[c]{@{}c@{}}1364\\ (24.2\%)\end{tabular} & \begin{tabular}[c]{@{}c@{}}1674\\ (48.5\%)\end{tabular} & \begin{tabular}[c]{@{}c@{}}1635\\ (60.3\%)\end{tabular} & \begin{tabular}[c]{@{}c@{}}5773\\ (28.6\%)\end{tabular} & \textless{}.0001* \\
Vasopressors & \begin{tabular}[c]{@{}c@{}}745\\ (8.9\%)\end{tabular} & \begin{tabular}[c]{@{}c@{}}598\\ (10.6\%)\end{tabular} & \begin{tabular}[c]{@{}c@{}}1273\\ (36.8\%)\end{tabular} & \begin{tabular}[c]{@{}c@{}}1139\\ (42.0\%)\end{tabular} & \begin{tabular}[c]{@{}c@{}}3755\\ (18.6\%)\end{tabular} & \textless{}.0001* \\
In-hospital mortality & \begin{tabular}[c]{@{}c@{}}353\\ (4.2\%)\end{tabular} & \begin{tabular}[c]{@{}c@{}}278\\ (4.9\%)\end{tabular} & \begin{tabular}[c]{@{}c@{}}748\\ (21.6\%)\end{tabular} & \begin{tabular}[c]{@{}c@{}}703\\ (25.9\%)\end{tabular} & \begin{tabular}[c]{@{}c@{}}2082\\ (10.3\%)\end{tabular} & \textless{}.0001* \\
\begin{tabular}[c]{@{}l@{}}90-day mortality$^b$\\(if in-hospital alive)\end{tabular} & \begin{tabular}[c]{@{}c@{}}994\\ (12.8\%)\end{tabular} & \begin{tabular}[c]{@{}c@{}}715\\ (13.8\%)\end{tabular} & \begin{tabular}[c]{@{}c@{}}614\\ (23.6\%)\end{tabular} & \begin{tabular}[c]{@{}c@{}}435\\ (23.0\%)\end{tabular} & \begin{tabular}[c]{@{}c@{}}2758\\ (15.8\%)\end{tabular} & \textless{}.0001* \\
\begin{tabular}[c]{@{}l@{}}365-day mortality$^c$\\ (if in-hospital alive)\end{tabular} & \begin{tabular}[c]{@{}c@{}}1962\\ (24.4\%)\end{tabular} & \begin{tabular}[c]{@{}c@{}}1375\\ (25.7\%)\end{tabular} & \begin{tabular}[c]{@{}c@{}}997\\ (36.8\%)\end{tabular} & \begin{tabular}[c]{@{}c@{}}709\\ (35.3\%)\end{tabular} & \begin{tabular}[c]{@{}c@{}}5043\\ (27.9\%)\end{tabular} & \textless{}.0001* \\ \cmidrule(l){1-7}
\multicolumn{7}{l}{Abbreviations: SD: standard deviation; ICU: intensive care unit.}\\ 
\multicolumn{7}{l}{$^a$Proportion ($\%$) is presented in parentheses below the count number.}\\
\multicolumn{7}{l}{$^b$Total number is 17,432 after excluding in-hospital death and missing.} \\
\multicolumn{7}{l}{$^c$Total number is 18,107 after excluding in-hospital death.}\\
\multicolumn{7}{l}{$^d$Chi-squared test \textit{p}-value for clinical endpoint.}\\
\multicolumn{7}{l}{*Significant difference exists among the distributions across the 4 clusters.}\\ \bottomrule
\label{tab:PostHocSeneca}
\end{tabular} }
\end{table}

%%%%%%%%%%%%%%%%%%%%%%%%%%%%%%%%%%%%%%%%
%%%%%%%%%%%%%%%%%%%%%%%%%%%%%%%%%%%%%%%%

%%%%%%%%%%%%%%%%%%%%%%%%%%%%%%%%%%%%%%%%
%%%%%%%%%%%%%%%%%%%%%%%%%%%%%%%%%%%%%%%%
%%% EDEN Trial Data Application - Supplemental Tables %%%

\begin{table}[]
\small\center
\caption{EDEN Trial data variable importance estimation by BFMM[EEI] given 3 clusters.}
\scalebox{.88}{
\begin{tabular}{l|ccccccc|c}
\toprule	
\multirow{3}{*}{\textbf{\begin{tabular}[c]{@{}l@{}}EDEN Trial Data\\ Variable\end{tabular}}} & \multicolumn{7}{c|}{\textbf{\begin{tabular}[c]{@{}c@{}}Variable importance weight by different values of threshold $k$\\ used to set slab vs. spike prior variance ratio $\omega$ \end{tabular}}} & \multirow{3}{*}{\textbf{\begin{tabular}[c]{@{}c@{}}Avg.\\Weight\end{tabular}}} \\ \cmidrule(l){2-8}
 & $k=60$ & $k=65$ & $k=70$ & $k=75$ & $k=80$ & $k=85$ & $k=90$ &  \\ \cmidrule(l){2-8} 
 & $\omega=67$ & $\omega=73$ & $\omega=82$ & $\omega=90$ & $\omega=100$ & $\omega=116$ & $\omega=135$ &  \\ \cmidrule(l){1-9}
\multicolumn{9}{l}{\textbf{Censored biomarker (i.e. all continuous)}} \\ \cmidrule(l){1-9}	
RAGE & 1.00 & 1.00 & 1.00 & 1.00 & 1.00 & 1.00 & 1.00 & 1.00   \\
TNF-$\alpha$ & 1.00 & 1.00 & 1.00 & 1.00 & 1.00 & 1.00 & 1.00 & 1.00   \\
IL-6  & 0.95 & 0.95 & 0.96 & 0.96 & 0.96 & 0.97 & 0.98 & 0.96   \\
PCT  & 0.91 & 0.91 & 0.91 & 0.92 & 0.93 & 0.94 & 0.94 & 0.92   \\
IL-8 & 0.89 & 0.88 & 0.90 & 0.90 & 0.90 & 0.92 & 0.93 & 0.90   \\
ST2 & 0.76 & 0.76 & 0.77 & 0.76 & 0.77 & 0.78 & 0.78 & 0.77   \\
Angiopoietin   & 0.75 & 0.75 & 0.76 & 0.75 & 0.76 & 0.76 & 0.77 & 0.76   \\
\cmidrule(l){1-9}	
\multicolumn{9}{l}{\textbf{Continuous variable}} \\\cmidrule(l){1-9}	
Blood urea nitrogen & 1.00 & 1.00 & 1.00 & 1.00 & 1.00 & 1.00 & 1.00 & 1.00   \\
Creatinine & 1.00 & 1.00 & 1.00 & 1.00 & 1.00 & 1.00 & 1.00 & 1.00   \\
PEEP & 0.83 & 0.83 & 0.84 & 0.84 & 0.85 & 0.86 & 0.87 & 0.85   \\
Minute ventilation & 0.78 & 0.78 & 0.80 & 0.80 & 0.82 & 0.82 & 0.82 & 0.80   \\
Mean airway pressure & 0.77 & 0.77 & 0.78 & 0.78 & 0.78 & 0.79 & 0.79 & 0.78   \\
PaO2/FiO2 ratio & 0.76 & 0.76 & 0.78 & 0.78 & 0.78 & 0.79 & 0.78 & 0.77   \\
APACHE III score & 0.76 & 0.76 & 0.76 & 0.77 & 0.77 & 0.78 & 0.77 & 0.77   \\
Plateau pressure & 0.74 & 0.74 & 0.74 & 0.74 & 0.74 & 0.74 & 0.74 & 0.74   \\
Age & 0.73 & 0.73 & 0.73 & 0.73 & 0.73 & 0.74 & 0.74 & 0.73   \\
Bicarbonate & 0.72 & 0.73 & 0.72 & 0.73 & 0.73 & 0.73 & 0.72 & 0.73   \\
Platelets & 0.71 & 0.71 & 0.72 & 0.73 & 0.73 & 0.74 & 0.74 & 0.73   \\
Heart rate & 0.67 & 0.67 & 0.68 & 0.69 & 0.69 & 0.69 & 0.70 & 0.68   \\
Hematocrit level & 0.60 & 0.61 & 0.63 & 0.64 & 0.65 & 0.66 & 0.66 & 0.64   \\
Temperature & 0.49 & 0.49 & 0.50 & 0.52 & 0.51 & 0.52 & 0.53 & 0.51   \\
Albumin & 0.45 & 0.47 & 0.49 & 0.50 & 0.51 & 0.54 & 0.55 & 0.50   \\
Sodium & 0.38 & 0.40 & 0.41 & 0.43 & 0.43 & 0.45 & 0.47 & 0.42   \\
Total protein & 0.36 & 0.38 & 0.39 & 0.40 & 0.40 & 0.43 & 0.44 & 0.40   \\
Glucose & 0.27 & 0.28 & 0.28 & 0.29 & 0.29 & 0.29 & 0.30 & 0.28   \\
Body mass index & 0.24 & 0.23 & 0.24 & 0.24 & 0.24 & 0.24 & 0.24 & 0.24   \\
Tidal volume & 0.20 & 0.21 & 0.20 & 0.20 & 0.21 & 0.19 & 0.20 & 0.20   \\
White blood cell & 0.18 & 0.19 & 0.19 & 0.18 & 0.18 & 0.18 & 0.18 & 0.18   \\
Mean arterial pressure & 0.19 & 0.18 & 0.18 & 0.18 & 0.18 & 0.18 & 0.18 & 0.18   \\
\cmidrule(l){1-9}
\multicolumn{9}{l}{\textbf{Categorical variable (i.e. all binary) }} \\\cmidrule(l){1-9}	
Diabetes & 0.68 & 0.67 & 0.68 & 0.68 & 0.68 & 0.68 & 0.69 & 0.68   \\
Sepsis-induced PLI & 0.65 & 0.64 & 0.65 & 0.65 & 0.64 & 0.64 & 0.66 & 0.65   \\
Pneumonia-induced PLI & 0.17 & 0.17 & 0.17 & 0.16 & 0.16 & 0.16 & 0.16 & 0.16   \\
Chronic pulmonary & 0.10 & 0.10 & 0.09 & 0.11 & 0.10 & 0.10 & 0.10 & 0.10   \\
Sex (female/male) & 0.10 & 0.10 & 0.09 & 0.10 & 0.10 & 0.10 & 0.10 & 0.10   \\
Smoker & 0.08 & 0.08 & 0.08 & 0.08 & 0.09 & 0.08 & 0.08 & 0.08   \\
Race (white/others) & 0.08 & 0.08 & 0.08 & 0.08 & 0.08 & 0.08 & 0.08 & 0.08   \\
Aspiration-induced PLI & 0.06 & 0.06 & 0.05 & 0.06 & 0.06 & 0.06 & 0.06 & 0.06  \\
\cmidrule(l){1-9}
\multicolumn{9}{l}{\small Abbreviations: Avg.: averaged; RAGE: receptor for advanced glycation end products; } \\ 
\multicolumn{9}{l}{\small TNF: tumor necrosis factor; IL: interleukin; PCT: procalcitonin; ST: suppression of tumorigenicity; } \\ 
\multicolumn{9}{l}{\small PEEP: positive end-expiratory pressure; PaO2: arterial oxygen partial pressure; }\\ 
\multicolumn{9}{l}{\small FiO2: fractional inspired oxygen; APACHE: acute physiology and chronic health evaluation; }\\ 
\multicolumn{9}{l}{\small PLI: primary lung injury.}\\  
\bottomrule
\end{tabular}}
  \label{tab:VarImpEDEN}
\end{table}

%%%%%%%%%%%%%%%%%%%%%%%%%%%%%%%%%%%%%%%%
%%%%%%%%%%%%%%%%%%%%%%%%%%%%%%%%%%%%%%%%

%%%%%%%%%%%%%%%%%%%%%%%%%%%%%%%%%%%%%%%%
\begin{table}[!htbp]\center
\caption{\fontsize{12pt}{12pt}\selectfont Distributions of all variables used to cluster EDEN trial data by BFMM[EEI].}
\scalebox{0.77}{
\begin{tabular}{l|ccccc|c}
\toprule
\multirow{2}{*}{\textbf{\begin{tabular}[c]{@{}l@{}}EDEN Trial Data\\ Variable\end{tabular}}} & \multicolumn{5}{c|}{\textbf{Cluster assigned by BFMM{[}EEI{]} with $G=3$ and $k=75$}} & \multirow{2}{*}{\textbf{\begin{tabular}[c]{@{}c@{}}Avg.$^d$\\Weight\end{tabular}}} \\ 
& \textbf{Cluster 1} & \textbf{Cluster 2} & \textbf{Cluster 3} & \textbf{Overall} & \textbf{\textit{P} value$^c$} &  \\ 
\cmidrule(l){1-7}
\textbf{Cluster size} & \textbf{\begin{tabular}[c]{@{}c@{}}386\\ (43\%)\end{tabular}} & \textbf{\begin{tabular}[c]{@{}c@{}}282\\ (32\%)\end{tabular}} & \textbf{\begin{tabular}[c]{@{}c@{}}221\\ (25\%)\end{tabular}} & \textbf{\begin{tabular}[c]{@{}c@{}}889\\ (100\%)\end{tabular}} &  & \\ 
\cmidrule(l){1-7}
\multicolumn{7}{l}{\textbf{Continuous variable$^a$, mean (SD) or median [IQR] at original scale}} \\ 
\cmidrule(l){1-7}

RAGE, $\times10^3$ pg/mL & 7.6 [6.3] & 22.5 [25.9] & 19.3 [43.0] & 12.6 [19.1] & \textless{}.0001* & 1 \\

TNF-$\alpha$, $\times10^2$ pg/mL & 36.4 [24.9] & 113.0 [85.6] & 41.1 [25.0] & 50.6 [52.6] & \textless{}.0001* & 1 \\

Blood urea nitrogen, mg/dL & 16 [13] & 41 [31] & 14 [11] & 21 [23] & \textless{}.0001* & 1 \\

Creatinine, mg/dL & 0.9 [0.5] & 2.5 [2.2] & 1.0 [0.7] & 1.1 [1.2] & \textless{}.0001* & 1 \\

IL-6, pg/mL & 54.2 [92.8] & 196.6 [604.7] & 184.7 [418.1] & 105.7 [251.1] & \textless{}.0001* & 0.96 \\

PCT, $\times10^2$ pg/mL $^b$ & 6.5 [16.3] & 33.0 [21.6] & 25.3 [26.9] & 19.5 [27.8] & \textless{}.0001* & 0.92 \\

IL-8, pg/mL $^b$ & 13.1 [16.6] & 51.9 [109.8] & 31.4 [46.5] & 23.0 [43.3] & \textless{}.0001* & 0.90 \\

PEEP, cm H$_2$O & 8.1 (2.9) & 8.9 (3.6) & 12.9 (4.8) & 9.5 (4.2) & \textless{}.0001* & 0.85 \\
Minute ventilation, L/min & 9.6 [3.7] & 11.3 [4.0] & 11.8 [4.7] & 10.7 [4.1] & \textless{}.0001* & 0.80 \\
Mean airway pressure, cm H$_2$O & 13 [4] & 14 [6] & 19 [7] & 15 [6] & \textless{}.0001* & 0.78 \\

PaO2/FiO2 ratio & 158.0 [83.3] & 132.5 [88.8] & 78.0 [42.0] & 125.0 [92.0] & \textless{}.0001* & 0.77 \\
APACHE III & 73.9 (19.7) & 110.7 (23.5) & 93.6 (23.6) & 90.5 (27.0) & \textless{}.0001* & 0.77 \\

ST-2, $\times10^4$ pg/mL $^b$ & 18.1 [21.0] & 52.0 [93.3] & 36.3 [48.1] & 28.7 [44.1] & \textless{}.0001* & 0.77 \\

Angiopoietin-2, $\times10^3$ pg/mL & 14.3 [12.7] & 37.0 [36.6] & 20.7 [20.1] & 20.2 [23.6] & \textless{}.0001* & 0.76 \\

Plateau pressure, cm H$_2$O & 21.5 (4.2) & 23.2 (4.5) & 28.1 (4.9) & 23.7 (5.2) & \textless{}.0001* & 0.74 \\
Age, y & 53.40 (15.2) & 57.2 (16.1) & 43.0 (14.0) & 52.0 (16.1) & \textless{}.0001* & 0.73 \\
Bicarbonate, mEq/L & 23.6 (4.5) & 18.6 (4.7) & 21.8 (5.3) & 21.6 (5.3) & \textless{}.0001* & 0.73 \\

Platelets, $\times10^9$/L & 190.0 [133.0] & 133.5 [131.8] & 169.0 [118.0] & 169.0 [131.0] & \textless{}.0001* & 0.73 \\

Heart rate, beats/min & 89.9 (18.3) & 94.0 (20.0) & 103.0 (19.7) & 94.5 (19.9) & \textless{}.0001* & 0.68 \\

Hematocrit level, \% & 30.4 (5.7) & 28.5 (6.1) & 31.6 (5.7) & 30.1 (5.9) & \textless{}.0001* & 0.64 \\
Temperature, $^{\circ}$C & 37.3 (0.9) & 37.3 (1.0) & 37.7 (1.0) & 37.4 (1.0) & \textless{}.0001* & 0.51 \\
Albumin, g/dL & 2.4 (0.6) & 2.2 (0.6) & 2.3 (0.6) & 2.3 (0.6) & \textless{}.0001* & 0.50 \\
Sodium, mEq/L & 138.1 (4.9) & 136.5 (5.7) & 136.7 (5.2) & 137.3 (5.3) & 0.0002* & 0.42 \\
Total protein, g/dL & 5.2 (1.1) & 4.9 (1.0) & 5.1 (1.0) & 5.1 (1.1) & 0.0002* & 0.40 \\
Glucose, mg/dL & 108.0 [37.0] & 104.0 [47.8] & 112.0 [42.3] & 108.0 [39.0] & 0.1520 & 0.28 \\

BMI, kg/m$^2$ & 28.4 [10.7] & 28.5 [10.3] & 30.1 [11.9] & 28.8 [10.8] & 0.0465 & 0.24 \\
Tidal volume, mL/kg & 424.2 (85.1) & 421.2 (80.1) & 413.4 (100.2) & 420.6 (87.6) & 0.3369 & 0.20 \\

White blood cell, $\times10^9$/L & 11.8 [6.7] & 13.1 [11.3] & 11.7 [10.5] & 12.0 [8.8] & 0.2781 & 0.18 \\

Mean arterial pressure, mm Hg & 76.9 (13.4) & 74.8 (12.1) & 75.9 (12.9) & 76.0 (12.9) & 0.1031 & 0.18 \\

\cmidrule(l){1-7}
\multicolumn{7}{l}{\textbf{Categorical variable, n (\%)}} \\ 
\cmidrule(l){1-7}
Diabetes & 97 (25.1\%) & 111 (39.4\%) & 34 (15.4\%) & 242 (27.2\%) & \textless{}.0001* & 0.68 \\
Sepsis-induced PLI & 23 (6.0\%) & 58 (20.6\%) & 28 (12.7\%) & 109 (12.3\%) & \textless{}.0001* & 0.65 \\
Pneumonia-induced PLI & 274 (71.0\%) & 175 (62.1\%) & 142 (64.3\%) & 591 (66.5\%) & 0.0391* & 0.16 \\
Chronic pulmonary & 54 (14.0\%) & 38 (13.5\%) & 20 (9.0\%) & 112 (12.6\%) & 0.1824 & 0.10 \\
Sex (female) & 197 (51.0\%) & 129 (45.7\%) & 109 (49.3\%) & 435 (48.9\%) & 0.3978 & 0.10 \\
Smoker & 225 (58.3\%) & 173 (61.3\%) & 135 (61.1\%) & 533 (60.0\%) & 0.6733 & 0.08 \\
Race (white) & 69 (17.9\%) & 60 (21.3\%) & 45 (20.4\%) & 174 (19.6\%) & 0.5185 & 0.08 \\
Smoker & 225 (58.3\%) & 173 (61.3\%) & 135 (61.1\%) & 533 (60.0\%) & 0.6733 & 0.08 \\
Aspiration-induced PLI & 34 (8.8\%) & 29 (10.3\%) & 20 (9.0\%) & 83 (9.3\%) & 0.7995 & 0.06 \\ \cmidrule(l){1-7}
\multicolumn{7}{l}{\small Abbreviations: SD: standard deviation; IQR: interquartile range, equal to Q3 - Q1; Avg.: averaged; } \\ 
\multicolumn{7}{l}{\small RAGE: receptor for advanced glycation end products; TNF: tumor necrosis factor; IL: interleukin; PCT: procalcitonin; } \\ 
\multicolumn{7}{l}{\small PEEP: positive end-expiratory pressure; PaO2: partial pressure of oxygen; FiO2: fractional inspired oxygen;}\\ 
\multicolumn{7}{l}{\small  APACHE: acute physiology and chronic health evaluation; ST: suppression of tumorigenicity; PLI: primary lung injury.}\\  
\multicolumn{6}{l}{$^a$Median [IQR] is reported for continuous variable with skewed distribution that needs log-transformation.}\\
\multicolumn{7}{l}{$^b$Biomarkers with summary statistics calculated using detection limits if given censored observations.}\\
\multicolumn{7}{l}{$^c$One-way ANOVA or $\chi^2$ test \textit{p}-value (with log-transformtion if needed); *significant difference across 3 clusters.}\\
\multicolumn{7}{l}{$^d$Variable importance weight averaged across 7 different Markov chains that have reached convergence.}\\  \bottomrule
\end{tabular} }
  \label{tab:AllVarEDEN}
\end{table}

%%%%%%%%%%%%%%%%%%%%%%%%%%%%%%%%%%%%%%%%
%%%%%%%%%%%%%%%%%%%%%%%%%%%%%%%%%%%%%%%%

\begin{table}[!htbp]
\small\center
\caption{Distributions of clinical endpoints by BFMM[EEI] clustering results of EDEN trial data.}
\scalebox{0.95}{
\begin{tabular}{lccccc}
\toprule	
\multirow{2}{*}{\textbf{\begin{tabular}[c]{@{}l@{}}Intervention and\\ Clinical Endpoint\end{tabular}}} & \multicolumn{5}{c}{\textbf{Cluster assigned by BFMM[EEI] with $G=3$ and $k=75$}} \\
 & \textbf{Cluster 1} & \textbf{Cluster 2} & \textbf{Cluster 3} & \textbf{Overall} & \textbf{\textit{P} value$^a$} \\ \cmidrule(l){1-6}
\textbf{Cluster size} & \textbf{\begin{tabular}[c]{@{}c@{}}386\\ (43\%)\end{tabular}} & \textbf{\begin{tabular}[c]{@{}c@{}}282\\ (32\%)\end{tabular}} & \textbf{\begin{tabular}[c]{@{}c@{}}221\\ (25\%)\end{tabular}} & \textbf{\begin{tabular}[c]{@{}c@{}}889\\ (100\%)\end{tabular}} & \textbf{} \\
\cmidrule(l){1-6}
\multicolumn{6}{l}{\textbf{Randomized Intervention, n (\%)}} \\
\cmidrule(l){1-6}
Full feeding & 181 (46.9\%) & 141 (50.0\%) & 115 (52.0\%) & 437 (49.2\%) & \multirow{2}{*}{0.4480} \\ 
Trophic feeding & 205 (53.1\%) & 141 (50.0\%) & 106 (48.0\%) & 452 (50.8\%) &  \\ \cmidrule(l){1-6}
\multicolumn{6}{l}{\textbf{Clinical Endpoint, n (\%)}} \\
\cmidrule(l){1-6}
60-day mortality (death) & 64 (16.6\%) & 94 (33.3\%) & 45 (20.4\%) & 203 (22.8\%) & \textless{}.0001* \\
\begin{tabular}[c]{@{}l@{}}Gastrointestinal intolerances \\  (developed)\end{tabular} & 154 (39.9\%) & 166 (58.9\%) & 119 (53.8\%) & 439 (49.4\%) & \textless{}.0001* \\
Bacteremia (developed) & 26 (6.7\%) & 31 (11.0\%) & 29 (13.1\%) & 86 (9.7\%) & 0.0249* \\ \cmidrule(l){1-6}
\multicolumn{6}{l}{$^a$Chi-squared test \textit{p}-value; *significant difference exists in the distributions among the 3 clusters.}\\ 
 \bottomrule
\label{tab:PostHocEDEN2}
\end{tabular} }
\end{table}

%%%%%%%%%%%%%%%%%%%%%%%%%%%%%%%%%%%%%%%%
\begin{table}[!htbp]
\small\center
\caption{Comparison of within-cluster clinical endpoint distributions between full and trophic feeding groups based on BFMM[EEI] clustering results of EDEN trial data.}
\scalebox{0.9}{
\begin{tabular}{lcccccccc}
\toprule	
\multirow{2}{*}{\textbf{}} & \multicolumn{8}{c}{\textbf{Cluster assigned by BFMM[EEI] with $G=3$ and $k=75$}} \\
 & \multicolumn{2}{c}{\textbf{Cluster 1}} & \multicolumn{2}{c}{\textbf{Cluster 2}} & \multicolumn{2}{c}{\textbf{Cluster 3}} & \multicolumn{2}{c}{\textbf{Overall}} \\
\textbf{Cluster size} & \multicolumn{2}{c}{\textbf{386 (43\%)}} & \multicolumn{2}{c}{\textbf{282 (32\%)}} & \multicolumn{2}{c}{\textbf{221 (25\%)}} & \multicolumn{2}{c}{\textbf{889 (100\%)}} \\ \cmidrule(l){1-9}
\textbf{\begin{tabular}[c]{@{}l@{}}Feeding\\ Intervention (n)\end{tabular}} & \textbf{\begin{tabular}[c]{@{}c@{}}Full\\ (181)\end{tabular}} & \textbf{\begin{tabular}[c]{@{}c@{}}Trophic   \\ (205)\end{tabular}} & \textbf{\begin{tabular}[c]{@{}c@{}}Full \\ (141)\end{tabular}} & \textbf{\begin{tabular}[c]{@{}c@{}}Trophic\\ (141)\end{tabular}} & \textbf{\begin{tabular}[c]{@{}c@{}}Full  \\ (115)\end{tabular}} & \textbf{\begin{tabular}[c]{@{}c@{}}Trophic   \\ (106)\end{tabular}} & \textbf{\begin{tabular}[c]{@{}c@{}}Full   \\ (437)\end{tabular}} & \textbf{\begin{tabular}[c]{@{}c@{}}Trophic   \\ (452)\end{tabular}} \\ \cmidrule(l){1-9}
\multicolumn{9}{l}{\textbf{Clinical Endpoint, n (\%)}} \\ \cmidrule(l){1-9}
\begin{tabular}[c]{@{}l@{}}60-day mortality\\ (death)\end{tabular} & \begin{tabular}[c]{@{}c@{}}26 \\ (14.4\%)\end{tabular} & \begin{tabular}[c]{@{}c@{}}38   \\ (18.5\%)\end{tabular} & \begin{tabular}[c]{@{}c@{}}48   \\ (34.0\%)\end{tabular} & \begin{tabular}[c]{@{}c@{}}46   \\ (32.6\%)\end{tabular} & \begin{tabular}[c]{@{}c@{}}25   \\ (21.7\%)\end{tabular} & \begin{tabular}[c]{@{}c@{}}20   \\ (18.9\%)\end{tabular} & \begin{tabular}[c]{@{}c@{}}99   \\ (22.7\%)\end{tabular} & \begin{tabular}[c]{@{}c@{}}104   \\ (23.0\%)\end{tabular} \\
\textit{P} value$^a$ & \multicolumn{2}{c}{\textit{P} = 0.3357} & \multicolumn{2}{c}{\textit{P} = 0.8995} & \multicolumn{2}{c}{\textit{P} = 0.7171} & \multicolumn{2}{c}{\textit{P} = 0.9634} \\ \cmidrule(l){1-9}
\begin{tabular}[c]{@{}l@{}}Gastrointestinal\\ intolerances\\ (developed)\end{tabular} & \begin{tabular}[c]{@{}c@{}}84\\ (46.4\%)\end{tabular} & \begin{tabular}[c]{@{}c@{}}70   \\ (34.1\%)\end{tabular} & \begin{tabular}[c]{@{}c@{}}90   \\ (63.8\%)\end{tabular} & \begin{tabular}[c]{@{}c@{}}76   \\ (53.9\%)\end{tabular} & \begin{tabular}[c]{@{}c@{}}69   \\ (60.0\%)\end{tabular} & \begin{tabular}[c]{@{}c@{}}50   \\ (47.2\%)\end{tabular} & \begin{tabular}[c]{@{}c@{}}243   \\ (55.6\%)\end{tabular} & \begin{tabular}[c]{@{}c@{}}196   \\ (43.4\%)\end{tabular} \\
\textit{P} value$^a$ & \multicolumn{2}{c}{\textit{P} = 0.0187*} & \multicolumn{2}{c}{\textit{P} = 0.1157} & \multicolumn{2}{c}{\textit{P} = 0.0757} & \multicolumn{2}{c}{\textit{P} = 0.0003*} \\ \cmidrule(l){1-9}
\begin{tabular}[c]{@{}l@{}}Bacteremia\\ (developed)\end{tabular} & \begin{tabular}[c]{@{}c@{}}11\\ (6.1\%)\end{tabular} & \begin{tabular}[c]{@{}c@{}}15\\ (7.3\%)\end{tabular} & \begin{tabular}[c]{@{}c@{}}11   \\ (7.8\%)\end{tabular} & \begin{tabular}[c]{@{}c@{}}20   \\ (14.2\%)\end{tabular} & \begin{tabular}[c]{@{}c@{}}14   \\ (12.2\%)\end{tabular} & \begin{tabular}[c]{@{}c@{}}15   \\ (14.2\%)\end{tabular} & \begin{tabular}[c]{@{}c@{}}36   \\ (8.2\%)\end{tabular} & \begin{tabular}[c]{@{}c@{}}50   \\ (11.1\%)\end{tabular} \\
\textit{P} value$^a$ & \multicolumn{2}{c}{\textit{P} = 0.7783} & \multicolumn{2}{c}{\textit{P} = 0.1278} & \multicolumn{2}{c}{\textit{P} = 0.8138} & \multicolumn{2}{c}{\textit{P} = 0.1900} \\ \cmidrule(l){1-9}
\multicolumn{9}{l}{$^a$Chi-squared test \textit{p}-value; *significant difference exists in the distributions between the 2 feeding groups.}\\ 
\bottomrule
\label{tab:PostHocEDEN3}
\end{tabular} }
\end{table}

%\lipsum[2]

%%%%%%%%%%%%%%%%%%%%%%%%%%%%%%%%%%%%%%%%
%%%%%%%%%%%%%%%%%%%%%%%%%%%%%%%%%%%%%%%%

\end{document}